\RequirePackage{fix-cm}
\documentclass{svjour3}                     
\smartqed  

\usepackage{soul}  

\usepackage{graphicx}
\journalname{Space Science Reviews}

\usepackage{amsmath}
\usepackage{amsbsy}
\usepackage{amsfonts}
\usepackage{amssymb}
\usepackage{bm}
\usepackage{placeins}
\usepackage{xcolor}
\usepackage{natbib}
\usepackage{wrapfig}
\usepackage{bbold}
\usepackage{url}
\usepackage{lineno}
\usepackage{upgreek} 

\newcommand{\aap}{Astron.\ Astrophys.}
\newcommand{\apj}{Astrophys.\ J.}
\newcommand{\apjs}{Astrophys.\ J.\ Suppl. Ser}

\newcommand{\apjl}{Astrophys.\ J.\ Lett.}

\newcommand{\jgr}{J.\ Geophys.\ Res.}

\newcommand{\ssr}{Space Sci.\ Rev.}

   \newcommand{\vct}[1]  {\ensuremath{\boldsymbol{#1}}}  

\newcommand{\zpm}{{\bm{z}}^\pm}

\newcommand{\zmp}{{\bm{z}}^\mp}

\newcommand{\zp}{{\bm{z}}^+}

\newcommand{\zm}{{\bm{z}}^-}

\newcommand{\bz}{\bm{z}}
\newcommand{\bu}{ {\bm{U}} }
\newcommand{\bv}{ {\bm{V}}_\mathrm{A} }

 \renewcommand{\epsilon}{\varepsilon}

\usepackage{xspace}
\newcommand{\Helios}{\textit{Helios}\xspace}
\newcommand{\Ulysses}{\textit{Ulysses}\xspace}
\newcommand{\Voyager}{\textit{Voyager}\xspace}
\newcommand{\NewHorizons}{\textit{New Horizons}\xspace}
\newcommand{\voy}[1]{\textit{V{#1}}\xspace}
\newcommand{\ACE}{\textit{ACE}\xspace}
\newcommand{\IBEX}{\textit{IBEX}\xspace}
\newcommand{\NH}{\textit{NH}\xspace}
\newcommand{\SC}{\mathrm{sc}}

\newcommand{\au}{AU\xspace}

\newcommand{\p}{\mathrm{p}}
\newcommand{\A}{\mathrm{A}}
\newcommand{\ci}{\mathrm{ci}}

\begin{document}

\title{Turbulence in the outer heliosphere
}


\author{F. Fraternale  \and
	L. Adhikari \and
    H. Fichtner*\and
	T.~K. Kim \and
	J. Kleimann\and
	S. Oughton \and
	N.~V. Pogorelov \and
	V. Roytershteyn\and
	C.~W. Smith \and
	A.~V. Usmanov \and
	G.~P. Zank \and
	L.-L. Zhao 
}

\authorrunning{Fraternale et al.} 

\institute{Federico Fraternale \at
Center for Space Plasma and Aeronomic Research, The University of Alabama in Huntsville, Huntsville, AL 35805, USA
\email{federico.fraternale@uah.edu}             \and
Laxman Adhikari\at
Center for Space Plasma and Aeronomic Research, The University of Alabama in Huntsville, Huntsville, AL 35805, USA
\and
*Horst Fichtner (corresponding author)\at
Theoretische Physik IV, Ruhr-Universit\"at Bochum, 44780 Bochum, Germany, 
\email{hf@tp4.rub.de}
\and
Tae K. Kim\at
Center for Space Plasma and Aeronomic Research, The University of Alabama in Huntsville 
Huntsville, AL 35805, USA
\and
Jens Kleimann\at
Theoretische Physik IV, Ruhr-Universit\"at Bochum, 44780 Bochum, Germany
\and
Sean Oughton\at
Department of Mathematics, University of Waikato, Hamilton 3240, New Zealand
\and
Nikolai V. Pogorelov \at
Center for Space Plasma and Aeronomic Research and Department of Space Science, The University of Alabama in Huntsville, Huntsville, AL 35805, USA
\and
Vadim Roytershteyn\at
Space Science Institute, Boulder, CO, USA 
\and
Charles W. Smith\at
Institute for the Study of Earth, Oceans, and Space, Morse Hall, University of New Hampshire, Durham, New Hampshire, USA
\and
Arcadi V. Usmanov\at
Department of Physics and Astronomy, Bartol Research Institute, University of Delaware, Newark, DE 19716, USA
\and
Gary P. Zank\at
Center for Space Plasma and Aeronomic Research and Department of Space Science, The University of Alabama in Huntsville,
Huntsville, AL 35805, USA
\and
Lingling Zhao\at
Center for Space Plasma and Aeronomic Research, The University of Alabama in Huntsville,
Huntsville, AL 35805, USA
}

\date{Accepted: July 1, 2022}
\maketitle
\begin{abstract}
The solar wind (SW) and local interstellar medium (LISM) are turbulent media. Their interaction is governed by complex physical processes and creates heliospheric regions with significantly different properties in terms of particle populations, bulk flow and turbulence. 
Our knowledge of the solar wind turbulence nature and dynamics mostly relies on near-Earth and near-Sun observations, and has been increasingly improving in recent years due to the availability of a wealth of space missions, including multi-spacecraft missions.  In contrast, the properties of turbulence in the outer heliosphere are still not completely understood. In situ observations by \Voyager and \NewHorizons, and remote neutral atom measurements by \IBEX strongly suggest that turbulence is one of the critical processes acting at the heliospheric interface. It is intimately connected to charge exchange processes responsible for the production of suprathermal ions and energetic neutral atoms.
This paper reviews the observational evidence of turbulence in the distant SW and in the LISM, advances in modeling efforts, and open challenges.

\keywords{Turbulence \and Solar wind \and Interstellar medium \and Heliosphere}
\end{abstract}



\section{Introduction}\label{sec:intro}
Turbulence is a critical player in the interaction of the solar wind (SW) with the local interstellar medium (LISM). Arguably, it can be considered as one of the most fundamental processes because of the vast range of scales involved and its ability to mediate various heliospheric events.  
For example, it plays a fundamental role in the SW acceleration at the solar corona. Moreover, it has long been recognized that turbulence and wave-particle interactions serve as the sources of effective viscosity and resistivity in weakly-collisional and magnetized SW plasma \citep{coleman1968,parker1969,griffel1969}. This makes it possible to study large-scale SW flows in the magnetohydrodynamic (MHD) formulation,  and investigate turbulence through concepts and analytic tools derived from hydrodynamics \citep[e.g.,][]{karman1938,taylor1938,kolmogorov1941a,kolmogorov1962,obukhov1962,monin_book,frisch1995}. Turbulence provides efficient channels for cross-scale transfer of energy injected either via large scales gradients or instabilities and for  energy dissipation on sub-ion scales. It affects the transport and acceleration of suprathermal particles and cosmic rays (CRs), the properties of neutral atoms detected at 1\,\au, and the structure of discontinuities. It is clear that turbulence must be taken into account in order to explain the observed thermodynamic properties of the distant SW flow, especially its non-adiabatic radial temperature profile.  It can be argued that the presence of turbulence and associated dissipation processes, including magnetic reconnection (inseparable from turbulence), affect the shape of the heliosphere on the global scale. 

A number of comprehensive reviews focus on theoretical aspects of the SW turbulence and related near-Earth observations \citep[e.g.,][]{parker1969,jokipii1973,tu1995,schlickeiser2002book,biskamp2003book,zhou2004,matthaeus2011, bruno2013, alexandrova2013,zank2014book,oughton2015,chen2016,beresnyak2019book,smith2021,lazarian2020}. Some other reviews address the astrophysical implications of turbulence \citep[e.g.,][]{ferriere2001, elmegreen2004, scalo2004}, see also the paper of \citet{Linsky_EA_this_volume} in this volume. The purpose of this review is to discuss the manifestations of turbulence in the outer heliosphere
  (beyond $\sim 10 $\,\au)
and very local interstellar medium (VLISM), its role in the SW--LISM interaction, and the major challenges that need to be addressed in the future.

The SW--LISM interaction creates a tangential discontinuity (the heliopause, HP). Deceleration of the supersonic SW due to the presence of the HP and counter pressure in the tail creates a heliospheric termination shock (HTS). The details of the global SW--LISM interaction determine the existence of a bow shock (BS) or a bow wave (BW) in front of the HP \citep{baranov1979,holzer1989}. The HTS plays a crucial role in the transmission and amplification of turbulence from the supersonic SW region into the inner heliosheath (IHS, the region between the HTS and the HP). Since the LISM is  weakly ionized, interstellar neutral atoms can penetrate deep into the heliosphere, where they experience charge exchange with the SW ions. As a result, non-thermal, pickup ions (PUIs) and secondary neutral atoms are born \citep[e.g.,][]{moebius1985}. The latter, especially H atoms born in the supersonic SW, which are often referred to as the neutral SW, can propagate back into the LISM and modify its properties by heating and decelerating ions in it  \citep{gruntman1982}, out to hundreds of \au from the Sun. 

The part of the LISM affected by the presence of the heliosphere, regardless of what processes are involved and which quantities are affected (LISM plasma, magnetic field, or CR fluxes), is now commonly called the very local interstellar medium (VLISM) \citep{zank2015, zhang2020, fraternale2021a}. The VLISM can extend to hundreds of AU into the LISM upwind direction and to thousands of AU into the heliotail and directions perpendicular to it \citep{zhang2020}.  
Observational data and numerical simulations indicate that the VLISM region that extends $\sim $300\,\au in roughly the nose direction is highly dynamic. 
It is also characterized by the presence of the enhanced neutral H and He densities, relatively strong gradients, interstellar magnetic field (ISMF) draping around the HP, enhanced turbulence, propagating shocks, CR flux anisotropy, and kinetic wave activity.  This region is often referred to as the outer heliosheath (OHS), and is the likely site where ENAs creating the \IBEX ribbon are generated.

There is an intimate coupling between the SW and the LISM through charge-exchange and turbulence. In fact, the waves driven by instabilities of the PUI distribution function strongly contribute to production of magnetic turbulence, which heats up the SW beyond $\sim$10\,\au \citep{wu1972,vasyliunas1976}. Multiple demonstrations of this are provided by the turbulence transport models and cascade rates computed  on the basis of the MHD extensions of the Navier-Stokes theory. Shears, shocks, and coherent structures also play important roles, thus creating multiple heating mechanisms. The canonical heliospheric current sheet (HCS) topology \citep{parker1958a} is disrupted by the turbulent motions and magnetic reconnection, and no longer exists in the  SW beyond 10\,\au \citep[e.g.,][]{burlaga2002}. Turbulence in the outer heliosphere coexists with compressible and incompressible waves, large-scale coherent structures, remnants of random fluctuations of solar origin, locally generated turbulence due to microinstabilities and, possibly, magnetic reconnection. Arguably, all these phenomena can be considered as part of the ``turbulence'' manifestations in the outer heliosphere. The complexity of turbulence dynamics reveals itself through the thermodynamic dominance of suprathermal ions, partial ionization and Coulomb collisionality of the VLISM, and the effect of heliospheric boundaries. 

Much of our current understanding of turbulence in the outer heliosphere and VLISM relies on the \textit{Voyager} (\voy1, \voy2), \textit{Interstellar Boundary Explorer} \IBEX, and \NewHorizons (\NH) missions. Launched in 1977, \voy1 and \voy2 crossed the HP at $\sim $120\,\au (in 2012 and 2018, respectively), and continue to provide us with unique in situ measurements in the heliosheath and VLISM.

Difficulties in the study of turbulence in these regions stem from the fact that one-dimensional in situ data cover just a tiny fraction of space, while the instruments onboard \Voyager were not specifically designed to study turbulence in the outer heliosphere. Moreover, combined magnetic field and PUIs measurements are not available. In spite of this, the \Voyager exploration resulted in truly remarkable discoveries, most of which being summarized in the papers constituting this volume. One of them is the observation of compressible turbulence in the heliospheric boundary regions \citep{burlaga2006,burlaga2015}.  The heliospheric community is eagerly expecting exciting new results before \textit{Voyagers} lose their contact with Earth.

The properties of turbulence in the outer heliosphere are very much different from those in the near-Earth environment. For this reason, further observational and theoretical studies are expected to shed light onto the nature of turbulence in space plasmas.

The review is organized as follows. Section~\ref{sec:supersonicSW} is focused on turbulence in the distant, supersonic SW and summarizes the principal methods of its analysis. Section~\ref{sec:transport_models} gives an overview of the turbulence transport models and their predictions, and describes the efforts undertaken to couple them with global, 3D  models.
Section~\ref{sec:TS}  describes turbulence and magnetic structures observed by \voy2 across the HTS and their implications for the transport of energetic particles. The observational evidence and our current understanding of turbulence, time-dependent structures and related scales in the IHS and VLISM are reviewed in Sections~\ref{sec:IHS} and~\ref{sec:VLISM}, respectively.  Finally, Section~\ref{sec:reconnection} provides a brief overview of the as yet unsolved problems involving  magnetic reconnection in these regions. Our conclusions are formulated in Section \ref{sec:conclusions}.


\section{Evidence of turbulence in the distant supersonic solar wind}\label{sec:supersonicSW}
The turbulent dynamics of the solar wind beyond 10\,\au can best be described as an evolution of what is seen at 1\,\au with the addition of a significant driving source in the form of waves excited by newborn interstellar pickup ions (PUIs). 
We can examine what has been learned from studies of 1\,\au observations as they form the bulk of solar wind turbulence studies for the purpose of obtaining a better understanding of turbulence beyond 10\,\au.

Turbulent dynamics is studied via data analysis using four separate analysis methods, i.e. (i) comparing the form of the power spectrum to predictions, (ii) third-moment predictions for the cascade rate, (iii) comparison of both above rates to spatial gradients and/or transport theory, and (iv) multi-s/c techniques.
 
Predictions for the power spectrum based on specific theories of the nonlinear dynamics result in a testable prediction as well as an associated energy cascade rate through the inertial range that equals the heating rate. 
This can be augmented by other spectral analyses that are related to features such as helicity, polarization, etc. 
Single-spacecraft third-moment calculations are based on fewer assumptions of the underlying dynamics and give a rate of energy cascade that is largely independent of dynamics apart from an assumption of geometry (the statistical distribution of wave vectors in 3D space). 
Comparison of the computed energy cascade rate to the rate of heating as obtained either from statistical spatial gradients of the plasma temperature or transport theory yield a measure of the correctness of the computed energy cascade rate. 
Multi-spacecraft techniques that include, but are not limited to, $k$-telescope methods attempt to resolve the underlying distribution of wave vectors and assign dynamics to their evolution.

\begin{figure}[t]
	\centering
	\includegraphics[width=0.7\columnwidth]{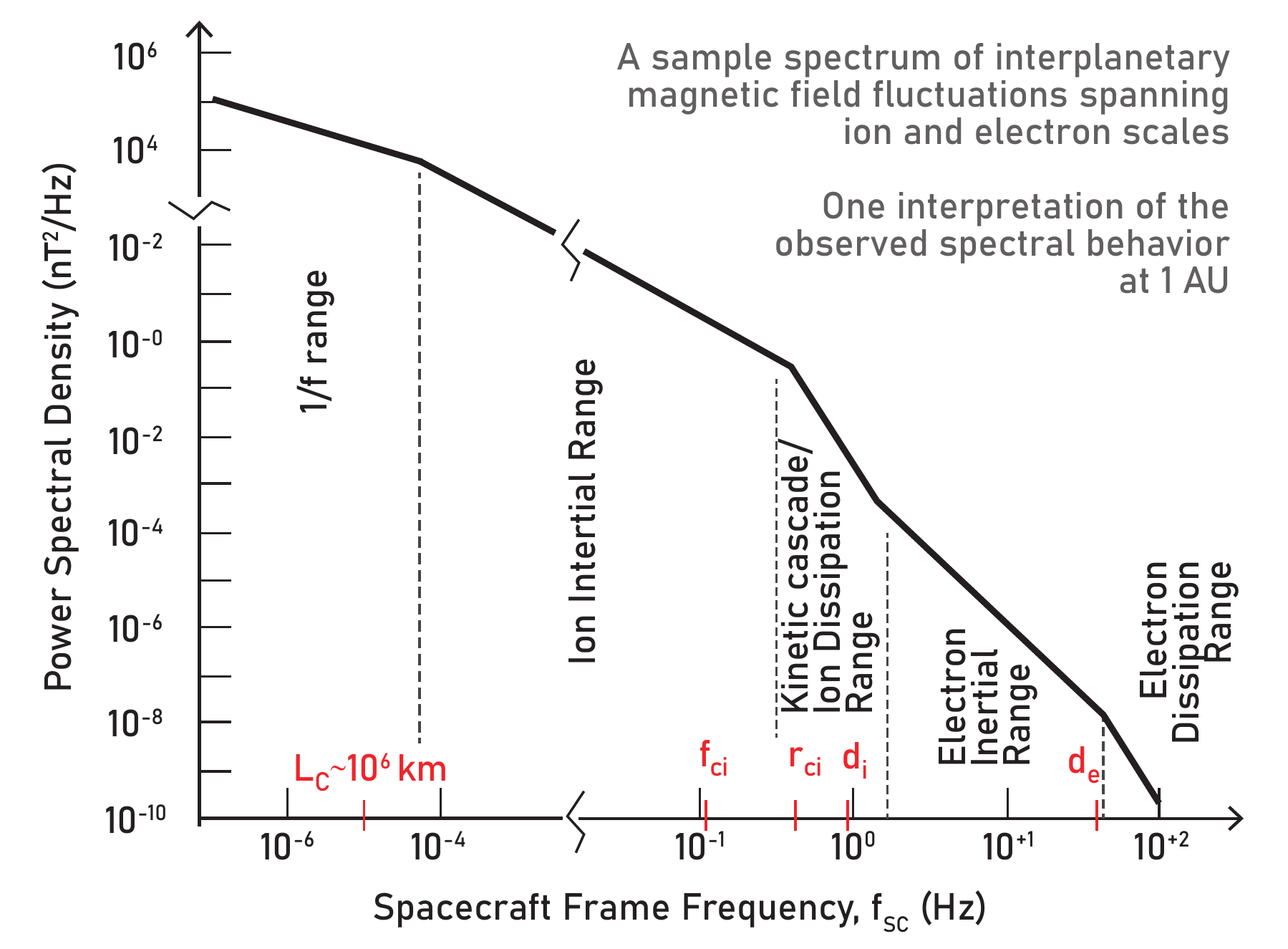}\\
	\caption{Schematic illustration of the magnetic spectrum of SW turbulence at 1\,\au \citep{smith2012}. {The correlation scale ($L_\mathrm{C}$, proton gyrofrequency ($f_\mathrm{ci}$), proton Larmor radius ($r_\mathrm{ci}$) and the proton and electron inertial lengths, ($d_\mathrm{i}$ and $d_\mathrm{e}$, respectively)  derived from the Doppler-shift relation are also shown.} \label{fig:spec}} 
\end{figure}
The power spectral density (PSD --- or simply the spectrum, $P$) of SW turbulence can be described as having five subranges with (approximate)
power law behavior $P\sim f^\alpha$, where $\alpha$ is the spectral index. 
Figure~\ref{fig:spec} illustrates this. 
At the lowest frequencies $f_{\SC} < f(L_\mathrm{C})$ where $f_{\SC}$ is the frequency as measured in the spacecraft frame and $L_\mathrm{C}$ is the correlation scale \citep{matthaeus1999c,smith2001,smith2006b}, the spectrum consists of unprocessed energy that originates in the acceleration region \citep{matthaeus1983,matthaeus1986}. By definition, the lifetime of these fluctuations is longer than the age of the SW plasma (causality condition).
At the intermediate frequencies $f(L_\mathrm{C}) < f_\SC < f(L_\mathrm{D})$ where $L_\mathrm{D}$ is the scale where dissipation sets in {- typically the larger of the Larmor radius and the ion inertial length -}  \citep{leamon1998a,markovskii2008,smith2012,woodham2018,pine2020a},
the evolution of the fluctuations is (almost) energy-conserving and the nonlinear dynamics of the turbulence remakes the energy so as to transport the energy to smaller scales \citep{kolmogorov1941a,matthaeus1982,leamon1998a}. 
This is the so-called inertial range where the fluctuations are unpolarized and the power spectra indices are reproducible \citep{matthaeus1982,pine2020b}. 
The ion ``dissipation range'' is described as $  f_\SC > f(L_\mathrm{D})  $ and at 1\,\au it is generally characterized as a steepening of the power spectrum starting at $f_\SC \simeq 0.2$\,Hz. {Various terminologies have been used in the literature. Our notation refers to the onset of dissipation effects at sub-ion scales. It is consistent with recent results that turbulent energy conversion into internal energy turns on at sub-ion scales, in weakly collisional plasmas \citep[e.g.,][and references therein]{matthaeus2020, matthaeus2021} and it agrees with observations. It has been shown that the rate of heating thermal protons in the SW is in good agreement with the rate of energy transport through the spectrum. What we know is that the spectrum steepens at a predictable scale,  and  that the polarization then changes in keeping with resonant dissipation removing one of the polarizations. Observations suggest that there must be dissipation of most of the transported turbulent energy in order to match in situ heating.  However, the cross-scale transfer is not precluded in the kinetic regime. In fact cascade phenomenology in the ion kinetic regime is predicted by a number of studies \citep[e.g.,][]{howes2011}. However, many aspects of how the cascade operates in the kinetic regime are unknown, especially in the outer heliosphere. }
Interestingly, the steepness of the ion dissipation spectrum depends on the strength of the inertial range energy transport \citep{smith2006a} and is generally absent beyond $\sim 2$\,\au \citep{pine2020a}.
At still higher frequencies there is an electron inertial range where the dynamics are supported by the thermal electrons until dissipation occurs 
\citep{bale2005,alexandrova2008,alexandrova2009,alexandrova2012}. 

\begin{figure}
	\centering
	\includegraphics[width=\columnwidth]{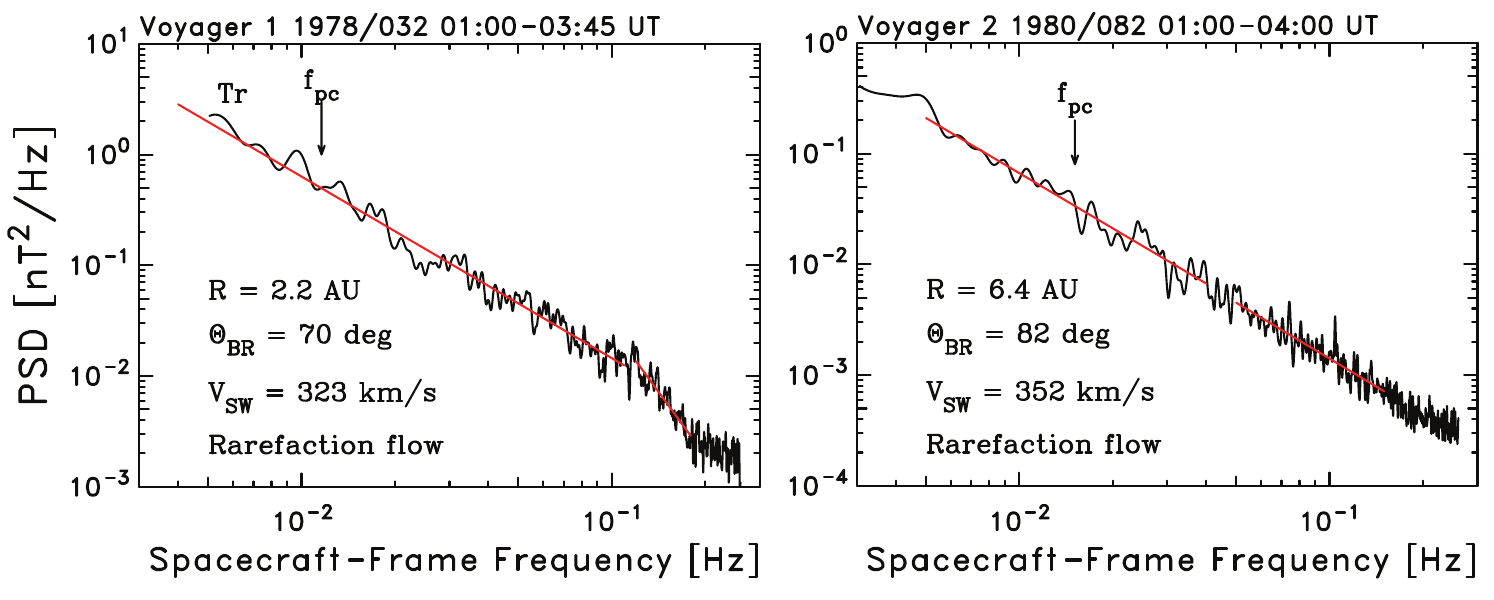}
	\caption{Two examples of solar wind magnetic turbulence power spectra \citep{pine2020a}.	(left) Example showing the spectral break associated with dissipation. The apparent flattening at $\sim 0.2$\,Hz is more likely due to noise in the data rather than the onset of the electron inertial range. (right) Example showing the absence of a spectral break as is most often seen in spectra from beyond $\sim 2$\,\au.}
	\label{fig:examples}
\end{figure}
Figure~\ref{fig:examples} shows two typical examples of magnetic power spectra from the \Voyager~1 \& 2 spacecraft at 2.2 and 6.4\,\au, respectively. 
Figure~\ref{fig:examples} (left panel) shows an example where the spectral break marking the onset of dissipation is evident at $\sim 0.12$\,Hz. 
The flattening of the spectrum at $\sim 0.2$\,Hz is most likely due to noise in the data rather than the onset of the electron inertial range. Figure~\ref{fig:examples} (right panel) shows an example where the spectral break is not observed. 
Analysis suggests that the onset of dissipation is resolved within the frequency range shown, but the spectral transport of energy through the inertial range is too weak to result in a observable break in the power spectrum when dissipation sets in. 

As is frequently seen at 1\,\au, the onset of dissipation is characterized by a bias of the polarization that can be interpreted to be a measure of the role of resonant dissipation \citep{leamon1998b,hamilton2008,pine2020a} and an increase in the relative amount of compressive fluctuations as measured by the spectrum of the magnetic field magnitude and parallel component.

\begin{figure}
	\sidecaption
	\resizebox{0.7\hsize}{!}{
		\includegraphics{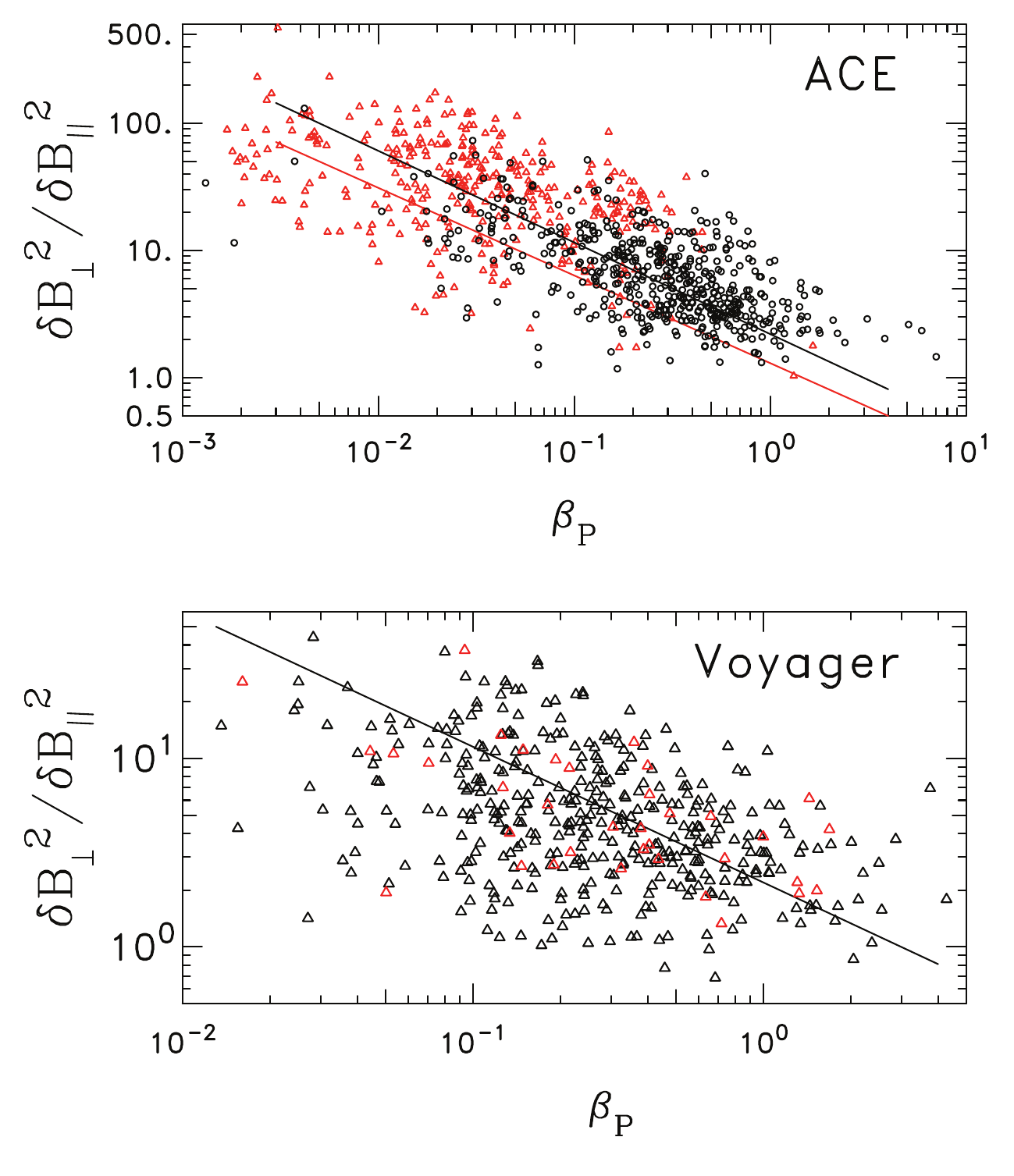}}  
	\caption{Analysis of the magnetic fluctuation anisotropy as studied by \citet{belcher1971} that is the ratio of the power spectral density matrix summing the two components 	perpendicular to the mean field over the component parallel to the mean field.  See text for details. 
		(top) Analysis of magnetic spectra from 960 intervals of \ACE data with black (red) symbols representing undisturbed (magnetic cloud) field lines. (bottom) Analysis of spectra from 438 intervals of \voy1 and \voy2 observations spanning launch in 1977 through 1990 and 1 to 45\,\au. Reproduced from \citet{smith2006c} and \citet{pine2020c}. \vspace{25pt}}
	\label{fig:Pine3a}
\end{figure}

There are many properties of solar wind turbulence beyond 1\,\au that are consistent with results obtained using 1\,\au observations, including the observation of inertial range spectral indices falling within a consistent range of values.   Broadband power spectra of plasma, magnetic field, and helicities have been computed in the MHD regime at 5\,\au and 20\,\au using \voy2 MAG and PLS data \citep{fraternale2016,gallana2016,iovieno2016, fraternale2017phd}.
Special care has to be taken when analyzing and interpreting \Voyager time series in the distant SW, due to the large fraction of missing data and the large noise-to-signal ratio at frequencies near ion scales.   The cross and magnetic helicity analysis of \voy2 data at 20\,\au by \citet{iovieno2016} confirmed a tendency towards the reduction of cross helicity, consistent with model predictions of \citet{matthaeus2004} and observations of \citet{roberts1987b}. 
At large inertial scales ($8.5\times10^{-7}\,\mathrm{Hz}\lesssim f_\SC\lesssim10^{-5}$ Hz)
speed fluctuations spectral indexes were found by \citet{burlaga2003c} to drop from the Kolmogorov-like value ($\alpha=-5/3$) at 5\,\au to $-2.5 \lesssim \alpha \lesssim -1.7$ at $15\lesssim R \lesssim 35$\,\au. This is the region where corotating interaction regions (CIRs) merge producing CMIRs where the plasma features pressure jumps of broad angular extent, shocks, and shock-like structures.	Interestingly, the spectral index further decreases at larger distances  $40\lesssim R\lesssim90$\,\au, which was associated with the observed slowly evolving jump-ramp profiles of the SW speed. Consistently, the spectral indexes of magnetic field fluctuations in the same frequency range were found to vary between -1.8 and -2.5 \citep{burlaga2003b}. 
		
Looking at smaller scales in the inertial regime,  magnetic field structures with quasi-2D, filamentary topology and multifractal statistics of increments are ubiquitous in the supersonic SW \citep[e.g.,][]{burlaga2001,burlaga2004}. As discussed by \citet{voros2006}, both the local interaction and the cross-scale interaction between these structures and shocks play an important role in the dynamics of SW turbulence and the 
evolution of intermittency.
In the inertial and dissipation regimes of turbulence, the distribution of magnetic field increments ($\Delta B(t) = B(t)-B(t+\tau)$) is not Gaussian  at frequencies higher than about the solar rotation frequency \citep[e.g.,][]{marsch1994,sorriso1999}. Indeed, the q-Gaussian distribution \citep[Gaussian core and fat power-law tails,][]{tsallis1988} was found to excellently fit the data \citep{burlaga2007b}, and is associated with intermittent behavior. A remarkable feature of turbulence in the distant SW is the significant decrease of intermittency observed by \citet{burlaga2007b} at $\sim$60\,\au, (see details in \citet{Richardson_EA_this_volume}, this volume), 
{and recently further investigated by \citet{parashar2019} and \citet{cuesta2022}. In these later papers the observed reduction of the small-scale intermittency of magnetic field increments with distance, out to 10\,\au, is associated with the decreasing bandwidth of the inertial range with distance (effective Reynolds number scaling as $R^{-2/3}$).}

Another consistent property is the apparent dependence of the fluctuation anisotropy upon both the ambient plasma parameters, spectral intensity, and other turbulence properties 
\citep{leamon1998a,smith2006a,hamilton2008,macbride2010,pine2020c}. 
There are two forms of fluctuation anisotropy that provide insight into the underlying dynamics. 
The first is the  {spectral ratio of $P_{xx}(f)$ and $P_{yy}(f)$, the two  diagonal components of the PSD perpendicular to the mean field.  In a powerlaw region of the spectrum this ratio varies with the angle of mean field to the radial (observation) direction and provides insight into the fraction of energy associated with parallel and perpendicular wave vectors \citep{bieber1996}.}
Efforts to extend this method beyond 10\,\au have not been satisfactory \citep{pine2020c}. 
The second {type of} anisotropy is the ratio of the energy of the perpendicular component relative to the parallel component that is a measure of the relative content of compressive fluctuations \citep{belcher1971}.

\begin{figure}
	\sidecaption
	\resizebox{0.717\hsize}{!}{\includegraphics{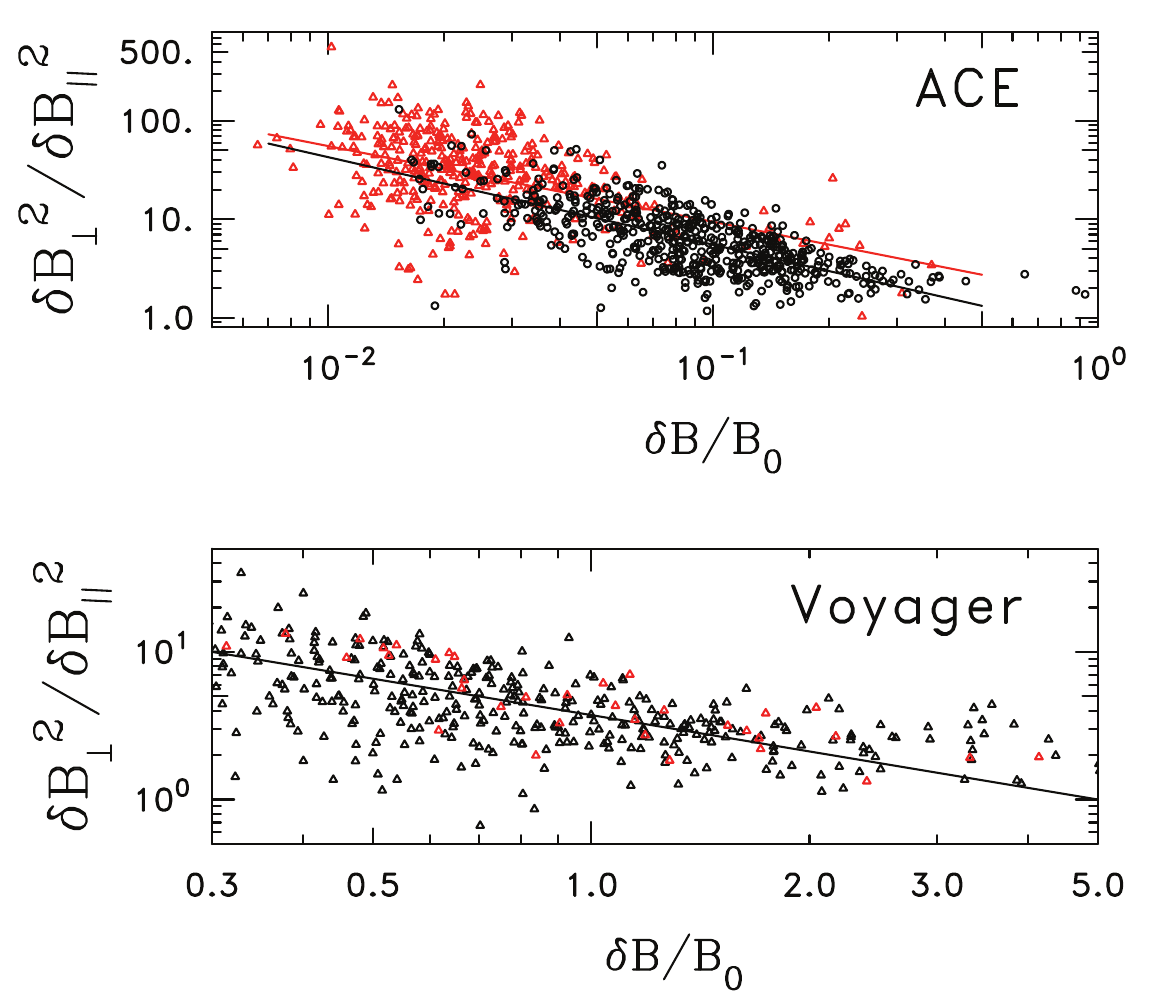}} 
	\caption{Analysis of same \ACE and \Voyager observations as were used in Fig.~\ref{fig:Pine3a}. Plot of fluctuation anisotropy as a function of the ratio of the square root of the 	power spectrum averaged over the same frequency ranges to the magnitude of the mean magnetic field. Reproduced from \citet{smith2006c} and \citet{pine2020c}.\vspace{30pt}}
	\label{fig:Pine3b}
\end{figure}

Figure~\ref{fig:Pine3a} (top) shows the latter form of spectral anisotropy computed for 960 intervals of \ACE observations at 1\,\au averaged over the frequency range $8\,{\rm mHz} < f_\SC < 100\,{\rm mHz}$. 
Figure~\ref{fig:Pine3a} (bottom) shows the spectral anisotropy computed for 438 intervals of \voy1 and \voy2 observations averaged over the spacecraft-frame frequency range $5\,{\rm mHz} < f_{sc} < 0.8~ f(10 d_{\rm i})$ where $d_{i} = V_\A / \omega_\ci=c/\omega_{\rm pi}$ is the ion inertial scale ($V_\A = B_0 (4 \pi \rho)^{-1/2}$ is the Alfv\'en speed, $B_0$ is the mean field strength, $\rho$ is the thermal proton mass density, and $\omega_\ci,\omega_{\rm pi}$ are the proton cyclotron and plasma frequencies, respectively). 
This analysis has its roots in the general behavior of compressive waves. 
At the same time, a strong scaling of the same anisotropy is seen as a function of the ratio of the fluctuation level to the mean field strength. 
Figure~\ref{fig:Pine3b} shows the analysis of the same \ACE and \Voyager spectra as a function of the square root of the integral of the power spectrum over the prescribed frequency range (the fluctuation amplitude) divided by the mean field strength. 
This analysis has its roots in nearly incompressible turbulence theory 
\citep{zank1992a,zank1992b,zank1993}. 
At present there is no resolution to the ambiguity presented in these two figures as both the proton beta and $\delta B/B_0$ are themselves strongly correlated. 
However, they point to the fundamental physics on which the nonlinear dynamics of turbulence is built.

Analyses based on the use of data from a single spacecraft suffer from uncertainty derived from the necessary application of the Taylor Frozen-In-Field assumption \citep{taylor1938}. 
This results in ambiguity regarding the actual orientation of the wave vector as it projects onto the SW velocity. 
The continuum of wave vectors that possess the same projection represent significantly different nonlinear dynamics leading to ambiguity in the resulting analysis. 
Assumptions are often made based on characteristics of the spectrum \citep{bale2005,alexandrova2008} and the resulting interpretations are often debated, but this represents a major source of uncertainty in many SW turbulence analyses. 
There are statistical tests that attempt to address this question using single-spacecraft data 
\citep{bieber1996,leamon1998a,dasso2005,hamilton2008,pine2020c}, but there also exist competing dynamics that can mask the effects of the nonlinear dynamics. 

{The leading theories for the inertial-range total energy power spectrum include the $k_\perp^{-3/2}$ and the $k_\perp^{-5/3}$ scaling for wavenumbers perpendicular to the mean magnetic field. The former is commonly referred to as the Iroshnikov--Kraichnan (IK) scaling due to their seminal works \citep{iroshnikov1964,kraichnan1965}. Using weak turbulence arguments, they first recognized the importance of the large-scale magnetic field in the turbulence dynamics, i.e., the role of interacting Alfv\'en-wave packets propagating in opposite directions. The rigorous weak turbulence theory predicts a $k_\perp^{-2}$ scaling \citep{galtier2000,schekochihin2012}, but the IK spectral scaling is recovered for the case of strong, globally isotropic, Alfv\'enic turbulence, if the time scale that determines the energy transfer is given by the Alfv\'en time scale. Naturally, the original isotropy assumption can be difficult to justify when there is a mean field and the IK model is no longer heavily used for these cases. The $-5/3$ scaling  \citep[Kolmogorov-type, after][]{kolmogorov1941a} arises in MHD in strong anisotropic turbulence scenarios when the nonlinear timescale determines the energy transfer \citep[][GS]{goldreich1995}.  Later theoretical developments for strong turbulence \citep{boldyrev2005,boldyrev2006} attempted to reconcile GS-style arguments with a $-3/2$ scaling, since both were claimed to be observed in data and numerical simulations  \citep[e.g.,][]{maron2001}, despite the undeniable difficulty to discriminate between them. The model of \citet{boldyrev2006} is based on a scale-dependent ``dynamic alignment'' of the polarizations of magnetic- and velocity-field fluctuations, according to which both the $-5/3$ and $-3/2$ scaling can be obtained, depending on the anisotropy level. We note that the Boldyrev strong turbulence $-3/2$ scaling is unrelated to the IK scaling (or at least not directly related).  Additionally, a (distinct) $-3/2$ scaling is also derived for the fast-mode cascade in compressible MHD turbulence \citep[e.g.,][]{cho2002}. Though observations and more recent simulations tend to favor the $-5/3$ scaling \citep[e.g.,][]{beresnyak2012}, the question is still debated \citep[see the extensive reviews of][]{zhou2004,beresnyak2015}, and many papers report one or the other in their study of specific events.
}
 
The energy cascade rates vary significantly between the theories within this range of spectral predictions 
\citep{leamon1999,vasquez2007,smith2009,matthaeus2011}.
One example of such predictions is the MHD generalization of traditional hydrodynamics 
\citep{kolmogorov1941a,leamon1999}:
\begin{equation}
E(k) = C_{\rm K} \epsilon^{2/3} k^{-5/3}
\label{eq:kolpower}
\end{equation}
where $E(k)$ is the total  inertial range power spectrum (magnetic + kinetic), $\epsilon$ is the rate of turbulent energy  transport through the inertial range, and $k$ is the wave vector magnitude. 
Evaluation of the constant $C_K$ to reach agreement with SW observations results in an expression for $\epsilon$: 
\begin{equation}
\epsilon_{\rm K} = \frac{f^{5/2} [E(f)]^{3/2} \cdot (21.8)^3}{U N_\p^{3/2}}
\label{eq:koleps}
\end{equation}
where $E(f)$ is the measured power spectrum as a function of frequency, $U$ is the bulk SW speed in units of km\,s$^{-1}$, and $N_\p$ is the proton number density in units of cm$^{-3}$. The above expression has roots in the hydrodynamic analog first used by  \citet{kovasznay1948}, and is based on dimensional arguments. 

When applying these theories to inferred heating rates, it is assumed that the energy that passes through the inertial range is converted into heat by various kinetic processes with the bulk of the energy going into thermal protons and a 
  smaller fraction
being absorbed by heavy ions and thermal electrons. 
Using the published radial dependence of \Helios thermal proton observations, \citet{vasquez2007} concluded that the MHD extensions of hydrodynamic theory 
\citep{kolmogorov1941a,leamon1999,matthaeus2011} 
provided a better description of the observed heating rates from 0.3 to 1\,\au once the universal constant was adjusted. 
They also obtained the following scaling for the average thermal proton heating rate as a function of heliocentric distance: 
\begin{equation}
	\epsilon_V \equiv \left(5.3 \times 10^{-5}\right) U T_\p / R_\au
\label{eq:epsv}
\end{equation}
in units of J\,kg$^{-1}$\,s$^{-1}$,
$T_\p$ is the temperature of thermal protons in Kelvin, and $R_\au$ is the heliocentric distance in \au.

Third moments, or third-order structure function, theory originates with hydrodynamics \citep{kolmogorov1941c}. 
The great advantage of applying this formalism to MHD is that the derivation and application does not make use of any particular model of the dynamics \citep{politano1998b,politano1998a}. 
In this way, it provides a formulation for the rate of energy cascade through the inertial range that is independent of any specific turbulence model. 
Instead, the MHD equations are combined with assumptions regarding compressibility, stationarity, homogeneity, and underlying geometry (i.e.  rotational symmetry). 
The combined assumptions of incompressibility, stationarity, and homogeneity along with specific assumptions of geometry produce expressions that are general and applicable to data from single spacecraft without further assumption 
\citep{politano1998a,politano1998b,macbride2005,sorriso2007,marino2008,marino2011,marino2012,macbride2008,stawarz2009,coburn2012,coburn2014,coburn2015,hadid2017,smith2018a}. 
Third-moment theory does not require specific power spectral forms or the assumption of a detailed nonlinear dynamic. 
Compressibility leads to expressions that include some terms that cannot be evaluated using single spacecraft data \citep{galtier2008,carbone2009,hadid2017,hellinger2018}. 
The de K{\'{a}}rm{\'{a}}n--Howarth  equation derived for incompressible MHD reads \citep{politano1998a}: 
\begin{equation}
{\bm \nabla_{\bm\ell} \cdot D_3^{\pm}}({\bm\ell}) = -4 \epsilon^{\pm} 
\label{eq:deldot}
\end{equation}
where 
\begin{eqnarray}
{\bm  D_3^{\pm}}({\bm\ell})&\equiv&\langle|\Delta {\zpm}|^2\Delta {\zmp}\rangle,\\
\Delta {\zpm} &\equiv& {\zpm}({\bm x}+{\bm \ell}) - {\zpm}({\bm x}). 
\end{eqnarray}
Here $\zpm = {\delta\bm  u} \pm {\delta\bm B}/\sqrt{\mu_0 \rho}$ are the Els\"asser field fluctuations \citep[][with ${\bm u}$ the plasma speed,
${\bm B}$ the magnetic field, 
$\rho$ the proton mass density, 
$\mu_0$ the magnetic permeability, 
and ${\delta\bm u} = {\bm u} - \langle {\bm u} \rangle$, etc]{elsasser1950} 
and $\epsilon^{\pm}$ denotes the rate of cascade of     $({ \bm z^{\pm}})^2$. 
The total energy cascade rate per unit mass, $\epsilon^T$, is given by 
\begin{equation}
  \epsilon^T = (\epsilon^+ + \epsilon^-) / 2. 
\label{eq:epstotal}
\end{equation}
When an assumed distribution of wave vectors 
 (e.g., 1D, 2D, or isotropic)
is applied to the divergence ${\bm \nabla_\ell \cdot}$, expressions are obtained that are applicable to single spacecraft observations. 
Assuming isotropy, 
  for example, 
we can write:
\begin{equation}
\left\langle \Delta z^{\mp}_\mathrm{R}(\tau) \sum_i^3 \left[ \Delta z^{\pm}_i(\tau) \right]^2 \right\rangle 
        = +(4/3) \epsilon_\mathrm{ISO}^{\pm} U \tau 
\label{eq:d3MHDt}
\end{equation}
where $\mathrm{R}$ denotes the radial component, $\tau$ is the time lag in the data, and $\langle \bullet \rangle$ denotes the ensemble average.

\begin{figure}
	\centering
	\includegraphics[width=0.67\columnwidth]{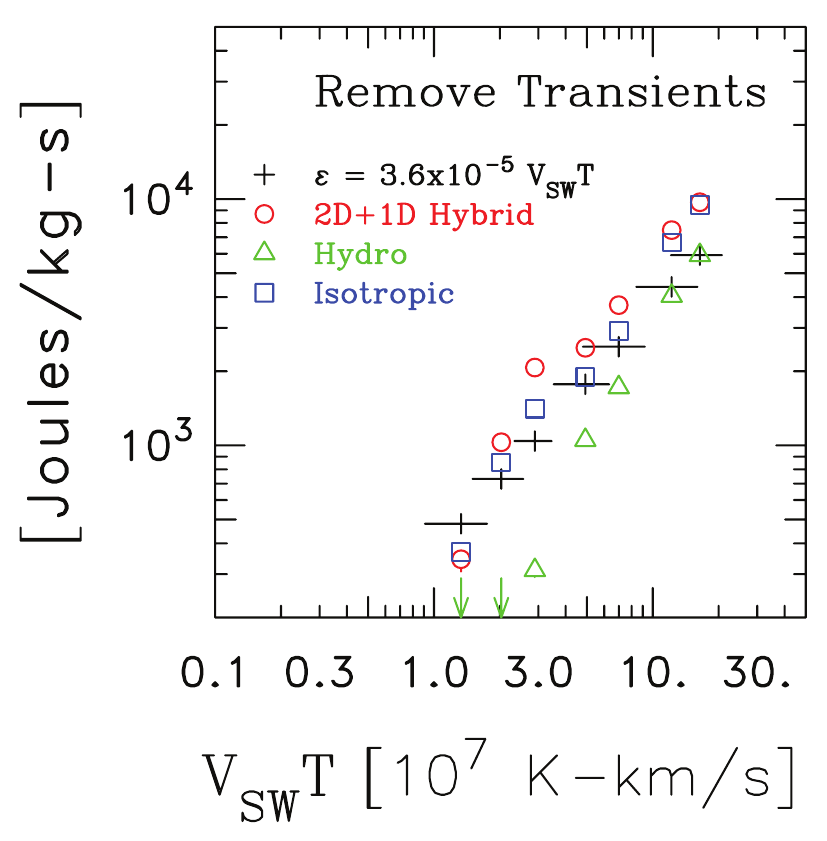}
	\caption{Rate of energy transport through the inertial range as computed at 1\, \au by 
		applying third-moment theory to 10 years of data from the \ACE spacecraft \citep{stawarz2009}. 
		Both the isotropic and hybrid (2D+1D) geometries are used along with the hydrodynamic 
		third-moment expression and the empirical scaling of \citet{vasquez2007}.}
	\label{fig:Stawarz}
\end{figure}

Third-moment theory has been shown to accurately reproduce 1\,\au heating rates in both the compressible and incompressible forms. 
Figure~\ref{fig:Stawarz} is taken from \citet{stawarz2009} showing the result of analyzing 10 years of \ACE data and compares the average energy spectral transfer rate (assumed to be the thermal proton heating rate) using the third-moment formalism under the isotropic and combined 2D+1D hybrid geometry assumptions and compares these results to the 
   \cite{vasquez2007}
prediction of Eq.~(\ref{eq:epsv}) and to the familiar hydrodynamic form.

Third-moment analyses using spacecraft data beyond 10\,\au are ongoing, but unpublished at this time. 
Turbulent transport theory can be used to propagate the turbulent dynamics into the outer heliosphere and obtain predictions for the turbulence level, temperature and heating rate, correlation scale, and cross-field correlation most notably called the ``cross helicity'' or  ``imbalance''.
The predictions of transport theory can then be compared to fluctuation levels and thermal proton heating rates which are assumed to be the same as the local energy cascade rate as obtained from theories based on the power spectrum. 

It is widely assumed that the PUIs generate turbulence in the distant SW due to their initial ring-beam velocity distributions which are unstable to the excitation of  Alfv\'en/ion cyclotron modes and should evolve into a spherical, filled-shell distribution \citep{wu1972,vasyliunas1976,lee1987,gary1988,gary1991,zank1999}. The fluctuation energy produced by PUIs adds to the existing turbulence and is assumed to dissipate at small spatial scales by supplying the energy to SW protons and electrons via a turbulent cascade.  Therefore, low-frequency magnetic turbulence heats the SW plasma and results in a non-adiabatic SW temperature profile and a slow temperature increase beyond 
  $\sim 30$\,\au
\citep[e.g.,][]{williams1995, matthaeus1999c,matthaeus2004, smith2001, smith2006b,isenberg2003, isenberg2005, chalov2006, breech2008,breech2009, isenberg2010, gamayunov2012}. 

It follows that in transport theory for SW fluctuations
a significant element is the inclusion of this outside energy source associated with wave excitation by newborn interstellar PUIs.  Indeed, it becomes the primary source of energy that drives the turbulence beyond 10\,\au.  This source term becomes critical to reproducing the observed turbulence and heating levels beyond 10\,\au
  \citep{matthaeus1999c,smith2001,breech2008,oughton2011,adhikari2017b}
and can be compared with the theories of interstellar neutral ionization and associated wave excitation. Further details on  PUI waves are provided in \citet{Sokol_EA_this_volume} in this volume.




\section{Modeling of the supersonic solar wind with turbulence transport}\label{sec:transport_models}

Numerous problems in space physics and astrophysics require a detailed understanding of the transport and dissipation of low-frequency turbulence in the expanding inhomogeneous magnetized solar wind plasma. For instance, knowledge of spatial distribution of turbulence intensity is an important input for computations of energetic particle propagation throughout the heliosphere. Coupling global heliospheric models and turbulence transport models provides not only mean-flow plasma and magnetic field parameters, but also the turbulence quantities,  which makes them useful also for calculation of diffusion coefficients and modulation of galactic cosmic rays (GCRs) \citep[see, e.g.,][]{florinski2013ssr,engelbrecht2013,wiengarten2016,chhiber2017,engelbrecht2017,zhao2018}. {This topic is reviewed by \citet{Engelbrecht_EA_this_volume} in this volume.}

Transport models for solar wind fluctuations in the supersonic and super-Alfv\'enic SW have advanced considerably since the presentation of the initial 1D Wentzel–Kramers–Brillouin (WKB) approach with prescribed (inhomogeneous) background fields \citep{Parker65-wkb}. 
The WKB theory \citep[see also][]{tu1984} can only describe the evolution of linearly interacting modes, whose typical time scale is much shorter than that of the overall cascading. 
The more general turbulence transport models are based on a few statistical parameters that characterize the turbulence in the supersonic SW. Most of them build upon the K\'arm\'an--Howarth kind of one-point closure models for local evolution of turbulence \citep{karman1938}.  
In the theory, a phenomenological description of the turbulent cascade is merged with transport equations obtained from a scale-separated decomposition (see Eq.~\ref{eq:ElsasserIncompress}) of the 
MHD equations \citep{zhou1989,marsch1989, zhou1990a,  tu1993} which supports coupling of the small-scale (turbulence) quantities to the large-scale quantities, e.g., the mean SW velocity $\vct{U}$, magnetic field $\vct{B}$, and mass density $\rho$
  \citep[for a review see][]{oughtonengelbrecht21}. 

In the turbulence transport theory, the Els\"asser variables represent  propagating modes moving parallel ($\vct{z}^-$) and antiparallel ($\vct{z}^+$) to $\vct{B}_0$,
provided that the wavevectors involved satisfy $ \vct{k} \cdot \vct{B}_0 \ne 0$
\citep[e.g.,][]{tu1984,zhou1990c,zhou1990b,beresnyak2015}. The interaction between these counter-propagating modes leads to the generation of quasi-2D turbulence in a plane perpendicular to the mean magnetic field, which by direct energy cascade eventually heats the SW. The time scales of such nonlinear interactions are $\tau_\mathrm{nl}^\pm \sim \lambda^{\pm}/\langle z^{\mp^2}\rangle^{1/2}$,  where $\lambda^{\pm}$ are the correlation lengths associated with the turbulent Els\"asser energies,  while the  time scale of linear interaction is $\tau_A^\pm\sim\lambda^\pm_\parallel/V_\A$. The characteristic time scale of the turbulent cascade (or spectral transfer time) can be expressed as $\tau^\pm_\mathrm{sp} \sim (\tau^\pm_\mathrm{nl})^2 / \tau_\mathrm{T}^\pm$, where $\tau^\pm_\mathrm{T}$ is the triple-correlation lifetime  \citep{matthaeus1989}. 
The evolution of incompressible ideal MHD fluctuations in the presence of (scale-separated) inhomogeneous large-scale fields can be written in the following form \citep{zhou1990c},
\begin{equation}\label{eq:ElsasserIncompress}
\begin{split} 
\frac{\partial \zpm}{\partial t} + ({\bf U} \mp\bv ) \cdot \nabla {\zpm } + \frac{1}{2} \nabla \cdot \bigg(\frac{\bu}{2} \pm\bv \bigg) \zpm +\\ \zmp \cdot \bigg[\nabla {\bu} \pm \frac{\nabla {\bf B}}{\sqrt{\mu_0 \rho}}  - \frac{1}{2}  \bm{I} \nabla \cdot \bigg(\frac{\bu}{2} \pm\bv \bigg) \bigg]  = {\bf NL}_\pm + {\bf S}^\pm,
\end{split} 
\end{equation}
where  ${\bf NL}^\pm = {\bf NL}^u \pm {\bf NL}^b/\sqrt{\mu_0 \rho}$ are the nonlinear terms, and $\textbf{S}^\pm$ are external sources. 
Constructing the moments of Els\"asser variables from Eq.~\ref{eq:ElsasserIncompress} introduces terms like $\langle z_i^+ z_j^- \rangle$, which are regarded as MHD analogs of the hydrodynamic Reynolds stress. Due to the presence of $\langle z_i^+ z_j^- \rangle$ terms, the backward and forward propagating modes interact through the small-scale fluctuations, large-scale SW speed and magnetic field.\footnote{Some definitions and nomenclature that will be used throughout in this paper are given: the Els\"asser energies $(Z^\pm)^2=\langle {z^\pm}^2 \rangle/2$, total turbulence energy density, $Z^2\equiv E_T =  E_u+E_b = \left(\langle {z^+}^2 \rangle + \langle {z^-}^2 \rangle\right)/2$, the residual energy $E_\mathrm{D} = \langle \delta u^2 - \delta b^2 \rangle  \equiv \langle \zp \cdot \zm \rangle $ and its normalized value,  $\sigma_\mathrm{D} = (E_u-E_b) / E_T$, the normalized cross helicity $\sigma_\mathrm{c} = \left( \langle {z^+}^2 \rangle - \langle {z^-}^2 \rangle \right) /E_T$, the Alfv\'en ratio $r_A = E_u/E_b$.  $E_u=\langle \delta u^2 \rangle /2$ and $E_b=\langle \delta B^2/(\mu_0\rho)/2 \rangle$ are the turbulent kinetic energy and magnetic energy densities in Alfv\'en units, respectively. }

Most of the fluctuation energy is associated with the `energy-containing range' of scales and transport models that follow energy-containing range quantities are of interest. 
These quantities include the Els\"asser energies $Z_\pm^2$, the residual energy (aka energy difference)
 $E_\mathrm{D}$, and characteristic lengthscales for each of these. 

A rough timeline of the development of these models is given here, and we note that they usually need to be solved numerically.
\citet{matthaeus1994} used the Reynolds decomposition approach to develop a  four-equation transport model for $ Z_\pm^2 $, $E_\mathrm{D}$, and a single characteristic lengthscale.
So-called `mixing' effects, due to gradients of the large-scale SW velocity and magnetic field, couple the turbulence quantities to each other and support for generic driving of the fluctuations is also included. \citet{zank1996} considered a zero cross helicity special case of this model with two dynamical equations (for the magnetic energy and its lengthscale) while also including turbulence sources due to shear, compression, and PUI heating. Their model was able to describe the observed radial decay of turbulence reasonably well. \citet{matthaeus1999c} extended this to include a transport model for the proton temperature with proton heating by PUIs, and a simple closure for local anisotropic MHD turbulence, and found excellent agreement with V2 data from 1 to 60\,\au. 
The effects of magnetic energy dissipation in the proton temperature were included by \citet{smith2001}. Later, \citet{smith2006b} improved the description of the PUI source term using the formalism of \citet{isenberg2003} (see Eqs.~\ref{eq:dZ2dr}--\ref{eq:dTdr}). A transport theory including cross helicity was formulated by 
  \cite[see also Matthaeus et al.\ 1994][]{matthaeus2004} 
and further developed by \citet{breech2005} and \citet{breech2008}. Using the model equations of \citet{smith2001} and \citet{isenberg2003}, \citet{ng2010} investigated the effect of IK cascade, finding similar or even higher heating rates than that obtained by using the Kolmogorov cascade.  
Later models included the electron heating \citep{breech2009}  and the deceleration of SW by PUIs \citep{isenberg2010}.  
Building on \citet{breech2008},  \citet{oughton2006,oughton2011} developed an anisotropic two-components model where PUIs can directly influence the quasi‐parallel wavenumber fluctuations. 

The above models are applicable to the super-Alfv\'enic SW ($U\gg V_\A$) under a number of assumptions and approximations, reviewed by \citet{oughtonengelbrecht21}. A six-equation incompressible MHD turbulence model applicable also to sub-Alfv\'enic flows was developed by  \citet{zank2012} by introducing separate correlation lengths associated with the forward- and backward-propagating modes  \citep[cf.][]{matthaeus1994}.  
\citet{adhikari2014} investigated the effects of solar cycle variability on the Zank et al.\ model, and  \citet{adhikari2015} obtained turbulence quantities at distances up to 100\,\au. With selection of appropriate shear constants, and boundary values, these transport theories have been able to account quantitatively for \textit{Helios} and \Ulysses proton temperature observations as well as \Voyager data from 1 to more than 60\,\au \citep{zank1996, matthaeus1999c, smith2001, smith2006b}. 
	
There are recent models that also support a 3D heliosphere with dynamically evolving background fields coupled to fluctuation quantities  \citep[e.g.,][]{kryukov2012,usmanov2012,usmanov2014,usmanov2018,wiengarten2016,shiota2017}. Attempts to include the heliosheath and VLISM in the global models have also been presented  \citep[e.g.,][]{usmanov2016,fichtner2020}. \citet{usmanov2014} improved their previous model by  using an eddy viscosity approximation for the Reynolds stress tensor and the mean turbulent electric field. They demonstrated that the effect of eddy viscosity and, correspondingly, of velocity shear on the mean-flow parameters manifests itself in the increased temperatures of SW protons. The turbulence energy and the correlation length are notably increased and the cross helicity decreased, especially near transitions between fast and slow SW flows. 

Simulation results based on these models give encouraging, if incomplete, agreement with outer heliosphere spacecraft observations  \citep{marsch1990b,williams1994,williams1995,richardson2003,richardson1995,richardson1996,usmanov2016}.

A relatively simple 
steady-state 
transport model \citep{smith2001,smith2006b,pine2020d} can be written as equations for the total average fluctuation energy density,
\begin{equation}
\frac{dZ^2}{dR} = - \frac{Z^2}{R} + \frac{C_{\rm sh}-M \sigma_\mathrm{D}}{R}Z^2
+ \frac{\dot{E}_\mathrm{PI}}U - \frac{\upalpha}{\lambda U} Z^3,
\label{eq:dZ2dr}
\end{equation}
the similarity scale,
\begin{equation}
\frac{d\lambda}{dR} = \frac{M\sigma_\mathrm{D} - \hat{C}_{\rm sh}}{R} \lambda
- \frac{\upbeta}{\upalpha} \lambda \frac{\dot{E}_{\rm PI}}{U Z^2}
+ \frac{\upbeta}{U}Z
\label{eq:dlamdr}
\end{equation}
and the   proton  temperature,
\begin{equation}
\frac{dT_\p}{dR} = - \frac{4}{3}\frac{T_\p}{R} + \frac{1}{3} \frac{m_\p}{k_\mathrm{B}}
\frac{\upalpha}{U}\frac{Z^3}{\lambda}.
\label{eq:dTdr}
\end{equation}
The parameters of the theory are heavily constrained by observations. 
\begin{figure}[t!]
	\centering
	\includegraphics[width=0.8\columnwidth]{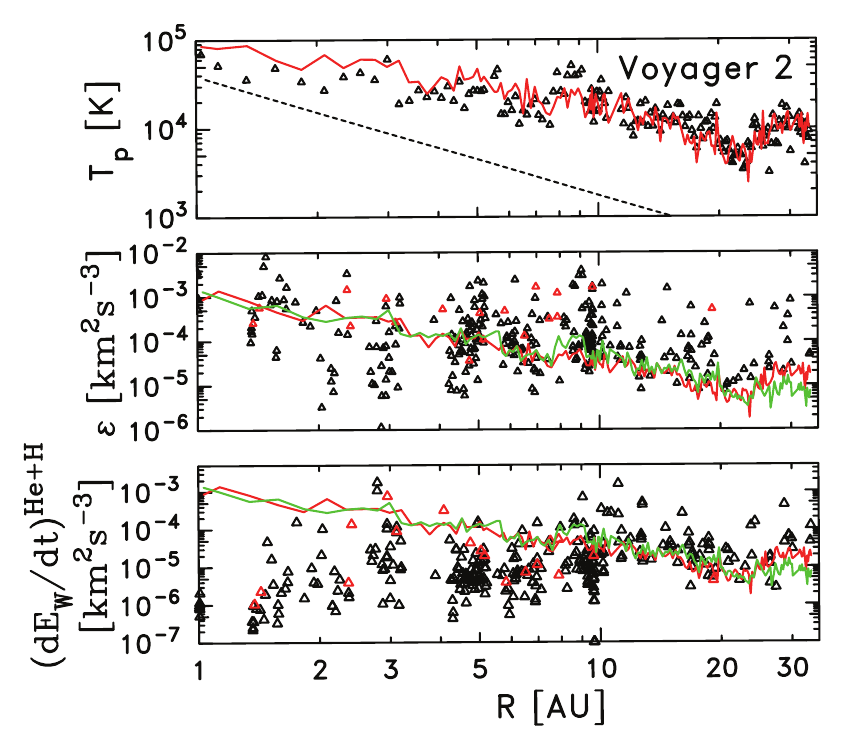}
	\caption{(top) Predicted average thermal proton temperature derived from transport theory (red curve) and average measured temperature (symbols). (Middle) Predicted average heating rate derived from transport theory (red curve), average heating rate derived from Eq.~\ref{eq:epsv} (green curve), and heating rate derived from the measured power spectra using Eq.~\ref{eq:koleps}. 	(Bottom) Predicted average heating rate derived from transport theory using Eqs.~\ref{eq:dZ2dr}--\ref{eq:dTdr} (red curve), average heating rate derived from Eq.~\ref{eq:epsv} (green curve), and rate of energy injection by pickup ion scattering for the intervals used in the spectral analysis in the middle panel (symbols). \citet{pine2020d}.}
	\label{fig:Pine4a}
\end{figure}

The above citations use $\upalpha = 0.8$, $\upbeta = 0.4$, $M = 1/3$, $\sigma_\mathrm{D} = -1/3$, 
$C_\text{sh} = 1.4$, and $\hat{C}_\text{sh} = 0$ 
  \citep{matthaeus1999c}. 
It is important to note that $Z^2$ represents the fluctuation energy in the large-scale, or energy-containing, range of scales. 
The inertial range is not explicitly represented here.
However, terms that scale as $Z^3$ represent the loss of energy in coherent turbulent fluctuations due to the spectral transport of energy from large to small scales and the conversion of that energy into heat. At large Reynolds numbers this spectral transport will involve energy transfer through an implied inertial range. 
The term $4 T_\p / (3R) $ represents expansive cooling. The term ${\dot{E}_\mathrm{PI}}$ represents the rate of energy injected into the turbulence via wave energy excitation by newborn interstellar pickup ions. This can be modeled using the rate of ion production obtained from the analytic Warsaw Test Particle Model (aWTPM) and the numerical Warsaw Test Particle Model (nWTPM) codes \citep[e.g.,][]{bzowski2013a,sokol2015,sokol2019},  or global MHD plasma/ kinetic neutrals simulations, reviewed by \citep{Kleimann_EA_this_volume} in this volume. The analysis of \Voyager data from launch through 1990 described below uses a photoionization rate model determined from series of solar EUV proxies, like F10.7, MgII core-to-wing index, and CELIAS/SEM correlated with the solar EUV measurements from TIMED \citep{bzowski2013a,bochsler2014,sokol2019}. The resulting ion production rates can be combined with the theory of wave excitation by pickup ions \citep{lee1987} to produce an estimate for ${\dot{E}_\mathrm{PI}}$.

The top panel of Fig.~\ref{fig:Pine4a}  from \citet{pine2020d} shows the result of comparing the predictions for the thermal proton temperature as derived from the above transport model (red line) with the measured average proton temperature (symbols). 
The dashed line represents adiabatic expansion from 1\,\au. The evidence for some form of in situ heating is undeniable. Transport theory accurately reproduces the observed average temperature of SW protons. Figure~\ref{fig:Pine4a} (middle) compares the rate of energy dissipation derived from the transport model (red line) and Eq.~\ref{eq:epsv} (green line) against the rate of heating obtained from Eq.~\ref{eq:koleps} as applied to magnetic spectra from 327 intervals of \textit{V2} observations.  Although agreement between transport theory and Eq.~\ref{eq:epsv} is good, the results derived from the magnetic spectra do show a broad distribution about the predictions for the average heating rate. 
This is partly due to the natural variation of the turbulence, partly due to the fact that the data intervals were selected to be used as controls in the analysis of waves due to PUIs, and partly due to rejection of low spectral levels with evidence of instrument noise in the data. Figure~\ref{fig:Pine4a} (bottom) compares the rate of thermal proton heating obtained from the transport model (red line) and eq.~\ref{eq:epsv} (green line) against the rate of wave energy excitation by newborn interstellar pickup ${\rm H^+}$ and ${\rm He^+}$ (symbols) using the above formalism applied to the same data intervals as the middle panel.  Wave excitation by newborn interstellar PUIs becomes the primary source of energy that drives the turbulence beyond 10\,\au. 

PUIs are thermodynamically different from thermal protons \citep{vasyliunas1976, isenberg1986}. While their number density is relatively low, their impact on the SW, including its heating and gradual deceleration, is significant. The very high effective temperature ($\sim$10$^7$~K) of pickup protons makes them the dominant component of the thermal pressure in the distant SW \citep{burlaga1996}. Speaking of global heliosphere numerical simulations, the most obvious problem with adopting the single-fluid description for SW plasma is that it implies an immediate assimilation of the newborn PUIs with thermal SW protons. As a result, single-fluid models predict a steep increase of the plasma temperature with radius beyond $ \sim $10\,\au, where the pickup protons play a major role. A modest increase in the temperature of SW protons is indeed present in \textit{V2} data beyond $\sim$30\,\au. However, the steep rise predicted by single-fluid models is in obvious disagreement with \textit{V2} observations. PUIs should, in principle,  be modeled as multiple populations  \citep{malama2006}. In the fluid approach, it is then important to model PUIs by a separate energy equation, as shown by  \citet{isenberg1986} and further elaborated by \citet{zank2014}. After the first 1D fluid model of \citet{isenberg1986},  3D models were developed by \cite{usmanov2006} and \cite{detman2011}, with PUIs treated as a separate fluid, but including only the supersonic SW region.
Later, the effects of pickup protons as a separate fluid were included in the 3D heliospheric models of \citet{pogorelov2016} (MS-FLUKSS code). For details on global models, see \citet{Kleimann_EA_this_volume}.

\begin{figure} [t]
	\centering
	\includegraphics[width=0.9\textwidth]{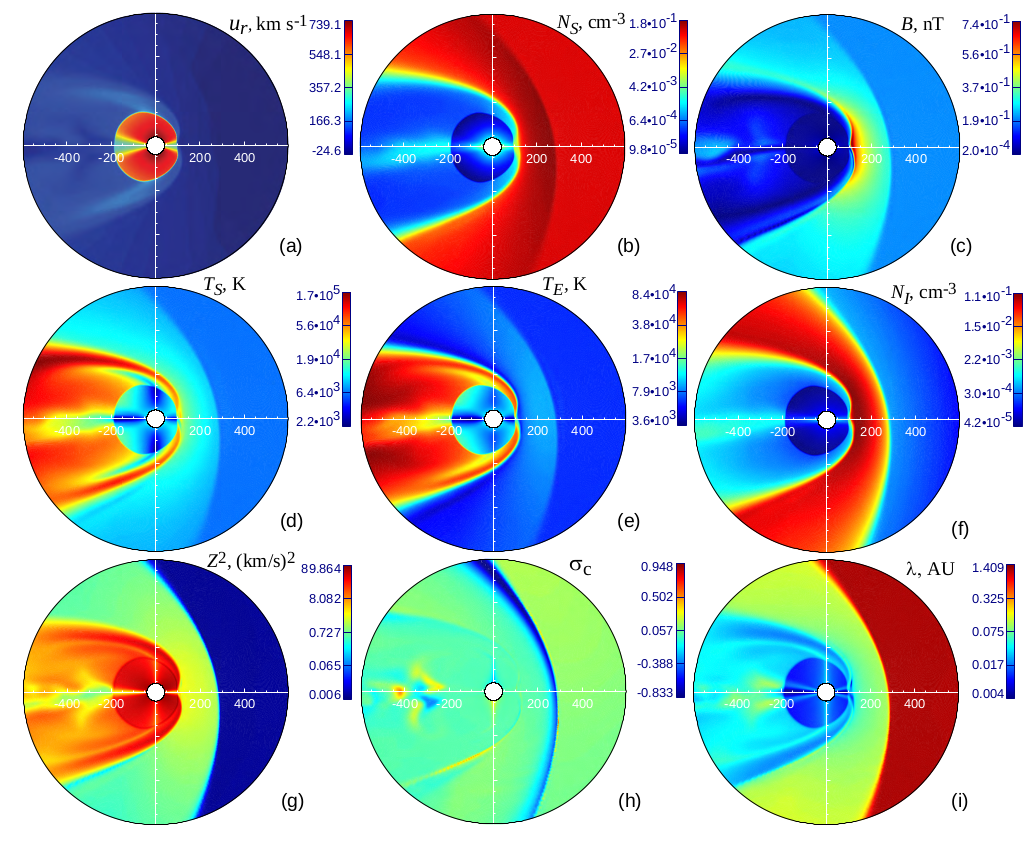}
	\caption{Distributions of computed mean-flow and turbulence parameters in the meridional plane containing the interstellar upwind direction from 40 to 600\,\au for an axisymmetric SW from a magnetic dipole on the Sun aligned with the solar rotation axis: (a)	radial velocity $u_\mathrm{r}$, (b) number density of thermal protons $N_\mathrm{S}$, (c) magnetic field magnitude $B$, (d) thermal proton temperature $T_\mathrm{S}$, (e) electron temperature $T_\mathrm{E}$, (f) pickup proton density $N_\mathrm{I}$, (g) turbulence
	energy per unit mass $Z^2$, (h) cross helicity $\sigma_\mathrm{c}$, and (i)	correlation length scale $\lambda$.  Reproduced from \citet{usmanov2016}.}
	\label{mp-0deg}
\end{figure}

The approach used by \cite{usmanov2014,usmanov2016} to modeling turbulence effects in the SW follows the transport theory, which describes the effects of transport, cascade, and dissipation of incompressible MHD turbulence. The time-dependent turbulence transport equations read

\begin{gather} 
\frac{\partial Z^2}{\partial t} +
(\bu\cdot\nabla)Z^2 +
\frac{Z^2}{2} \nabla\cdot \bu + 
\frac{2}{\rho}\boldsymbol{\mathcal{R}}:\nabla\bu -
\frac{\sigma_\mathrm{D} Z^2}{2} \nabla\cdot\bu +
2\boldsymbol{\varepsilon}_\mathrm{m}\cdot(\nabla\times \bv)\nonumber\\ -
(\bv\cdot\nabla)(Z^2\sigma_\mathrm{c})+
Z^2\sigma_\mathrm{c}\nabla\cdot\bv+
\frac{m_\p Z^2}{2\rho}\left[2q_\mathrm{T}(1+\sigma_\mathrm{D})+q_\mathrm{ph}(1-\sigma_\mathrm{D})\right] \nonumber\\
=-\frac{\upalpha f^+(\sigma_\mathrm{c})Z^3}{\lambda} + \dot{E}_\mathrm{PI},
\label{usm-1}
\end{gather}
\begin{gather} 
\frac{\partial(Z^2\sigma_\mathrm{c})}{\partial t}+ (\bu\cdot\nabla)(Z^2\sigma_\mathrm{c})+
\frac{Z^2\sigma_\mathrm{c}}{2} \nabla\cdot \bu +
\frac{2}{\rho}\boldsymbol{\mathcal{R}}:\nabla\bv
+ 2\boldsymbol{\varepsilon}_\mathrm{m}\cdot(\nabla\times\bu)\nonumber\\
 - (\bv\cdot\nabla)Z^2 + (1-\sigma_\mathrm{D})Z^2\nabla\cdot\bv +
\frac{Z^2\sigma_\mathrm{c}}{2\rho}(2q_\mathrm{T}+q_\mathrm{ph})m_\p =
  -\frac{\upalpha f^-(\sigma_\mathrm{c})Z^3}{\lambda},
\label{usm-2}
\end{gather}
\begin{gather} 
\frac{\partial \lambda}{\partial t} + (\bu\cdot\nabla)\lambda = \upbeta f^+(\sigma_\mathrm{c}) Z -\frac{\lambda\dot{E}_\mathrm{PI}}{\upalpha Z^2}.
\label{usm-3}
\end{gather}
where $\upalpha$ and $\upbeta$ are the K\'arm\'an--Taylor constants. The term 
 $\boldsymbol{\varepsilon}_\mathrm{m} = \langle \delta\textbf{u} \times \delta\textbf{B} \rangle/\sqrt{4\pi\rho}$ 
is the average induced fluctuation electric field, and $f^\pm(\sigma_\mathrm{c})=(1-\sigma_\mathrm{c}^2)^{1/2}[(1+\sigma_\mathrm{c})^{1/2}\pm(1-\sigma_\mathrm{c})^{1/2}]/2$ is
a function of cross helicity that modifies the nonlinear decay phenomenology if $\sigma_\mathrm{c}\neq 0$.  $\boldsymbol{\mathcal{R}}$ is the Reynolds stress tensor, and $q_\mathrm{T}$ and $q_\mathrm{ph}$ are source terms due to charge exchange and photoionization.
This model assumes the local incompressibility of fluctuations, and a single characteristic lengthscale, $\lambda$. 

\begin{figure} [t]
   \centering
	\includegraphics[width=\textwidth]{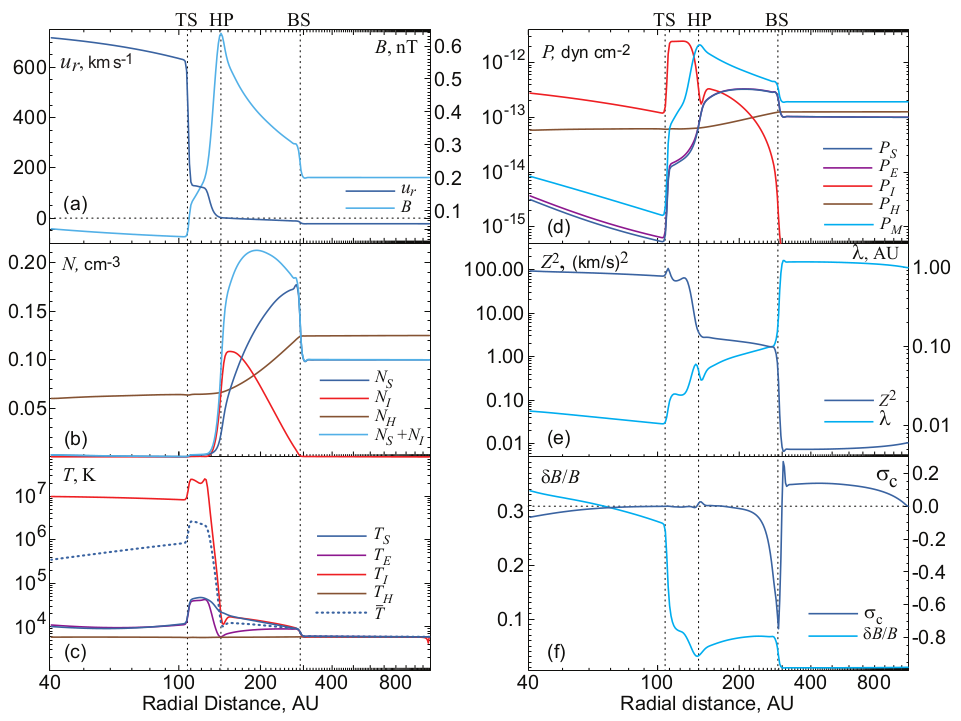}
	\caption{Radial profiles of the flow, magnetic field, and turbulence parameters along the \textit{V1} trajectory for the 0$^\circ$-model from 40\,\au to 1200\,\au: (a) radial velocity $u_\mathrm{r}$ and magnetic field strength; (b) thermal proton $N_\mathrm{S}$, pickup proton $N_\mathrm{I}$, interstellar hydrogen	$N_\mathrm{H}$, and total proton $N_\mathrm{S} + N_\mathrm{I}$ densities; (c) proton $T_\mathrm{S}$, electron
		$T_\mathrm{E}$, pickup proton $T_\mathrm{I}$, interstellar hydrogen $T_\mathrm{H}$, and mean $\bar T$	temperatures, where $\bar T = (T_\mathrm{S} N_\mathrm{S} + T_\mathrm{E} N_\mathrm{E} + T_\mathrm{I} N_\mathrm{I})/(N_\mathrm{S} + N_\mathrm{E} +	N_\mathrm{I})$; (d)  proton $P_\mathrm{S}$, electron $P_\mathrm{E}$, pickup proton $P_\mathrm{I}$, interstellar hydrogen $P_\mathrm{H}$, and magnetic $P_\mathrm{M}$ pressures; (e) turbulent energy density $Z^2$ and correlation length $\lambda$; (f) cross helicity $\sigma_\mathrm{c}$ and relative amplitude of the magnetic field fluctuations $\delta B/B$. The vertical lines mark the locations of the HTS, HP, and BS at $\sim$105/140/290\,\au, respectively. Reproduced from \citet{usmanov2016}.}
	\label{1d-0deg-v1}
\end{figure}

Having their focus on the heliospheric interface region, the existing global models of the outer heliosphere typically employ simplified patterns for the SW and interplanetary magnetic field parameters at their inner boundaries, which are usually placed between 10 and 50\,\au. The most frequent assumption is that the SW is spherically symmetric \citep[e.g.,][]{washimi1996, pogorelov1998, ratkiewicz1998, opher2003, izmodenov2005,ratkiewicz2002, borrmann2005, pogorelov2006, opher2009, izmodenov2014}. Latitudinal variations at the inner boundary consistent with \Ulysses observations of the bimodal SW near solar minimum have been included, e.g., by \cite{pauls1997, linde1998, pogorelov2013a, provornikova2014}.  The first global heliospheric models that used observations of solar magnetograms to extrapolate time-dependent inner boundary conditions at 0.1\,\au is that of \citet{detman2011}. Later,  \citet{usmanov2016} were able to carry solar corona/SW computations from the coronal base. 

\begin{figure} [t]
   \centering
	\includegraphics[width=0.85\textwidth]{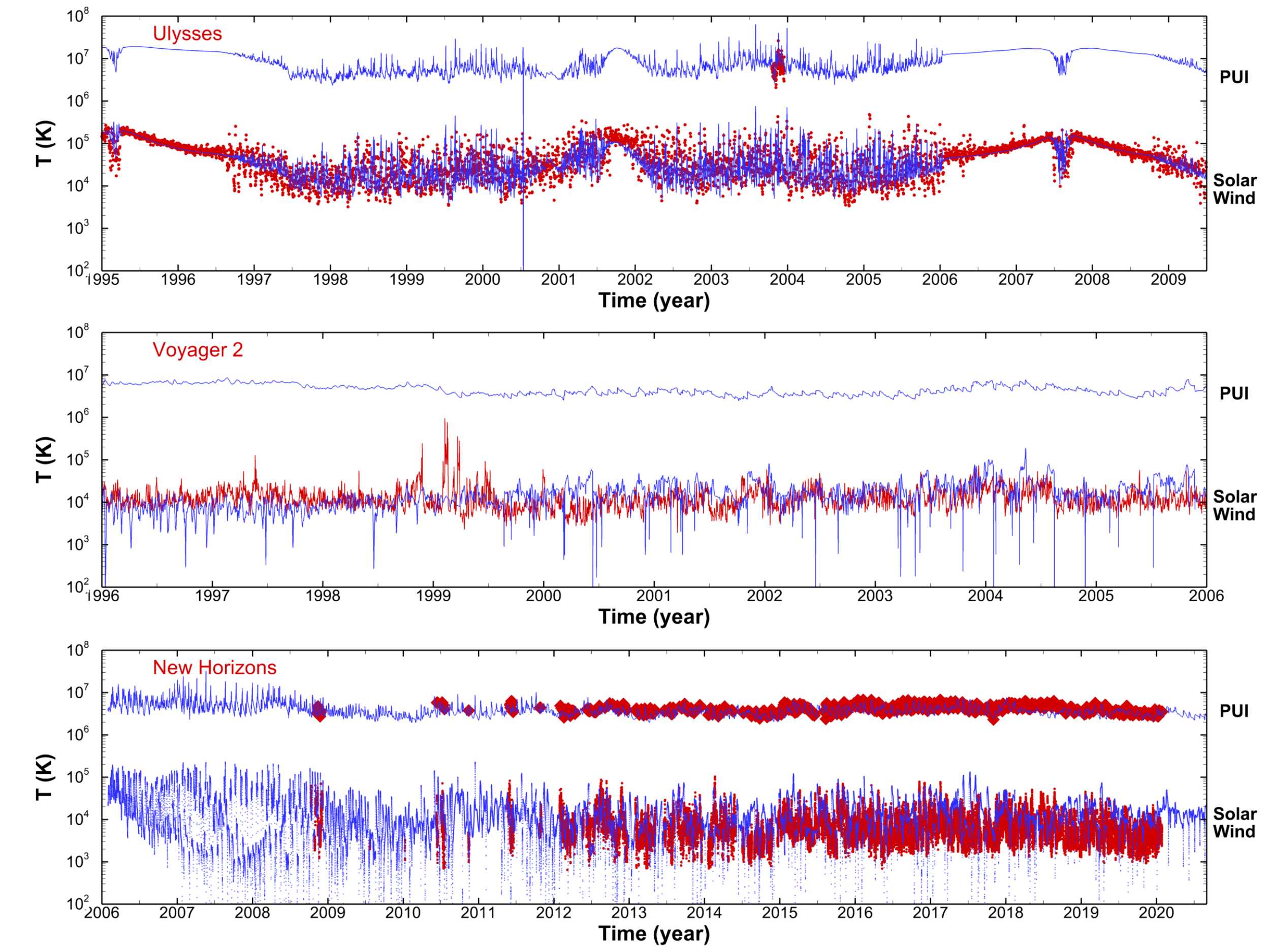}
	\caption{Simulated SW and interstellar PUI temperatures are shown in blue, compared with \Ulysses/SWOOPS/SWICS (top), \voy2/PLS (middle), and \NH/SWAP data in red. Adapted from \citet{kim2018}.}
	\label{fig:Kim_model}
\end{figure}
\citet{usmanov2016} further extended their three-fluid SW model by including the heliospheric interface region. This is the first attempt to include turbulence effects in  global simulations including the heliosheath and the VLISM. They constructed such a 3D model taking into account turbulence transport and separate energy equations for thermal protons, electrons, PUIs, and interstellar hydrogen, and then, using this model, studied the formation of the heliospheric interface region.  We note here that the \citet{usmanov2016} turbulence model is applicable to both super-Alfv\'enic and sub-Alfv\'enic flows.  A shortcoming of the model, as stated by the authors, is that it is suitable for incompressible turbulence, while compressibility is a prominent feature of turbulence in the inner and outer heliosheath. Modeling of these regions still represents a major challenge.
Figures \ref{mp-0deg} and \ref{1d-0deg-v1} show some results from  \citet{usmanov2016} for plasma, magnetic field and turbulence parameters in the outer heliosphere.

 A turbulence model for the supersonic SW \citep{breech2008} has also been included in a global, 3D data-driven simulation by \citet{kryukov2012}, assuming spherically symmetric SW at the inner boundary. While the data-driven, time-varying simulation produced mostly realistic SW variations along the \textit{V2} trajectory out to 80\,\au, there were still some systematic discrepancies during certain periods when the assumption of spherically symmetric SW using near-Earth data was clearly inappropriate away from the ecliptic plane. To alleviate such discrepancies, \citet{kim2016,kim2017b} introduced spatial variations across the model inner boundary using \Ulysses data as constraints, and the results showed excellent agreements with \Voyager and \NH data. Subsequently, \citet{kim2017a} used the improved boundary conditions along with the \citet{breech2008} turbulence model to reproduce the temporal/spatial variations of SW and PUI between 1 and 80\,\au. The simulated SW and PUI temperatures are shown compared with the \Ulysses \citet[PUI from][]{intriligator2012}, \textit{V2} (SW only), and \NH data  \citep[PUI from][]{mccomas2021} in Fig. \ref{fig:Kim_model}.

Another class of turbulence transport models is based on the nearly incompressible (NI) MHD  phenomenology \citep[][]{zank1992a,zank1993,hunana2010,zank2017b}. 
In contrast to 
incompressible MHD models of the fluctuations, NI MHD theory assumes that the fluctuations are weakly compressible. 
The compressible MHD equations are separated into a set of core incompressible equations and a weakly compressible  fluctuating part. Core equations are obtained using the bounded time derivatives method given by Kreiss \citep{kreiss1980},  
a constraint that is imposed to ensure 
that the fast-timescale magnetoacoustic waves vanish.  The normalized equations for the fluctuations are expanded with respect to the low Mach number, then terms of similar order are collected \citep[][]{zank1992a,zank1993}.  
\cite{zank1993} considered three plasma beta regimes,  $\beta\ll 1$, $\beta \sim 1$, and $\beta \gg 1$, respectively ($\beta=P/(B^2/8\pi)$). They showed that the leading order incompressible MHD description is fully 3D for  $\beta \gg 1$, while it reduces to 2D in the plane perpendicular to the mean magnetic field for  $\beta \ll 1$ and $\beta \sim 1$. Higher-order corrections  
to the leading-order 
NI fluctuations are fully 3D. 
Based on the observed values of the Mach number (NI expansion parameter) \cite{zank1993} predicted that SW turbulence in the $\beta \sim 1$ or $\beta \ll 1$ regimes is a superposition of the dominant ($\sim80\%$) 2D turbulence and a minority ($\sim20\%$) slab turbulence.    

{Incompressible MHD turbulence models for SW fluctuations are formally applicable in the high plasma beta regime ($\beta \gg 1$) \citep[although][suggest, based on 2D hybrid simulations, that a weakly collisional high beta plasma can possess a self-induced pressure anisotropy not contained in the standard MHD closure]{squire2017}} while the NI MHD turbulence model of \cite{zank2017b} is applicable in the low-beta regime ($\beta \ll 1$) or when $\beta\sim 1$. An important and practical distinction between the $\beta  \gg 1$ description and the $\beta \ll 1$ and $\sim 1$ descriptions is that the latter allows for a clear decomposition into a distinct majority quasi-2D turbulence component and a distinct minority slab turbulence component that responds dynamically to the majority component. This is the theoretical underpinning of the well-known 2D+slab model \citep[e.g.,][]{bieber1996,forman2011}. By contrast, the incompressible $\beta \gg 1$ description can allow for both quasi-2D and slab components but now on an equal footing and both dynamically coupled, for which descriptions such as critical balance \citep{goldreich1995} or 2D + wave-like \citep{oughton2011} have been developed. This renders the study of anisotropy throughout the heliosphere \citep{adhikari2017b}, provided the plasma beta regime is appropriate, rather more straightforward than use of the incompressible MHD model, provided in this case that $\beta \gg 1$. In addition, according to incompressible MHD model, turbulence turns off for the unidirectional Alfv\'en waves (in the homogeneous case), 
while it is not so in the NI MHD model \citep{adhikari2019}. The model was able to reproduce the recent finding \cite{telloni2019} that the unidirectional Alfv\'en wave can exhibit a Kolmogorov-type power law \citep{zank2020,zhao2020b}.

In the NI MHD phenomenology for a $\beta \sim 1$ or $\ll 1$ plasma, the total Els\"asser variables can be written as a summation of the majority quasi-2D and a minority NI/slab Els\"asser variables, i.e., $\zpm = \bz^{\infty \pm} + \bz^{* \pm}$, where $\bz^{\infty \pm} = {\bm u}^{\infty} \pm {\bm B}^\infty/\sqrt{\mu_0 \rho}$ and $\bz^{* \pm} = {\bm u}^{*} \pm {\bm B}^*/\sqrt{\mu_0 \rho}$ \citep{zank2017b}. Here, ``$\infty$'' denotes the quasi-2D turbulence, and ``*'' the NI/slab component. The transport equations for the quasi-2D Els\"asser fields fluctuation $\bz^{\infty \pm}$ read \citep{zank2017b}, 
\begin{equation}\label{eq:NI2D}
\begin{split} 
\frac{\partial \bz^{\infty \pm}}{\partial t} + {\bu} \cdot \nabla \bz^{\infty \pm} + \bz^{\infty \mp} \cdot \nabla \bz^{\infty \pm} + \bz^{\infty \mp} \cdot \nabla {\bu} + \frac{\bz^{\infty \pm} - \bz^{\infty \mp}}{4} \nabla \cdot {\bu} \\- \frac{\bz^{\infty \pm} - \bz^{\infty \mp} }{4 \rho} \bz^{\infty \mp} \cdot \nabla \rho = - \frac{1}{\rho} \nabla\bigg(P^\infty + \frac{B^{\infty 2}}{2 \mu_0} \bigg). 
\end{split} 
\end{equation}
The difference between Eq.~\ref{eq:ElsasserIncompress} and Eq.~\ref{eq:NI2D} is that the latter describes the convection of locally quasi-2D Els\"asser variables, and does not include Alfv\'en propagation effects. The NI MHD approach is suitable for studying turbulence in SW, which is also supported by      
\cite{zhao2020a} and \cite{chen2020}, who found several quasi-2D structures in SW, namely magnetic flux ropes. In addition, pressure-balanced structures (PBSs) or flux tubes 
\citep{burlaga1968,burlaga_book,vellante1987,bavassano1991,borovsky2008,sarkar2014}, 
are commonly observed in the SW, are equilibrium solutions of NI MHD \citep{zank1992a}. PBSs/flux tubes are highly dynamical structures in the presence of quasi-2D turbulence \citep{zank2004}. \cite{zank2012} developed 6 coupled turbulence transport equations by taking moments of Eq.~\ref{eq:ElsasserIncompress}, and \cite{zank2017b} developed 12 coupled quasi-2D and NI/slab turbulence transport equations to describe the transmission of energy in forward and backward propagating modes, residual energy, and the corresponding correlation lengths.

\begin{figure*}[t]
	\centering
	\includegraphics[width=\textwidth]{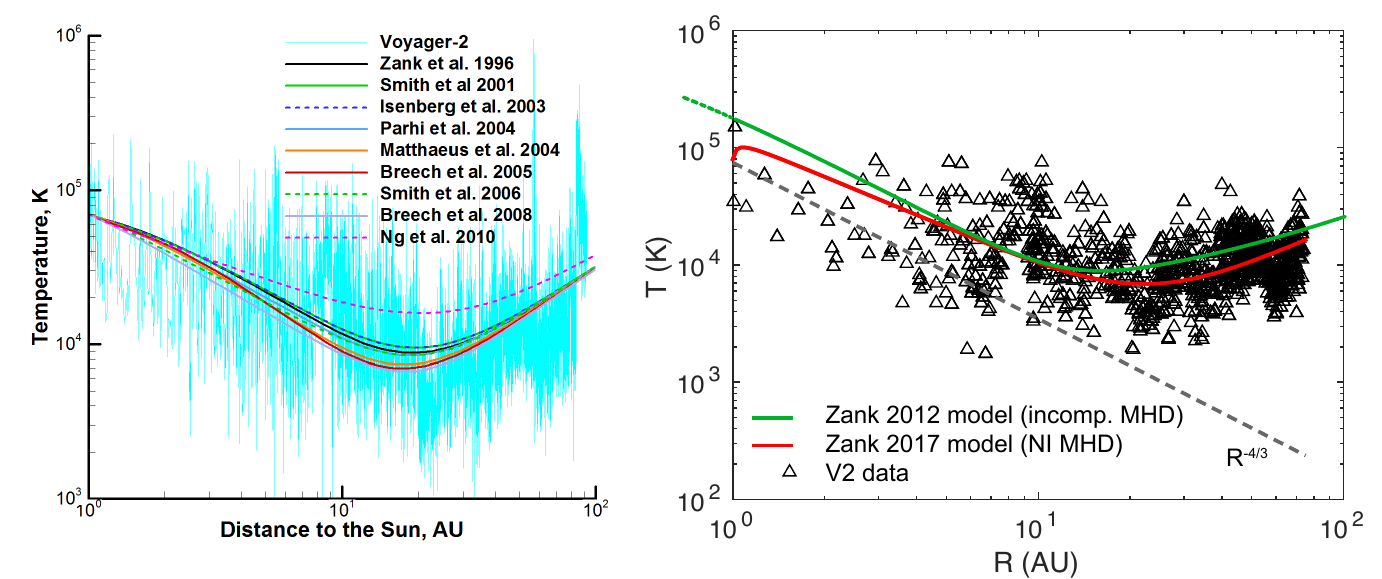}
\caption{Comparison between the thermal proton temperature observed by the \Voyager 2 spacecraft and the theoretical temperature from different models, as a function of heliocentric distance. The left panel compares results from different incompressible MHD turbulence models in a 3D, global simulation  \citep[partially published in ][]{pogorelov2012b}. The right panel \citep[adapted from][]{adhikari2015,adhikari2017b} shows the comparison between an incompressible MHD turbulence model \citep{zank2012} and a NI MHD turbulence model \citep{zank2017b}. The dashed gray curve shows an adiabatic temperature profile, $T\sim r^{-4/3}$.} 
	\label{fig:Adhikari_1}
\end{figure*}

Most of the turbulence transport models mentioned in this section 
address the SW proton heating. Figure  \ref{fig:Adhikari_1} (left panel), confronts the SW proton temperature measured by \voy2/PLS (blue line) with the result from different turbulence models implemented in a global, 3D unsteady simulation of the heliosphere.
The proton temperature equation can be expressed as 
\begin{equation}\label{eq:temperature}
\frac{\partial T_\p}{\partial t} + {\bu} \cdot \nabla T_\p + (\gamma-1)~ T_\p~ \nabla \cdot {\bu} = \frac{\gamma-1}{2} S_t,
\end{equation}
where $\gamma$ is a polytropic index, and $ S_t $ is a turbulent heating term derived from a von K\'arm\'an--Taylor phenomenology \citep{karman1938}. 
For example, the heating term $S_t$ using can be expressed as \citep{verdini2010,zank2012,adhikari2015}, 
\begin{equation}\label{eq:ST}
\begin{split}  
S_t & = \alpha \frac{m_\p}{k_\mathrm{B}} \bigg[ 2 \frac{\langle {z^+}^2 \rangle  \langle {z^-}^2 \rangle^{1/2}}{\lambda^+} +  2 \frac{\langle {z^-}^2 \rangle  \langle {z^+}^2 \rangle^{1/2}}{\lambda^-}  \bigg],
\end{split} 
\end{equation}
where $\alpha$ is a constant. Inside the square brackets of 
  Equation~(\ref{eq:ST}), 
the first term is the nonlinear dissipation term corresponding to the energy in forward propagating modes, and the second term the nonlinear dissipation term corresponding to the energy in backward propagating modes. In \cite{verdini2010}, their Eq.~(5) denotes the heating term, which was derived by using the nonlinear dissipation terms corresponding to the energy in forward and backward propagating modes. The first and second terms inside the square brackets of   Eq.~(\ref{eq:ST}) are larger than that of \cite{verdini2010}, resulting in a larger heating rate, and therefore a larger SW temperature. In  \cite{adhikari2015}, this is slightly ameliorated by the inclusion of the residual energy term. 

Using Eq.~\ref{eq:temperature} and Eq.~\ref{eq:ST}, \cite{adhikari2015} investigated the heating of SW plasma from 0.29\,\au to 100\,\au. Similarly, \citet{adhikari2017b} studied the proton heating from 1\,\au to 75\,\au using the NI MHD turbulence model \citep{zank2017b}. The temperature profile from these two models is shown in Fig. \ref{fig:Adhikari_1} (right panel). The stream-shear source is found to be important within 4--5\,\au, while the PUI-related turbulent source term is important beyond the ionization cavity boundary at $\sim 10$\,\au. 
 These sources drive turbulence throughout the heliosphere, and offset the decay of turbulence energy. The dissipation of turbulence, plus the additional driving of turbulence by the distributed heliospheric sources, yields a plasma temperature profile that is significantly different and of course higher than would be expected if only adiabatic cooling of the SW occurred (see the dashed curve in the right panel of Fig. \ref{fig:Adhikari_1}). Naturally, adiabatic cooling is included in the SW models with turbulent heating. The increase of $T$ beyond 20\,\au can be considered due to the presence of PUIs in the outer heliosphere. The results show that the theoretical proton temperature (solid red curve) obtained by using incompressible MHD and NI MHD turbulence models produce radial temperature profiles similar to the observed ones. 
 
\begin{figure}[t]
	\centering
	\includegraphics[width=\textwidth]{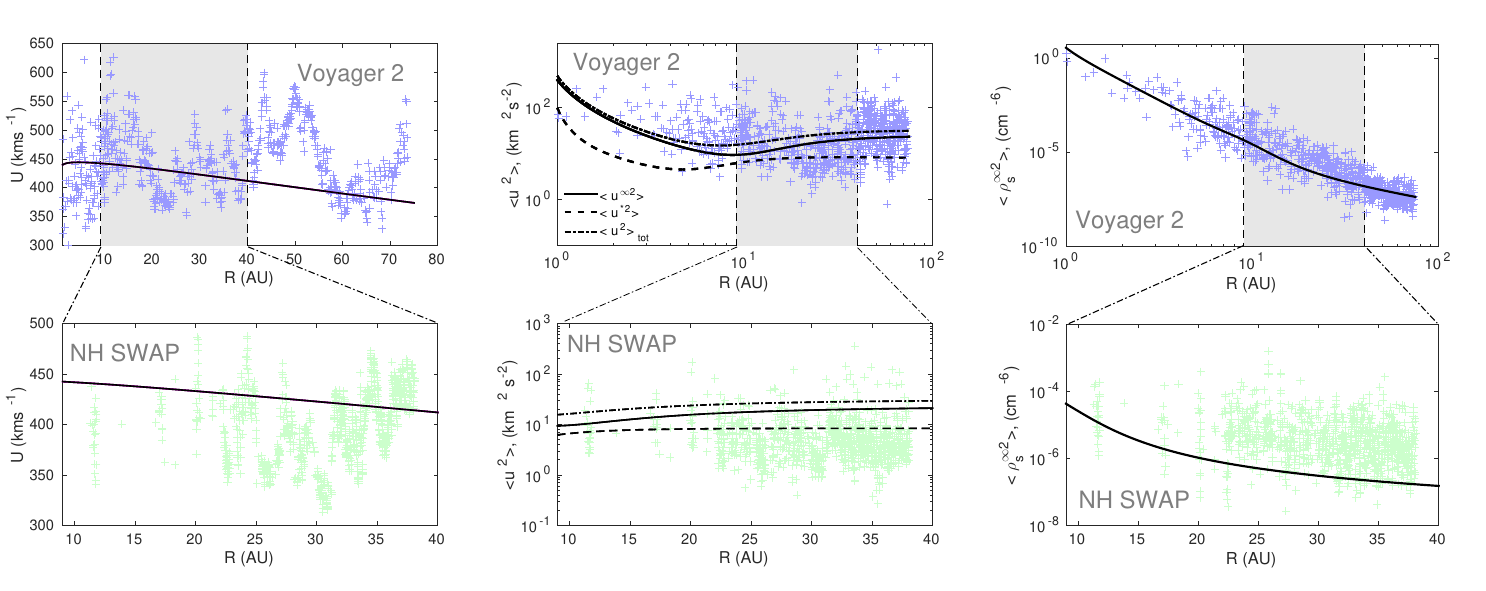}
	\caption{Comparison between theoretical (NI MHD) and observed results derived from \textit{V2} measurements between 1\,\au and 75\,\au (top panel), and \NH/SWAP measurements between 11.26\,\au and 38\,\au (bottom panel). Left: SW speed $U$. Middle: The fluctuating kinetic energy $\langle u^2 \rangle$ for the 2D (solid), slab (dashed), and total (dash-dotted) components. Right: Thermal SW density variance $\langle {\rho_s^{\infty}}^2 \rangle$. Reproduced from \citet{zank2018}.	\label{fig:Adhikari_2}}
\end{figure}     

Pickup ions not only produce turbulence in the outer heliosphere, but also influence the SW properties.
\citet{zank2018} extended the classical models of \citet{holzer1972} and \citet{isenberg1986} by coupling a NI MHD turbulence model of \citet{zank2017b} to a multi-fluid description of the SW plasma to properly examine the feedback between SW plasma heating, the modified large-scale SW velocity due to the creation of PUIs, and the driving of the turbulence by SW and interstellar PUI sources. The theoretical model of \citet{zank2018} describes the evolution of the large-scale SW, PUIs, and turbulence from 1--84\,\au. As shown in the left panel of Fig.  \ref{fig:Adhikari_2}, the theoretical SW speed (solid curve) gradually decreases with increasing heliocentric distance from 1\,\au to 75\,\au because the PUIs ions lead to the decrease of the momentum of the SW \citep[see also,][]{richardson2003b,elliott2019}. The theoretical speed is compared with \textit{V2} measurement (blue plus symbols, the top panel of Fig. \ref{fig:Adhikari_2}) and \NH SWAP measurements (green plus symbol, the bottom panel of Fig.  \ref{fig:Adhikari_2}).

The middle panel of Fig. \ref{fig:Adhikari_2} shows the quasi-2D, NI/slab, and total fluctuating kinetic energy with increasing heliocentric distance. The observed fluctuating kinetic energy exhibits quite a large scatter. The theoretical results are slightly higher than the observed \NH SWAP values, although not significantly. This difference may be due to the fact that a single boundary conditions is used to compare with the \textit{V2} and \NH data sets. The \textit{V2} observations are taken from 1983 - 1992 and those of \NH from 2008 - 2017 for the radial heliocentric distance interval 11--38\,\au, and the solar cycle observed by \NH was much weaker than that observed over this distance interval by \voy2 \citep{lockwood2011,zhao2014}.
\begin{figure}[t]
	\includegraphics[width=\linewidth]{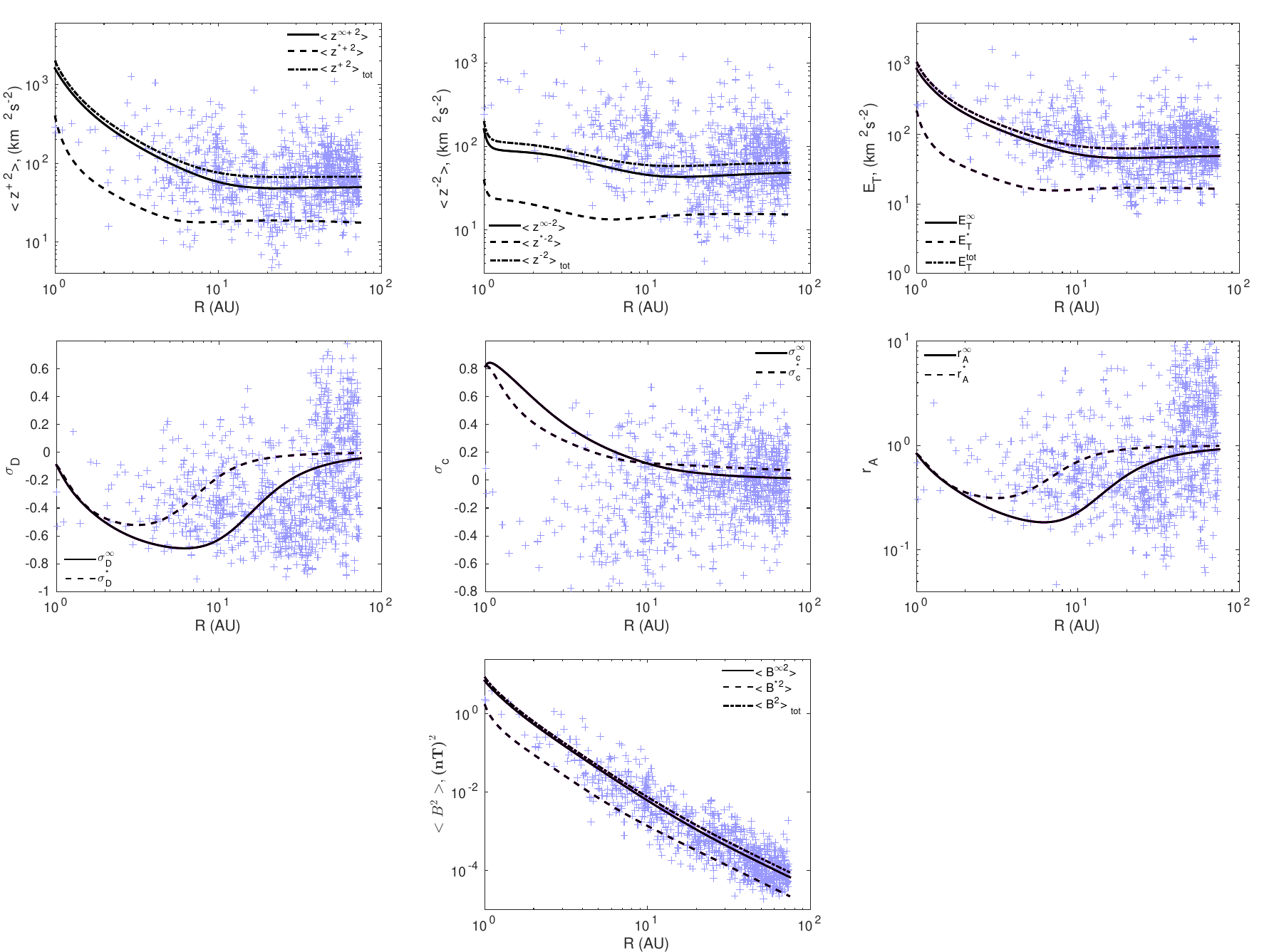}
	\caption{Comparison between the theoretical (NI MHD) and \textit{V2} derived turbulence quantities as a function of heliocentric distance. Solid and dashed curves correspond to 2D and slab turbulence quantities respectively, and the dash-dotted curve to the sum of 2D and slab quantities. Top left: Variance of the outward propagating  Els\"asser variable energy density, $\langle {z^+}^2 \rangle$. Top middle: Variance of the inward propagating  Els\"asser variable energy density, $\langle {z^-}^2 \rangle$. Top right: Total energy in turbulent fluctuations $E_\mathrm{T}$. Second panel, left: Normalized residual energy $\sigma_\mathrm{D}$. Second panel, middle: Normalized cross helicity $\sigma_\mathrm{c}$. Second panel, right: Alfv\'en ratio $r_\A$. Bottom middle: The variance in magnetic field fluctuations $\langle B^2 \rangle$.  Reproduced from  \citet{zank2018}.}
	\label{fig:zank_2}
\end{figure}
The variance of the fluctuating thermal plasma density is displayed in the right panel of Fig.  \ref{fig:Adhikari_2}. The theoretical fluctuating density variance $\langle {\rho_s^{\infty}}^2 \rangle$ shows good agreement with \textit{V2} observations (Fig.  \ref{fig:Adhikari_2}, top right panel), but underestimates the SWAP derived values (bottom right panel). 

The PUI mediated model of \cite{zank2018} predicted various turbulence quantities from 1--75\,\au, as shown in Fig. \ref{fig:zank_2}, which also shows the corresponding values derived from \textit{V2} observations (see \cite{adhikari2017b}). These results are slightly different from the results predicted by \cite{adhikari2017b} assuming that the background radial SW speed $U$ is constant. Pickup ions lead to a gradual decrease of the SW speed, and the background density and magnetic field are modified accordingly. The radial dependence of the background flow, density, and magnetic field influences the evolution of turbulence throughout the heliosphere. 

\begin{figure}[h!]
	\centering  
	\includegraphics[width=\textwidth]{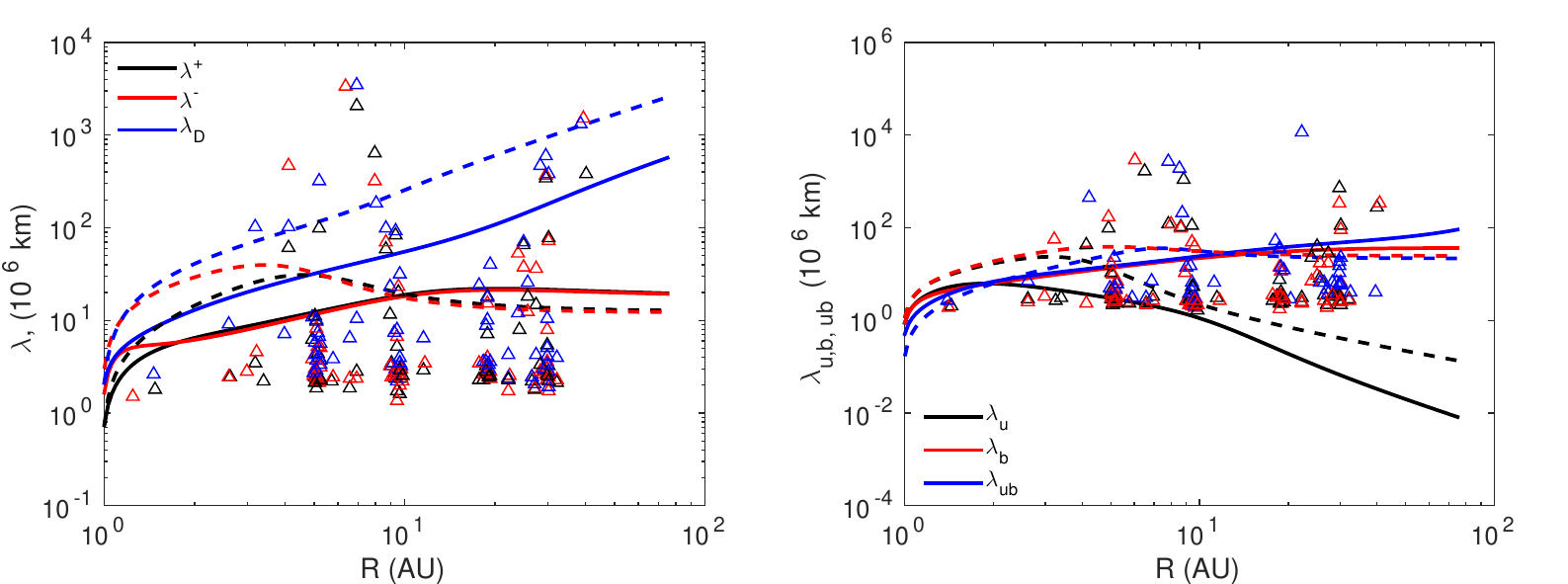}
	\caption{(Left panel) Correlation length corresponding to the energy in forward propagating modes (black curves and triangles), the energy in backward propagating modes (red curves and triangles), and the residual energy (blue curves and triangles). (Right panel) Correlation length corresponding to the fluctuating kinetic energy (black curves and triangles), the fluctuating magnetic energy (red curves and triangles), and the cross-correlation between covariance of velocity and magnetic field fluctuations (blue curves and triangles). The solid curves denote 2D modes and the dashed curves are used for the slab component. Scatter triangles indicate \textit{V2} observations. Reproduced from \citet{adhikari2017b}. }\label{fig:correlation_length}
\end{figure} 
\begin{figure}[h!]
	\centering  
	\includegraphics[width=\textwidth]{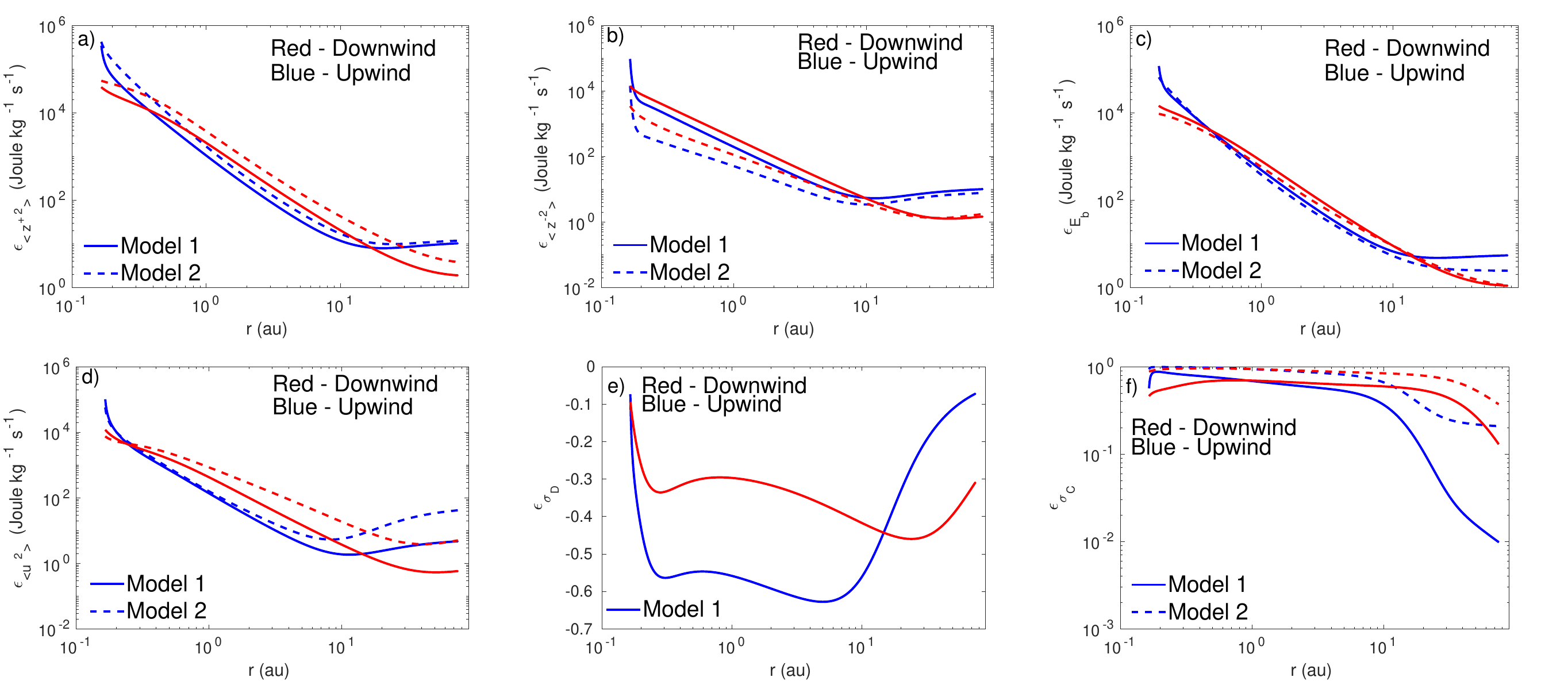}
	\caption{Turbulence cascade rate as a function of heliocentric distance.  Panels (a) to (f)  show the turbulence cascade rate corresponding to the energy in forward propagating modes,  energy in backward propagating modes,  fluctuating magnetic energy,  fluctuating kinetic energy,  normalized residual energy, and normalized cross-helicity, respectively. The blue curves denote the turbulence cascade rate corresponding to the upwind direction, and the red curves the turbulence cascade rate corresponding to the downwind direction. The solid and dashed curves represent the turbulence cascade calculated by models 1, and 2, respectively. The upwind results are along the trajectory of \voy2. The downwind results are along the trajectory of \textit{Pioneer} 10 (from 10\,\au to 60\,\au, HGI longitude from  320$^\circ$ to 360$^\circ$, HGI latitude from 7$^\circ$ to 3$^\circ$). Reproduced from \citet{adhikari2021}. }
	\label{fig:cascade_rates}
\end{figure} 
The energy density in the forward and backward Els\"asser variables is displayed in the top two plots of Fig. \ref{fig:zank_2}. Here, the solid curves denote the majority quasi-2D component, the dashed curves the minority NI/slab component, and the dashed-dotted curves the quasi-2D + NI/slab component. The  \textit{V2} observations are shown with blue plus symbols. Although the observed values have considerable dispersion, the predicted evolution in the forward, backward, and total Els\"asser energy densities is consistent with observations. The normalized residual energy $\sigma_\mathrm{D}$ shows that both the theoretical quasi-2D and slab components decrease towards a magnetically dominated state within $\sim 7$\,\au. However, as the PUI-driven turbulence becomes more important beyond the ionization cavity,  $\sigma_\mathrm{D}$ increase towards zero with increasing heliocentric distance, i.e., turbulence becomes increasingly Alfv\'enic with $\langle u^2 \rangle \simeq \langle B^2 /(\mu_0 \rho) \rangle$. The normalized cross-helicity $\sigma_\mathrm{c}$ monotonically decreases to zero as distance increases, indicating that the energy flux in forward and backward propagating directions gradually becomes approximately equal, in accord with observations. Although PUIs in the outer heliosphere drive the NI/slab component of turbulence, it remains a minority component. 

The fluctuating magnetic energy is displayed in the bottom panel of Fig. \ref{fig:zank_2}, indicating that theory and observations are consistent \citep{zank1996, matthaeus1999c,smith2001}. Figure \ref{fig:zank_2} provides a fairly complete characterization of the macroscopic (energy-containing scale) turbulence state throughout the heliosphere from 1--75\,\au.

The correlation length is an important quantity in turbulence because it helps control the energy decay rates. Figure \ref{fig:correlation_length} shows the comparison between theoretical and observed correlation lengths as a function of the heliocentric distance. The theoretical 2D correlation length corresponding to the energy in forward propagating modes (solid black curve) and the energy in backward propagating modes (solid red curve) increases gradually from 1\,\au to $\sim 20$\,\au, and then flattens with distance. However, since there is no turbulent shear source in the NI/slab turbulence transport equation, the theoretical slab correlation length corresponding to the energy in both forward and backward propagating modes increases initially, and then decreases slightly due to the presence of pickup ions in the outer heliosphere. The theoretical 2D and slab correlation length of the residual energy increases gradually as distance increases. In the right panel of Fig. \ref{fig:correlation_length}, the theoretical 2D and slab correlation length corresponding to the fluctuating kinetic energy and the cross-correlation between covariance of velocity and magnetic field fluctuations increase with distance. However, the opposite behavior is shown by the theoretical 2D and slab correlation length of the velocity fluctuations. 

Recently, \cite{adhikari2021} found that the turbulence property of the SW in the upwind direction is different from that in the downwind direction (along the \textit{Pioneer} 10 trajectory), due to the different PUI production rates \citep{nakanotani2020}. Therefore, the turbulent heating rates in the upwind and downwind directions are different. Figure \ref{fig:cascade_rates} shows the turbulence cascade rates as a function of heliocentric distance, in both the upwind and the downwind directions, and compares results of two transport models.
The solid curve corresponds to the turbulence cascade rate obtained from the turbulence transport theory \citep[Model 1, incompressible MHD,][]{zank2012}, and the dashed curve corresponds to the turbulence cascade rate obtained from the dimensional analysis between the power spectrum in the energy-containing range and the inertial range \citep[Model 2, NI MHD,][]{adhikari2017c}. The turbulence cascade rates corresponding to the (fluctuation) Els\"asser energies, magnetic energy,and kinetic energy all decrease gradually until $\sim 20$\,\au and $\sim 30$\,\au in the upwind and downwind directions, respectively. However, these turbulence cascade rates increase or flatten after $\sim 20$\,\au in the upwind direction, and flatten or slowly decrease after $\sim 30$\,\au in the downwind direction.



\section{Turbulence at the heliospheric termination shock}\label{sec:TS}
\begin{figure}[t!]
	\centering
	\includegraphics[width=\linewidth]{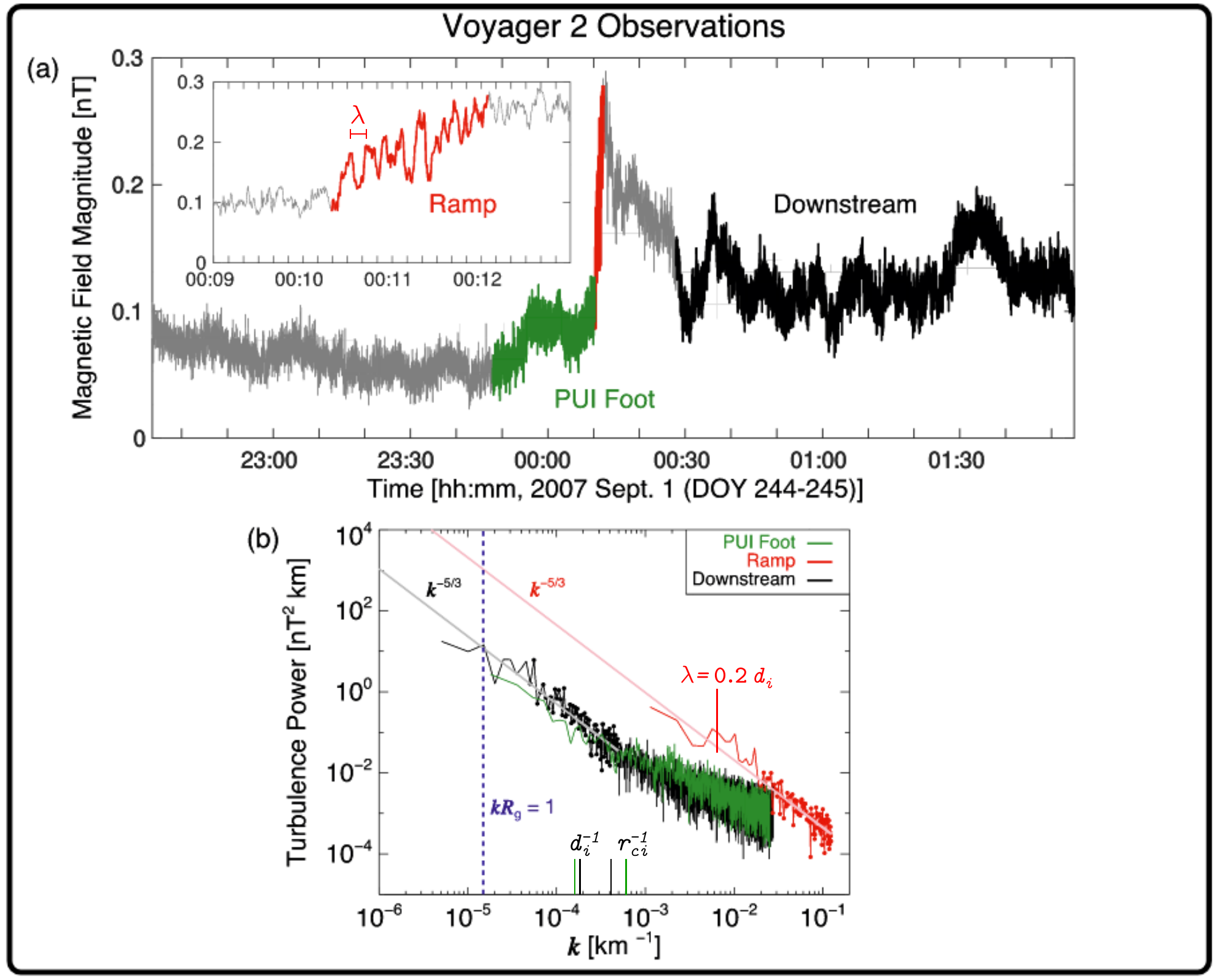} 
	\caption{\voy2 observations of magnetic field at the HTS \citep[third crossing;][]{burlaga2008}. (b) Turbulence power spectra in the PUI foot (green), ramp (red),and downstream (black) regions. The wavenumber at the upstream PUI Larmor scale is shown in blue, while the (inverse) inertial length and Larmor radius of thermal protons in the upstream and downstream regions are shown in green and black, respectively. Adapted from \citet{zirnstein2021b}.}
	\label{fig:zirnTS}
\end{figure}
The HTS plays a fundamental role in shaping the nature of turbulence in the inner heliosheath \citep{zank2006,zank2010,zank2018}. In front of the HTS, turbulence consists of both preexisting SW fluctuations advected at the shock and locally generated fluctuations due to kinetic processes, and reflected fluctuations.  

\begin{wrapfigure}[18]{r}[0pt]{0.55\textwidth}\vskip-15pt
	\centering
	\includegraphics[width=0.5\textwidth]{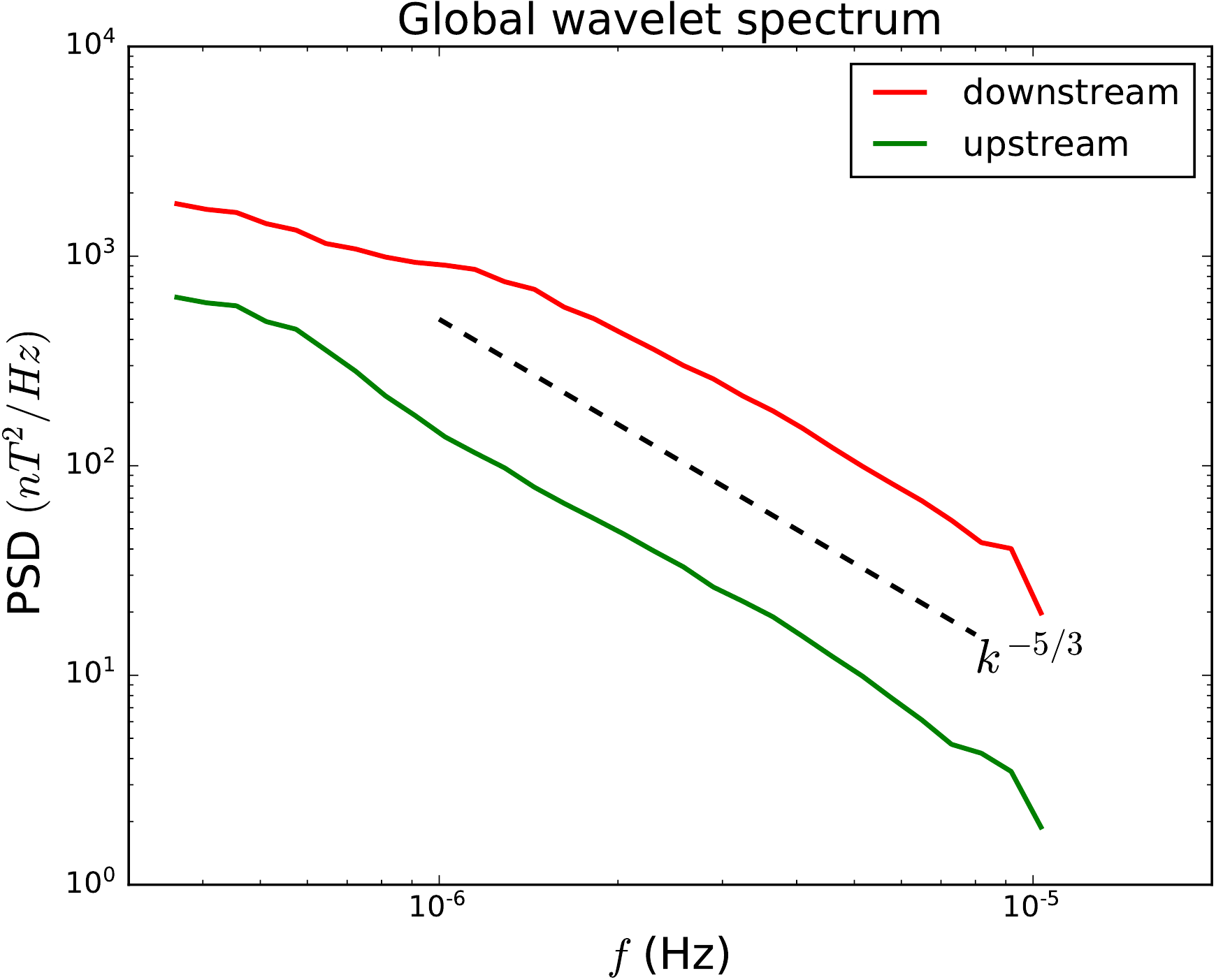}
	\caption{PSD of magnetic field upstream (green curve) and downstream (red curve) of the HTS. The dashed line displays a $k^{-5/3}$ spectrum, as a reference. Adapted from \citet{zhao2019b}.\label{fig:PSDzhao}} 
\end{wrapfigure}

\begin{figure*}{l}{}
	\centering
	\includegraphics[width=\textwidth]{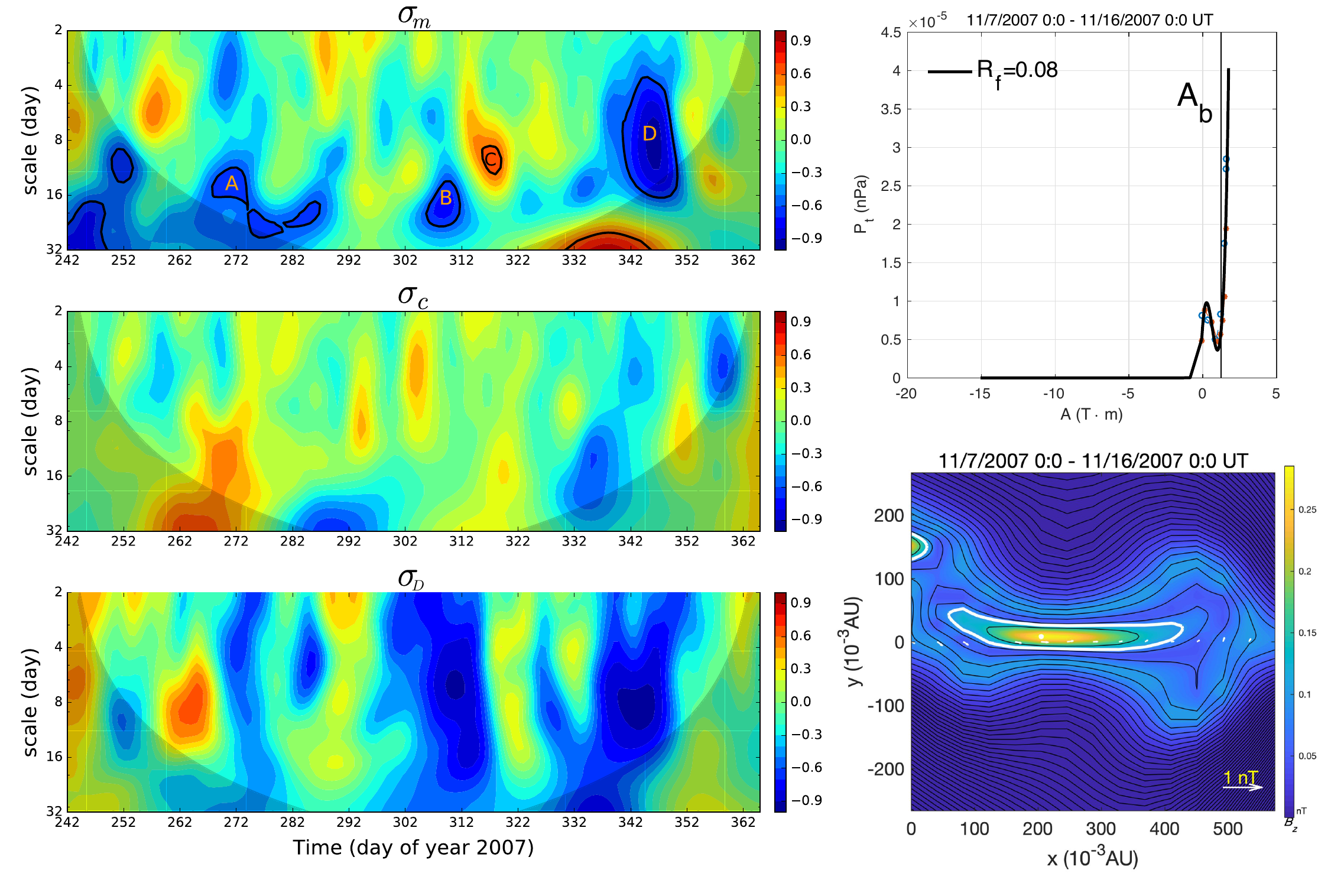}
	\caption{The left panels show the normalized reduced magnetic helicity $\sigma_\mathrm{m}$, normalized cross helicity $\sigma_\mathrm{c}$, and normalized residual energy $\sigma_\mathrm{D}$ spectrograms from a wavelet analysis for 124~days after the HTS crossing.  The right panels show the reconstruction map (top) and $P_\mathrm{t}$ vs.\ $A$ curve (bottom) of the GS reconstructed magnetic flux rope behind the HTS. Adapted from \citet{zhao2019b}. \label{fig:zhaoHelicity} }
\end{figure*}

The third HTS crossing observed by \voy2 \citep[hereinafter TS3, 2007 August 30,  84 AU;][]{stone2008} allowed to detect magnetic field fluctuations inside the shock structure \citep{burlaga2008}. Since the shock thickness was estimated to be $\sim 6,000 $\,km ($\sim d_\mathrm{i}$), these observations are only possible when using data of the highest resolution (0.48\,s). Intense quasi-periodic fluctuations of $B$ with wavelength $\sim 0.2\,d_\mathrm{i}$ characterize the shock ramp (see Fig.~\ref{fig:zirnTS}). Spectral analysis of turbulence for this specific crossing were recently conducted by \citet{zhao2019b} and \citet{zirnstein2021b}.
Figures~\ref{fig:zirnTS}--\ref{fig:zhaoHelicity} are reproduced from these studies.

\citet{zhao2019b} analyzed turbulence at relatively large scales using \voy2 data of 1 day resolution. Two intervals of 122~days were selected immediately upstream and downstream of the HTS. Power spectra are obtained from wavelet analysis, and show that magnetic turbulence on both sides exhibits a $f^{-5/3}$ power law in the frequency range $10^{-6}\,\mathrm{Hz}\lesssim f_\SC\lesssim 10^{-5} $\,Hz, as shown in Fig.~\ref{fig:PSDzhao}. The downstream spectrum is enhanced by a factor of $\sim 10$ with respect to the upstream spectrum. 

\citet{zirnstein2021b} computed magnetic turbulence spectra using \voy2 data at the highest resolution of  0.48\,s  and considered short time intervals that include the shock's PUI foot (green curves in Fig.~\ref{fig:zirnTS}), the ramp (red), and the downstream region (black), excluding the overshoot. The wavenumber range for the spectra shown in Fig.~\ref{fig:zirnTS}(b) include the gyroscale of PUIs with speed $\sim 335$\,km\,s$^{-1}$ ($R_\mathrm{g}$, in Fig.~\ref{fig:zirnTS}), the inertial length, and the Larmor radius of the thermal protons ($d_\mathrm{i}, r_\ci\approx 5300,1600 $\,km in front of the HTS and 5800, 2300~km behind of it, respectively). The flatter upstream spectrum for 
$k \gtrsim 10^{-3}$\,km$^{-1}$ 
may be due to the presence of $1/f$ noise from the fluxgate magnetometers. Interestingly, the turbulence power in the ramp is higher by a factor of $\sim100$ as compared to the upstream $k^{-5/3}$ spectrum. \citet{zirnstein2021b} investigate the effect of turbulence at the HTS on PUIs acceleration using a test-particle model. According  to these simulations,  the turbulence intensity observed by \voy2 at the PUI gyroscale  upstream of the HTS, $\langle \delta B^2 \rangle /B_0^2 \approx 0.01$, is not sufficient to explain the observed suprathermal tail in the proton spectrum measured by \IBEX at energies $\gtrsim2$ keV. Values similar to those extrapolated using the ramp spectrum, $\langle \delta B^2\rangle/B_0^2\approx0.1$, may be necessary to explain \IBEX observations. 
\citet{giacalone2010}  and \citet{giacalone2021} investigated particle acceleration at the HTS up to energies of 50\,keV via 2D hybrid simulations, and included background upstream SW turbulence in the form of random circularly polarized Alfv\'en waves with a Kolmogorov spectrum and normalized variance $\langle \delta B^2\rangle/B_0^2=0.5$. Note that this intensity refers to the total background turbulence spectrum and is a function of the chosen value for the correlation length (0.17\,AU). Their results suggest that large-scale turbulence ($k < R_g^{-1}$) may also affect the particle distribution \citep[see also][]{giacalone2005}. 

It should be noted that even without the background turbulence, 
fluctuations at the ion scale can be self-generated due to temperature anisotropy in the downstream region immediately behind the shock \citep[e.g.,][]{wu2009,liu2010,wu2010,kumar2018,gedalin2020}, by instabilities of the shock front 
\citep[e.g.][]{burgess2016rippling}, and by proton beams in the upstream region that are associated with protons reflected from the shock if the angle between the shock normal and the magnetic field is not too large \citep[e.g.][]{gedalin2021}.

The HTS crossing was also investigated via three-fluid simulations by \citet{zieger2015}. They highlighted the role of nonlinear, dispersive fast-magne\-tosonic modes associated with PUIs and thermal SW. Their coupling may results in a large-amplitude wave train downstream of the HTS that can evolve into shocklets. It would be interesting to compare these solutions with the results of hybrid simulations.   The relative importance of self-generated and background turbulence at the HTS and collisionless shocks is a topic of great interest and still an open challenge.  

Remarkably, \citet{gutynska2010} were able to investigate the $B$--$n$ cross correlations from \voy2 MAG and PLS data in two regions downstream of TS3, and found surprisingly large  cross correlation coefficients ($\sim0.8$). These values are large as compared to the typical correlation found in the Earth's magnetosheath ($\sim0.4$), at frequencies near
$10^{-4}$\,Hz to $4\times10^{-3}$\,Hz. 
However, the mixed signs of the correlations suggest that there was no preferred wave mode in those intervals.
   
\subsection{Flux ropes and their role in the transport of energetic particles at the HTS}
\citet{zhao2019b} also identified magnetic flux rope or magnetic island structures behind the HTS. 

These structures are identified as patches with enhanced magnetic helicity ($|\sigma_\mathrm{m}|\ge 0.7$), labeled as  A--D in the top panel of Fig.~\ref{fig:zhaoHelicity}. Since the feature of enhanced magnetic helicity is shared with circularly polarized Alfv\'en waves, a small cross helicity ($\sigma_\mathrm{c} \sim 0$) and negative residual energy ($\sigma_\mathrm{D} < 0$) are required for identifying magnetic flux ropes confidently.
The Grad--Shafranov (GS) reconstruction technique further confirms the finding by providing a reconstructed 2D cross section for one of the flux ropes as an example (bottom left panel).
The bottom right panel shows the $P_\mathrm{t}-A$ curve from the GS reconstruction, where $P_\mathrm{t}=p+B_z^2/2\mu_0$ is the total pressure
and $A(x,y)$ is the magnetic flux function. The double-folding pattern with a fitting residual of $R_f = 0.08$ indicates a good fit quality.
The flux rope appears to have a scale size of $\sim 0.4$\,AU.
The origin of flux ropes is unknown. In the supersonic SW small-scale magnetic flux ropes are often recognized as a representation of quasi-2D turbulence that is a majority component of the SW turbulence in this region \citep[e.g.,][]{zank2017b}.
The flux ropes identified in the IHS may be evidence for 2D fluctuations in the outer heliosphere as they are transmitted and amplified downstream of the HTS. The compression at the HTS may also lead to enhanced magnetic reconnection, which generates multiple magnetic flux ropes. 
\citet{zank2018} described the transmission of MHD turbulence across the PUI-modified HTS using the NI MHD model. On the shock passage, the model predicts a strong amplification of the 2D component of turbulence, consistent with observations of flux ropes. The model also predicts a downstream state in which the turbulent kinetic energy dominates over the magnetic energy. However, the observed increase of magnetic turbulence variance is larger than predicted, which is possibly due to the presence of compressive modes not accounted for in \cite{zank2018}.  \\ 
\citet{zank2021} have recently analyzed in detail the transmission of 2D MHD modes (including acoustic, entropy, vortical, and magnetic island modes) across collisionless shocks. The agreement with \voy2 observations presented above suggest that these structures are an important component of MHD turbulence at the HTS. 

\begin{wrapfigure}[32]{r}[0pt]{0.51\linewidth}\vskip-10pt
	\centering
	\includegraphics[width=\linewidth]{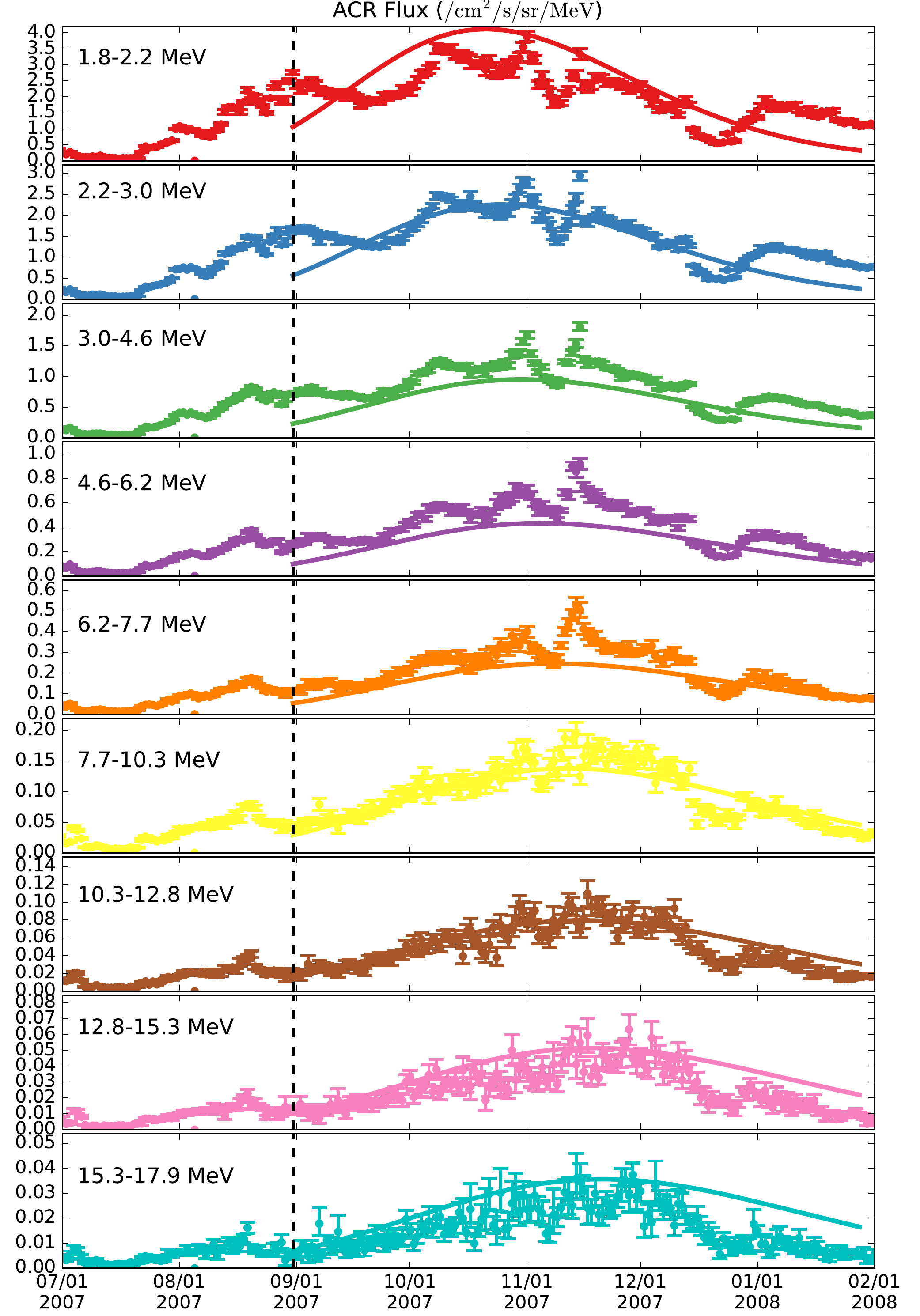}
	\caption{ACR proton flux evolution during the period from 2007 July 1 to 2008 February 1. The uncertainties of the observed proton flux are plotted as error bars. The dashed vertical line represents the HTS crossing, and the smooth curves behind the HTS show our theoretical modeling results. Adapted from \citet{zhao2019b}. \label{fig:flux}}
\end{wrapfigure}
It has been suggested that magnetic flux ropes generated by magnetic reconnection may accelerate particles in a stochastic fashion, which may be partially responsible for the generation of anomalous cosmic rays in the inner heliosheath \citep{drake2010, drake2017, zank2015b, leroux2016}. 
The idea is that enhanced magnetic reconnection in the IHS due to the compression associated with crossing the HTS leads to the generation of multiple magnetic flux ropes.
The interaction between flux ropes then leads to particle acceleration.
\cite{zank2014b} presented a theoretical framework describing the acceleration and transport of particles in regions of interacting magnetic flux ropes, taking into account processes of magnetic island contraction and merging.
The theory was applied by \cite{zhao2018, zhao2019a} to an energetic particle event observed by \textit{Ulysses} near 5\,\au. In the event, the energetic particle fluxes are found to be strongly enhanced after the crossing of an interplanetary shock and the peak enhancement occurs $\sim$ 5 days after the shock crossing.

The Zank et al. theory was also applied to the ACR proton flux enhancement behind the HTS by \cite{zhao2019b}.
Fig.~\ref{fig:zhaoHelicity} presents evidence of magnetic flux ropes after the HTS crossing, obtained from  \voy2 LECP data.
In Fig.~\ref{fig:flux}, the ACR fluxes are shown for different energy channels between 1.8 MeV and 17.9 MeV.
The measurements are fitted quantitatively to the \cite{zank2014b} theory using a Monte Carlo Markov Chain (MCMC) technique, as shown by the solid lines in the figure.
The fitted lines successfully reproduce  that (i) there is enhancement of the ACR flux behind the HTS; (ii) the enhancement is stronger for higher energy particles within the considered energy channels; and (iii) the location of peak flux enhancement is further away from the shock for higher energy particles. We note that the analysis here applies only to the region very close to the HTS (within $\sim 1$\,AU).
It is likely that other acceleration mechanisms such as the diffusive shock acceleration are active deeper in the heliosheath. For recent reviews on ACRs and particle acceleration processes at collisionless shocks, see \citet{Giacalone_EA_this_volume} and \citet{Perri_EA_this_volume} in this volume.

\FloatBarrier



\section{Turbulence in the inner heliosheath}
\label{sec:IHS}

In the IHS the solar wind plasma is subsonic, having been decelerated at the HTS. 
Much of our current knowledge of turbulence in the IHS has been acquired via in situ  (\Voyager) observations.  In particular, analyses of \Voyager measurements suggest that  significant levels of compressible fluctuations are present in the IHS \citep{burlaga2006, burlaga2008, fisk2008, burlaga2009, burlaga2012, burlaga2012b,burlaga2014d, richardson2013, fraternale2017phd,fraternale2019a}.
Thus, the nature of the turbulence there differs from that in the supersonic SW upstream of the termination shock, where the fluctuations are predominantly incompressible \citep[e.g.,][]{tu1994, roberts2018}.

Unfortunately,  this compressible turbulence in the IHS is still poorly understood, from both the observational and the theoretical perspectives. Observationally, this is mainly due to the well-known  limits of 1D measurements, unavailability of PUI measurements and plasma data (\voy1), and the level of noise. Consequently, to date, the dissipation regime of turbulence has been inaccessible to our investigations.
On the theory side, compressible MHD turbulence is clearly richer than its incompressible counterpart having additional parameters such as the sonic Mach number and the plasma beta,  $\beta \approx c_\mathrm{s}^2 / V_\mathrm{A}^2 $. 
Aspects that can play important roles include the sub or supersonic character of the system, the size of the $ \beta $ relative to unity, and the nature of any driving of the velocity field (e.g., is it the solenoidal velocity that is driven, the compressive component, or a combination).
Simulation studies investigating various features of compressible MHD, such as energy transfer (both across scales and between magnetic, internal energy, incompressible $\vct{v}$, and compressive $\vct{v}$ components), and variance and spectral anisotropy have been reported on and these may provide starting points for understanding IHS \Voyager observations
   \citep[e.g.,][]{GhoshMatt90,cho2002,cho2003a,VestutoEA03,carbone2009,KowalLazarian10, banerjee2013,OughtonEA16, GreteEA17-Exfer,hadid2017,YangEA21-budget}.
However, further studies are surely needed, including ones tailored to IHS conditions.

\subsection{An overview of the observed structures in the IHS}
In general, turbulence is present in the IHS and consists of both random fluctuations and coherent structures. The HTS is certainly the major source of turbulence in the IHS, at least in the direction of the heliospheric ``nose'', as it transmits and possibly amplifies the full spectrum of fluctuations from the supersonic SW to the IHS \citep[e.g.,][]{zank2018,zank2021}.

\subsubsection{Large-scale fluctuations}
\label{sec:IHS-large}
\begin{figure}[t]
	\centering
	\includegraphics[width=\textwidth]{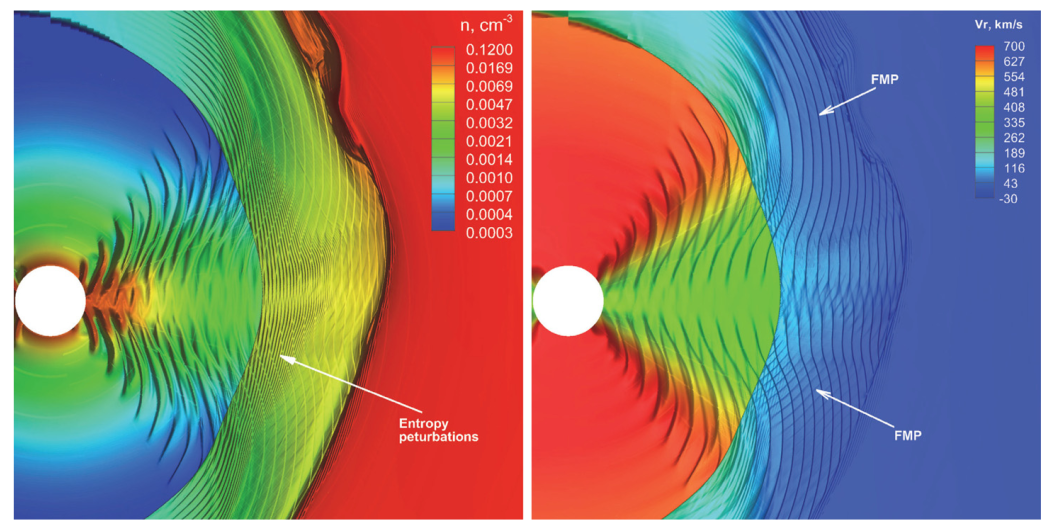}
	\caption{Density distribution (left panel) and radial component of the velocity (right panel) in the meridional plane for the solar cycle 23 minimum as the result of corotating streams propagating into the IHS. Entropy and fast magnetosonic perturbations are shown. Reproduced from \cite{borovikov2012}. \label{fig:borovikovCIR}}
\end{figure}

On large scales, transient structures of solar origin such as global merged interaction regions (GMIRs) have been observed in the IHS \citep{burlaga2011b,burlaga2016b,richardson2017}. They are associated with strong magnetic fields and enhancements of plasma density and temperature, and may be 
moving
fast enough to generate shocks or pressure pulses. When GMIRs interact with the HP, they produce transmitted shocks or compression waves in the VLISM, and reflected perturbations in the IHS. Figure \ref{fig:borovikovCIR} from \cite{borovikov2011} shows an example of the complex patterns and the different fluctuation modes that can arise as a consequence of corotating streams interacting with the HTS. Time dependent and data-driven 3D simulations are necessary to understand the dynamics and time-space evolution of the large-scale structures in the outer heliosphere. A recent study by \citet{pogorelov2021}, provides animations of magnetic and thermal pressure along \voy1 and \voy2 trajectories from earlier numerical solutions of \citet{pogorelov2017b,kim2017b}.   Large-scale features include the sector structure, however it is well known that the periodic sector structure no longer exists beyond $\sim 20$\,AU \citep[e.g.,][]{burlaga1994,pogorelov2017a}. In fact, in the IHS, sectors and regions with mixed polarity show complicated polarity patterns. \voy1 and \voy2 have observed regions of mostly unipolar fields and ``sector regions'' of mixed polarity, statistically characterized by \citet{richardson2016}. The sector region is the region swept by the HCS.  The presence of two topologically different regions within the IHS suggests that the properties of turbulence should also change across these regions, with implications for the transport of energetic particles \citep{burlaga2009b,opher2011,florinski2013,hill2014}. {The properties of turbulence in the heliotail are unknown and will not be discussed here. Numerical models show the strong, likely dominant effect of the solar cycle variations on the generation of large-scale structures in the tail \citep{pogorelov2017a} and the possible onset of large scale instabilities of various nature of the collimated SW lobes at high latitudes, a prominent feature of steady-state, spherically-symmetric solutions \citep{yu1974,pogorelov2015,opher2021}.}

Besides the structures of solar nature,  instabilities at the HP may also play an important role in the injection of fluctuation energy in the heliosheath at large scales ($\gtrsim 5$\,AU) and the generation of structures in the IHS. Long before \voy1 actually observed signatures of HP instability \citep[August 2012, DOY 210 -- 238]{burlaga2013}, \citet{fahr1986} first suggested that the HP may undergo Kelvin--Helmholtz (KH) instability at the HP flanks, and may also be Rayleigh--Taylor (RT) unstable due to local accelerations of the HP, especially when the HMF is weak. Early studies on this topic were conducted by \citet{baranov1992,chalov1994,chalov1996,liewer1996,zank1996b,belov1999,zank1999b,pogorelov2000,ruderman2000,florinski2005,ruderman2006,borovikov2008}. In particular,  \citet{liewer1996} and \citet{zank1996}  highlighted the role of the ion-neutral drag on the possible development of RT near the nose. \citet{zank1999b} have shown that charge exchange between neutrals and ions acts essentially as an effective gravitational term that can trigger the RT instability.   \citet{borovikov2008} further identified a mixed RT--KH form of instability on the HP flanks, assisted by hot secondary neutrals created by charge exchange in the SW. As shown by \citet{borovikov2008} the absence of HMF can increase the instability dramatically.  
\cite{borovikov2014} simulated a 3D heliosphere and demonstrated that the RT instability is considerably suppressed near the HP nose by the unipolar, unrealistically large, HMF in steady-state models, but is triggered in the presence of solar cycle effects. This is why accounting for solar cycle in simulations is not important per se, but because it creates favorable conditions for the instability to develop, when $B$ is small due to the presence of sector regions  \citep[see also][]{pogorelov2021}.
Analytic studies of the RT instability have also been carried out \citep{avinash2014,ruderman2015}. 
Results in \citet{pogorelov2015} indicate that only the KH instability is 
found in the tail region. Eventually, solar cycle simulations of \citet{pogorelov2017b} suggested that at the time of the HP crossing the KH/RT instability was more likely to occur at northern  latitudes, while magnetic reconnection may reveal itself as a tearing mode (plasmoid) instability at southern latitudes (see Fig.~\ref{fig:HCS}). 
A situation similar to the RT instability is found in the magnetic flux tube interchange instability, proposed by \cite{krimigis2013} to explain \voy1 observations and investigated theoretically by \citet{florinski2015}. This instability may develop on smaller scales ($\sim 0.2 $--5\,AU), and results from the positive plasma pressure gradient and the IMF curvature. 
 
\subsubsection{Fine-scale fluctuations}  
\label{sec:IHS-fine}

Ion-scale coherent features observed in situ in the IHS include current sheets (CS) described by \cite{burlaga2009,burlaga2011} as proton boundary layers, magnetic holes, and magnetic humps. Their typical size lies in the range of 
$ \sim 1$--30\,$r_\mathrm{ci, {mix}}$, 
where $r_\mathrm{ci, {mix}}$ is the proton cyclotron radius of the plasma mixture ($T \sim 5 \times 10^6$\,K). The exact nature of these structure is unknown, but an hypothesis was advanced by \citet{burlaga2009} that they originate from mirror mode instability and evolve nonlinearly as solitons, and eventually as pressure balanced structures. A theory of solitary waves has been proposed by \citet{avinash2007}. Mirror mode instabilities have been identified near the HTS \citep{liu2007,genot2008,fahr2007} and within the IHS by \citet{tsurutani2011a} and \citet{tsurutani2011b}, who also pointed out the role of PUIs in the amplification of this instability.  

As discussed in Sec.~\ref{sec:transport_models}, models aiming at a quantitative treatment of turbulence in the entire inner heliosheath are sparse. One of the first  was presented by
\citet{usmanov2016} who used a four-fluid model to simulate the heliosphere and nearby LISM, including the transport of turbulence in this whole region.  The turbulence is modeled using a one-component model for incompressible MHD turbulence, and thus, as the authors explicitly state, cannot address the presence or role of compressive fluctuations. 
Nonetheless, the work demonstrates the principal feasibility of such extensions of previous inside-the-HTS transport modeling to regions beyond the HTS. 
Despite its limitations the model provides a valuable reference case for forthcoming simulations.
A second model by  \citet{fichtner2020} has recently addressed the problem of generating compressible fluctuations throughout the whole IHS. It applied the quasilinear theory to the initial evolution of compressible wave modes using initial values obtained from a simulation of a 3D model of the SW/LISM interaction. The basic idea is that the compressible fluctuations are a result of the proton mirror instability, which is a consequence of a perpendicular temperature anisotropy $A := T_\perp/T_\| > 1$ (with $\|,\perp$ referring to the orientation relative to the local magnetic field direction $\vct{B}/B$). The resulting fluctuations $\delta n_\mathrm{p}$ in proton number density $n_\mathrm{p}$ are anticorrelated with the associated magnetic fluctuations $\delta B$ via the relation \citep[e.g.,][]{liu2007}
\begin{eqnarray}
\frac{\delta n_\mathrm{p}}{n_\mathrm{p}} = -(A -1)\,\frac{\delta B} {B} .
\label{eq:dn_from_A_dB}
\end{eqnarray}
To compute these fluctuations throughout the IHS the structure 
of the latter was determined from a numerical simulation of the 3D large-scale heliosphere using the MHD code \textsc{Cronos} \citep{kissmann2018} for the model equations formulated in \citet{wiengarten2015}. These equations are the usual MHD equations for the large-scale quantities and  Eqs.~\ref{usm-1} to~\ref{usm-3} with the simplifying assumptions $\boldsymbol{\varepsilon}_\mathrm{m} = 0$, $q_\mathrm{T} = q_\mathrm{ph} = 0$, and the choice $\upalpha = 2\upbeta =0.8$. Using the solutions of this model obtained for the IHS \citet{fichtner2020} applied the theory of temperature anisotropy-driven kinetic instabilities reviewed by \citet{yoon2017}. 

\begin{figure*}[h!]
	\begin{center}
		\includegraphics[width=0.85\textwidth]{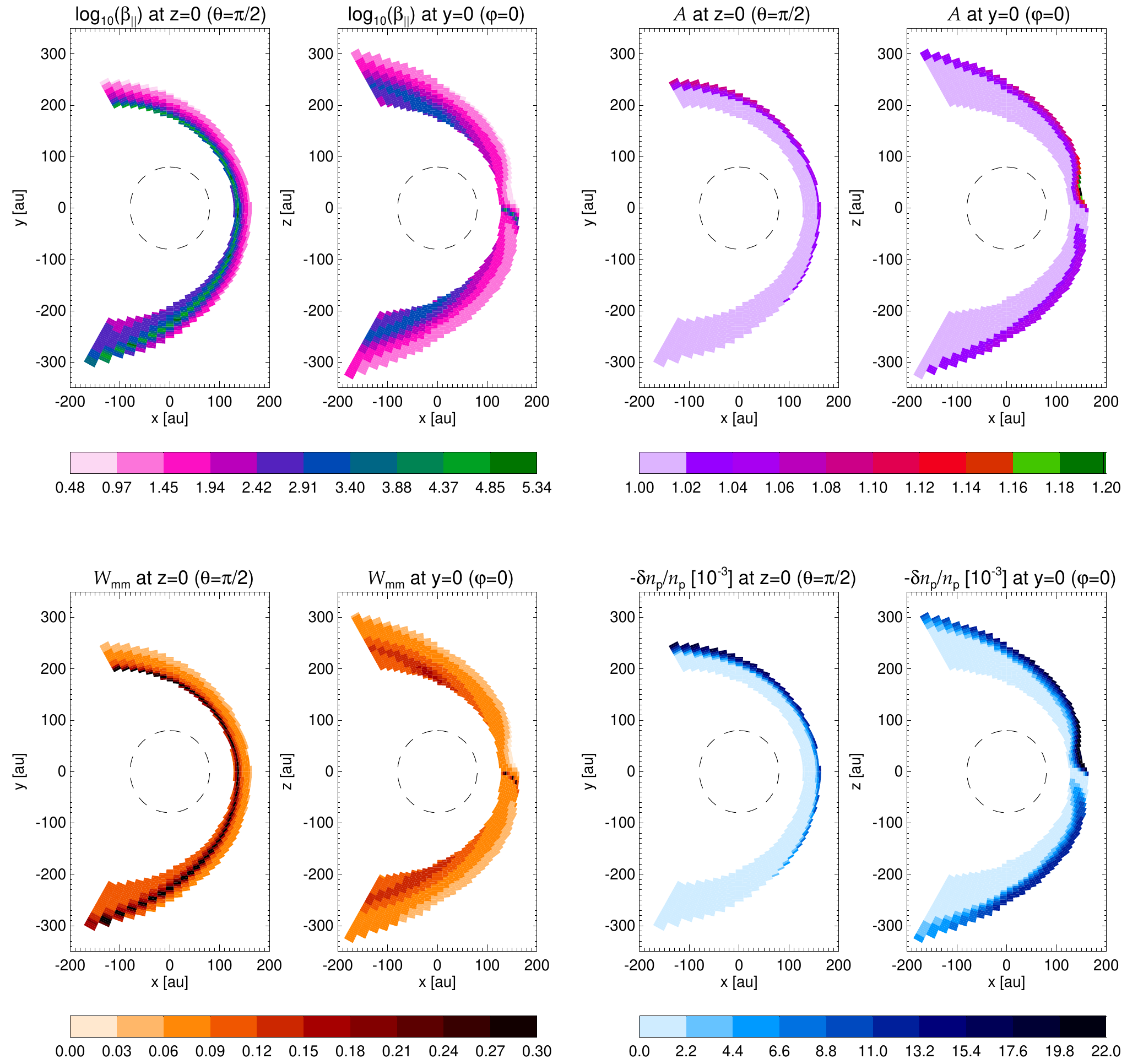}
	\end{center}
	\caption{The quasilinear evolution of the mirror-mode instability in the inner heliosheath (the inner/outer boundary of the color-coded region is the HTS/HP) in the equatorial plane ($\vartheta=\pi/2$, left subpanels) and in the meridional plane ($\varphi=0$, right subpanels): The final $\beta_\|$ (upper left panel) and the final temperature anisotropy $A$ (upper right panel), of the quasilinear evolution and the (normalized) energy density ${\cal W}_\textrm{mm}$ of the magnetic fluctuations (lower left panel), and the corresponding (normalized) amplitude of the corresponding fluctuations in number density (lower right panel).  The thin equatorial streak of very high beta values (green stripe in the upper left panel) occurs in an interface region slightly beyond the HTS characterized by very small magnetic field, which is likely to be an artifact of finite numerical resolution. Therefore, correspondingly high values of ${\cal W}_\textrm{mm}$ of up to $\sim2.3$	are neglected in the color bar of the panel below to allow the spatial structure to become fully discernible.	\citep[Reproduced from][]{fichtner2020}.
	}	\label{fig:ihsresults}
\end{figure*}

The central results are summarized with Fig.~\ref{fig:ihsresults}. 
The upper right panel reveals that the quasilinear evolution results in anisotropy values (that were initially   $A \gtrsim 1.25$) a little above unity across most of the IHS, which is consistent with the findings by, e.g., \citet{liu2007} and \citet{fahr2007}.
The corresponding magnetic energy density of the mirror mode-induced compressive fluctuations (lower left panel) and, via Eq.~(\ref{eq:dn_from_A_dB}), the associated, locally generated density fluctuations (lower right panel) are significant in large regions of the IHS, particularly also below a latitude of about $45^{\circ}$, i.e., in the region probed by the \Voyager spacecraft. At high northern latitudes and within the equatorial plane,  the energy density is decreasing towards the heliopause, which appears to reflect the distribution of the plasma beta. A comparison of the magnetic energy density associated with the mirror-mode instability with that associated with the MHD fluctuations as obtained from the \textsc{Cronos} simulation revealed that the compressible fluctuations indeed dominate in most parts of the inner heliosheath: only in the equatorial and high-latitude downstream vicinities to the termination shock do the incompressible fluctuations prevail. 

The computed the energy density of the locally generated magnetic fluctuations and the associated density fluctuations may serve as source terms in forthcoming models of the actual turbulence transport in the IHS.

\subsection{A statistical description of IHS fluctuations}
\label{sec:IHS-stat}

The description of turbulence obviously also requires a statistical approach. A series of studies \citep{burlaga1991,burlaga2003a,burlaga2004,burlaga2006b,burlaga2010,macek2012,macek2013,burlaga2013,macek2014} demonstrated the existence of a multifractal scaling symmetry for large-scale fluctuations ($>1$ day) in the distant SW and IHS. The multifractal formalism is a powerful classical tool to describe the structure of the dissipation rate in turbulence 
  \cite[see, e.g.,][]{meneveau1987,frisch1995}. 
The analysis of the fluctuations of $\vct{B}$ led to remarkable outcomes, such as demonstrations of 
(i) the existence of a  $P\sim f^{-1}$ power spectral regime on scales in the range of 1--100 days \citep[e.g,][]{burlaga2013b} and (ii) the presence of large-scale intermittency, although at lower levels with respect to the supersonic SW \citep[see][]{Richardson_EA_this_volume}.
\citet{burlaga2013b} applied the ``$q$-triplet'' concept from the nonextensive statistical mechanics  \citep{tsallis2009book} to provide evidence that fluctuations in the IHS are in a quasi-stationary, metaequilibrium state. 
Other studies
\citep{burlaga2009,burlaga2013,burlaga2019}  
indicate that magnetic field increments for time lags from 48\,s to 1\,day are well described by the $q$-Gaussian distribution with parameters $q$, $A_q$, $\beta_q$:
\begin{equation}
  f_q(\Delta B/\sigma)
         = 
    A_q [1+(q-1)\beta_q (\Delta B/\sigma)^2]^{\frac{1}{1-q}},
\end{equation} 
with the parameter $q$ is related to the kurtosis (or flatness) and is found to be as large as 1.6 in the IHS ($q=1$ corresponds to Gaussian statistics).  \citet{burlaga2009b,burlaga2010b,burlaga2017} have also shown that the daily distributions of $B$ are typically log-normal in the sector regions, but Gaussian in unipolar regions. 
 
\begin{figure}[h]
	\centering
	\includegraphics[width=\textwidth]{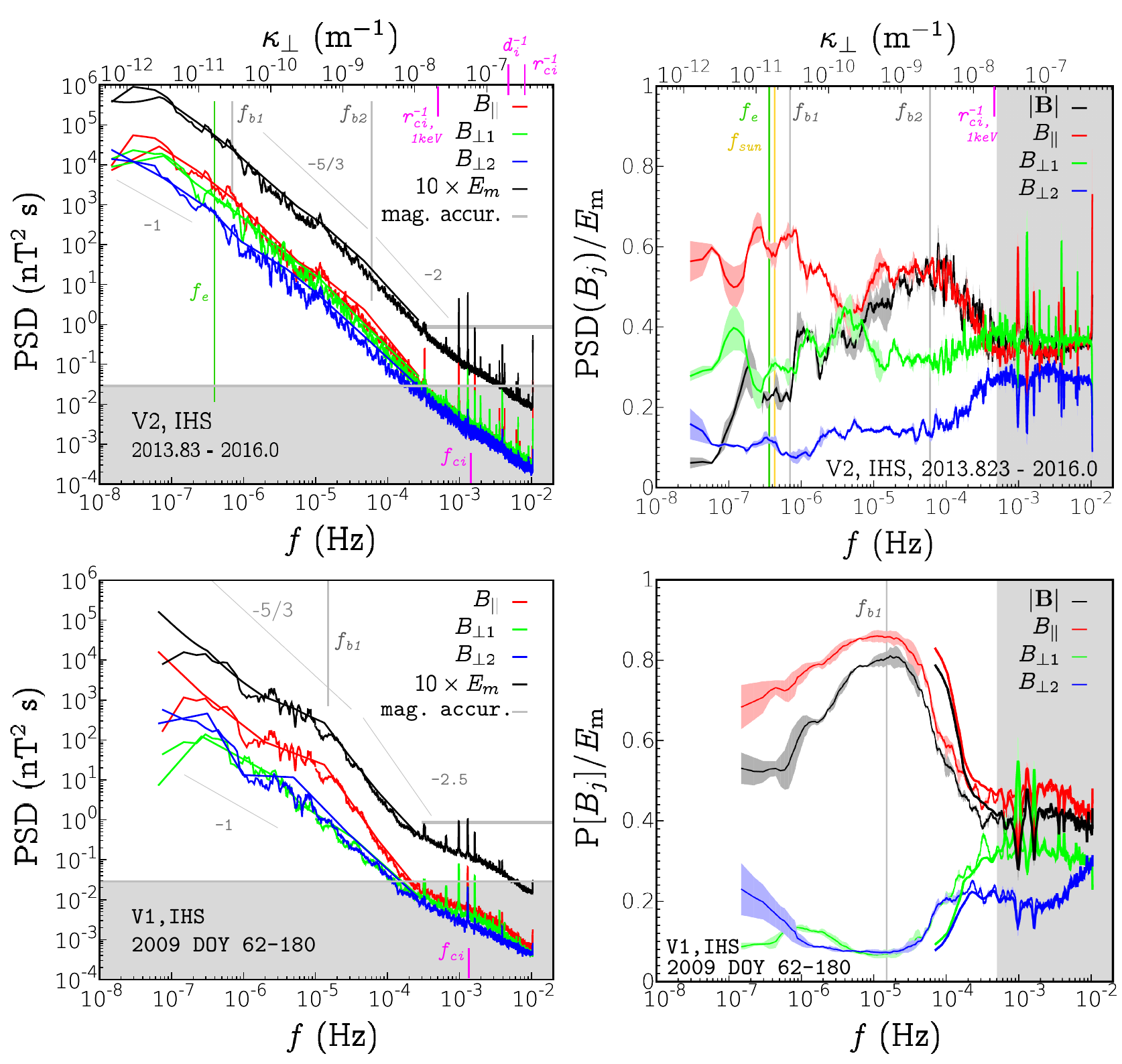}%
	\caption{(Left panels) Examples of magnetic turbulence spectra in the IHS at \voy2 (top) and \voy1 (bottom). (Right panels) Scale-dependent variance anisotropy and magnetic compressibility (black curve) for the same intervals. The gray bands indicate the range that may be affected by noise in the measurements. Top panels show results from \voy2 data in a sector region, bottom panels show analysis of \voy1 data in a unipolar zone. The green vertical line indicates the frequency corresponding to fluctuations that experienced one ``eddy-turnover''  if they were generated at the HTS. Adapted from \cite{fraternale2019a,fraternale2019b}. \label{fig:PSD_IHS}}
\end{figure}

The presence of intermittency and scaling laws are key ingredients of turbulence. To the best of our knowledge, evidence for the existence of different power-law regimes throughout the IHS was provided for the first time by \citet{fraternale2017phd} and \cite{fraternale2019a,fraternale2019b}. They conducted a spectral analysis of \voy1 and \voy2 high-resolution (48\,s) magnetic field data in several sector and unipolar regions identified by \citet{richardson2016} and \citet{burlaga2017}. Results are set forth in terms of power spectra and structure functions up to the fourth order.   Such analysis is challenging because of the sparsity of the 48\,s data. After the HTS, about 70\% of data points are missing due to tracking issues, noise, instrumental interference, and other reasons. Therefore, the combined use of different spectral estimation techniques becomes mandatory to rule out the numerous artifacts in the statistics. In particular,  \textit{compressed sensing}, a recent paradigm in signal processing \citep{donoho2006,candes2006a,candes2006b}, 
was successfully applied to SW turbulence analysis for the first time by \cite{gallana2016} and \cite{fraternale2017phd}, and subsequently applied to the analysis of IHS and VLISM \Voyager data, and magnetospheric turbulence \citep{sorriso2019}. 
\begin{figure}[t]
 	\centering
 	\includegraphics[width=\textwidth]{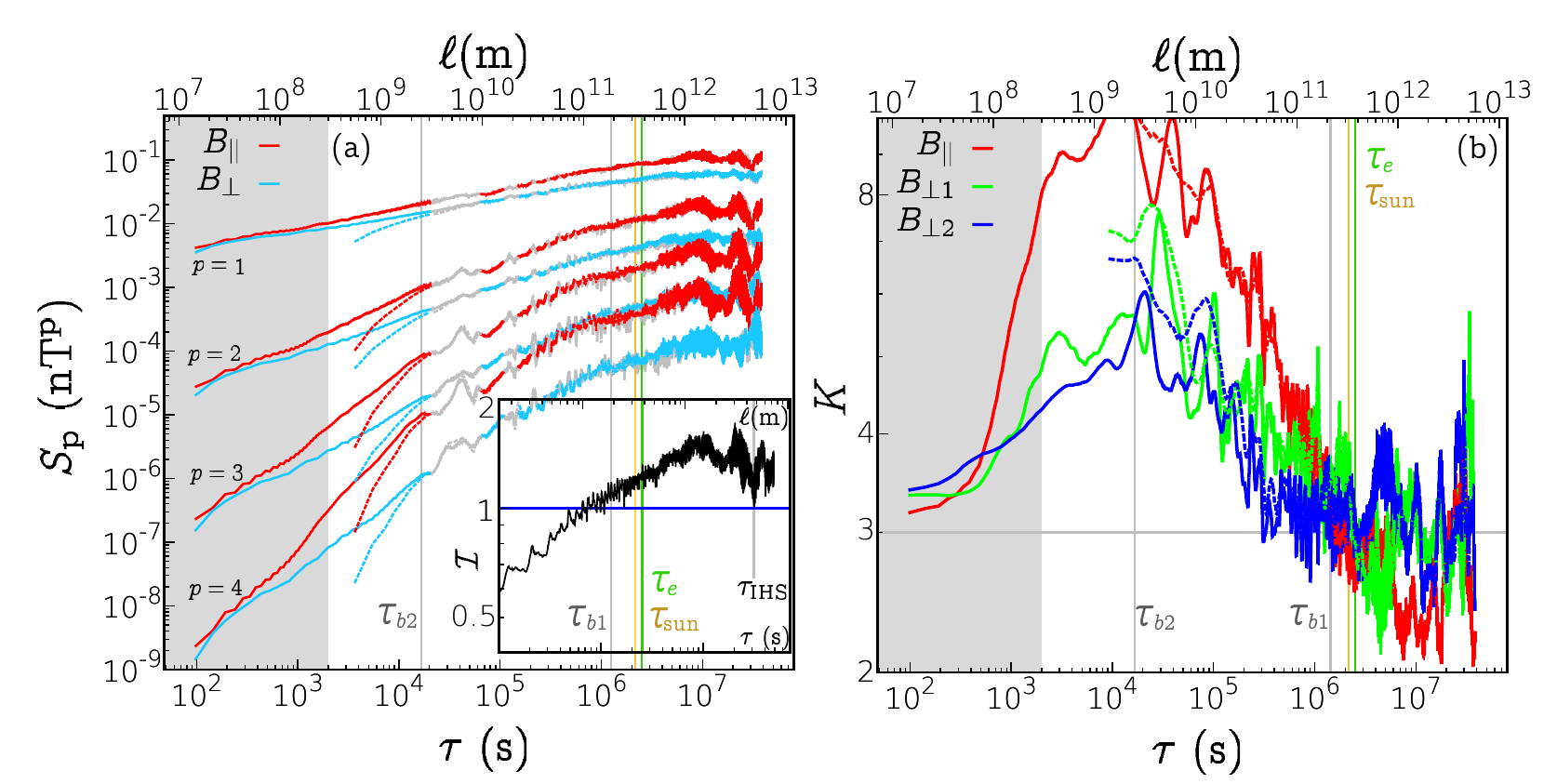}
 	\caption{(Left panel) Analysis of structure functions of magnetic field turbulence in the IHS (\voy2, 2013.83--2016, sector region). (Right panel) Scale dependent kurtosis, $K=S_4/S^2_2$. Adapted from \cite{fraternale2019b}. \label{fig:SF_IHS}}
\end{figure}

Figure \ref{fig:PSD_IHS} shows two examples of magnetic field turbulence spectra computed using 2 years of \voy2 observations in a sector region at 
$106 \pm 3.5$\,AU (top panels) 
and  120 days of \voy1 data at $109 \pm 0.5 $\,AU (bottom panels). 
The PSD is shown in the left panels, while the right panels show the scale-dependent variance anisotropy and the magnetic compressibility. The Larmor radius of thermal protons, 
$r_\ci \approx 2900 $\,km, the ion inertial length 
$ d_\mathrm{i} \approx 5200 $\,km, and the Larmor radius of  
1\,keV PUIs ($r_\mathrm{ci,~{1keV}}\approx45,000 $\,km) are shown in the top panels. The Taylor (frozen flow) hypothesis is applied only at \voy2, to obtain a rough estimate of the wavenumber perpendicular to $\bm B$, under the assumption of wavenumber anisotropy, $k_\perp > k_\parallel$. Clearly, 1D spectra contain contributions from all wave vectors. However, as pointed out by \citet{fraternale2019a} and \citet{zieger2020}, its application in the IHS is more critical than in the supersonic SW. The main reasons for this are the lower bulk flow speed  
($U \sim 150 $\,km\,s$^{-1}$ at \voy2), 
the presence of shocks and, possibly, finite-amplitude fluctuations of dispersive nature, whose propagation speed is affected by the energetically dominant population of PUIs. 
 
At the large scales,   \citet{fraternale2017phd} and \citet{fraternale2019a} confirmed the existence of a $1/f$ spectral regime in the frequency spectrum,  which is typically observed for $f_\SC<10^{-5} $\,Hz. They discussed the extension of this regime in the frequency space. A new finding was that a spectral break (labeled as $f_{b1}$ in Fig.~\ref{fig:PSD_IHS}) separates the $1/f$ regime from other power-law regimes observed at smaller scales.   The $1/f$ regime may be considered as a reservoir of energy for the turbulent cascade. Its frequency 
span
was found to decrease with distance and to be larger in unipolar regions than in sector regions. In the example of Figs.~\ref{fig:PSD_IHS}(top) and \ref{fig:SF_IHS}, the spectral break is seen at $f_{b1}\approx10^{-6} $\,Hz ($\ell_\perp\equiv2\pi/k_\perp\approx1 $\,AU). There is a possibility that the $1/f$ regime originates from SW turbulence being processed by the HTS (see also Fig.~\ref{fig:PSDzhao}). It may include perturbations such as those shown in 
Fig.~\ref{fig:borovikovCIR} and all large-scale structures described previously, but its nature remains unclear. 

At higher frequencies ($f_{b1}\lesssim f_\SC\lesssim 5\times10^{-4} $\,Hz) an inertial cascade regime was identified. In the unipolar regions sampled by \voy1, the (absolute) spectral index becomes large,   $\alpha\approx-2.5$, which is mainly ascribed to the dominant $\delta B_\parallel$ component.
At \voy2, a second spectral knee ($f_{b2}$, in the figures) is observed at  $f_\SC\sim10^{-4}$\,Hz ($\ell_\perp\approx0.015$\,AU, $\sim50~r_{\ci,\mathrm{~1keV}}$). This separates a 
Kolmogorov-like 
regime where $\alpha\approx-5/3$  from a steeper,  power-law regime where $\alpha\approx-2$.  A remarkable finding is that both the magnetic compressibility (black curves in the right panels of Fig.~\ref{fig:PSD_IHS}) and the intermittency (Fig.~\ref{fig:SF_IHS}) reach a maximum at this scale. 
\citet{fraternale2019a} and \citet{fraternale2019b} have suggested the possibility that the spectral steepening may be ascribed to  PUI-scale kinetic processes, even though these studies did not provide a model. In compressible turbulence, the presence of a variety of compressible and incompressible coherent structures, dispersive waves \citep[e.g.,][]{perrone2016} and magnetic reconnection \citep{loureiro2017} may all contribute to an increased turbulence
cascade rate at scales smaller than the typical size of CSs, $\sim 10 \, r_{ci}$.  Although the gyroscale of thermal protons lies in the last frequency decade in \Voyager measurements, PUI-related  effects might be observed at smaller frequencies near $f_\SC\gtrsim10^{-4}$\,Hz. The spectral knee may then be ascribed to the typical size (passing times, at \Voyager) of the kinetic structures described previously in this section \citep{burlaga2006}. This may represent a ``transition scale'' from the MHD to the kinetic regime of turbulence, similar to what was observed \citep[e.g.,][]{alexandrova2008} near the ion scale in the supersonic SW. However, here the transition would occur between the PUI gyroscale and the thermal ion scale.  The reduction of compressibility and intermittency at higher frequencies is also a known feature in the SW near 1 AU, ascribed to kinetic wave activity \citep[e.g.,][]{kiyani2009,wu2013, alexandrova2013,sorriso2017,bale2019}. Here, it may be related to PUI kinetic activity but it may also be an artifact of noise in the measurements, and deserves further investigations.

At finer scales in the range $f_\SC\gtrsim5\times10^{-4}$\,Hz, all spectra flatten significantly. Certainly, this regime is affected by the presence of noise in the measurements 
\citep[$1\sigma$ errors are in the range of 0.02--0.05\,nT,][]{berdichevsky2009}, including the $1/f$ noise from the fluxgate magnetometers \citep{behannon1977}. It should also be noted that the spikes at harmonics of $3.25\times10^{-4}$\,Hz are certainly instrumental artifacts. Some spectral techniques may make them appear broader then they are. Interestingly, during some intervals,  a further spectral steepening can be observed near the cyclotron frequency of the thermal plasma ($f_\SC\gtrsim2\times10^{-3}$\,Hz). A clear example is provided in the bottom panels of Fig.~\ref{fig:PSD_IHS} for \voy1 data in 2009. Using multi-ion fluid simulations, \citet{zieger2020} suggested that fast-mode turbulence originating from nonlinearly steepened fast-magnetosonic waves may explain \voy2 observations in the high-frequency regime near the HTS. Further investigations are needed to understand which  processes dominate the transitional and the dissipation regimes of turbulence in the IHS. \citet{gutynska2010} observed a damping of higher frequencies with distance, which supports the idea that fluctuations are generated and/or amplified at the HTS. However, the persistence of large magnetic compressibility deeper in the IHS,  as shown in Fig.~\ref{fig:PSD_IHS}, poses a number of questions about the physical processes responsible for the production of high-frequency fluctuations and the turbulent energy transfer across scales.

Second order statistics (i.e. PSDs) are not sufficient to assess the presence of turbulence. The evidence of turbulence in the IHS is supported by the existence of clearly defined power-law regimes in higher order statistics such as the structure functions, $S_p[B_j](\tau)=\langle \Delta B_j^p(\tau; t)\rangle$. An example is provided in the left panel of Fig.~\ref{fig:SF_IHS} for \voy2 data in a sector region. \citet{fraternale2019b} also shown that the \textit{extended self-similarity} principle \citep{benzi1993} can be successfully used to obtain the SF relative scaling exponents from \Voyager data. The black curve in the insert shows that at \voy2 the low-frequency spectral break occurs when the condition $\mathcal{I}\equiv\langle |{\bf \Delta B}|/B_0\rangle\sim 1$ is met. As described by \citet{matteini2018} in the case of supersonic SW turbulence, the ``saturation'' of the turbulent cascade at $f<f(\mathcal{I}=2)$ is  the result of incompressible fluctuations being bounded on a sphere of radius equal to $B$, which introduces another scale in the system. This condition is not reached in unipolar regions at \voy1, where the maximum intensity is about 0.7 in the energy injection regime. 

The above observations are suggestive of the presence of a highly structured, anisotropic magnetic turbulence cascade in the IHS, but are far from being conclusive on the nature of turbulence in this region. 
What happens in the transitional regime near the ion scale and at sub-ion scales is a topic of particular interest in theoretical analysis \citep[e.g.,][]{chen2017,chen2020,bowen2020a,bowen2020b}. Further analyses of compressible,  IHS turbulence are necessary to improve our theoretical understanding of MHD turbulence and its impact on the global shape of the heliosphere.

 \newpage

\section{Turbulence in the very local interstellar medium}
\label{sec:VLISM}

\begin{wrapfigure}[31]{l}[0pt]{0.5\linewidth}
	\centering
	\includegraphics[width=1.0\linewidth]{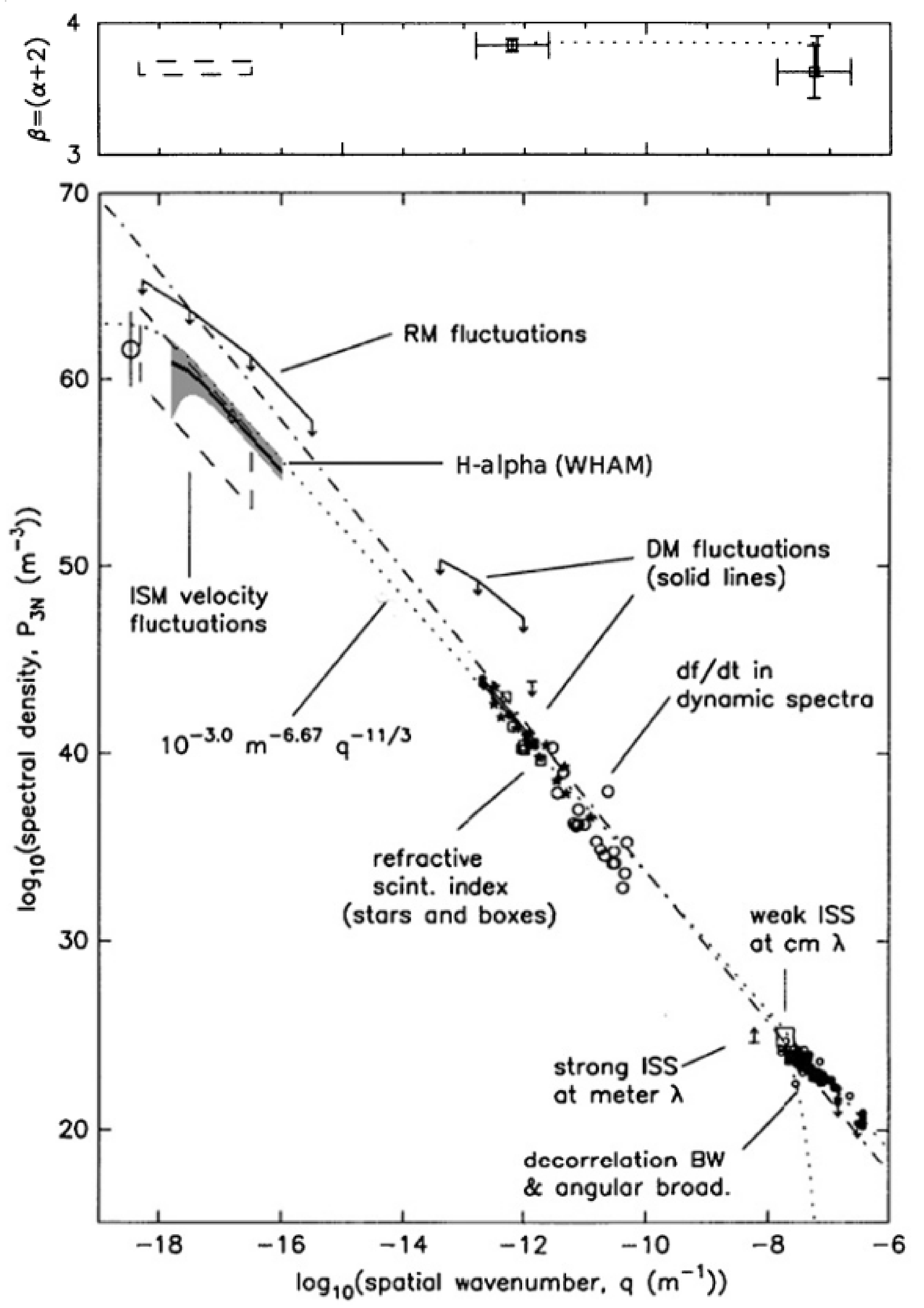}
	\caption{Electron density spectrum in the ISM obtained from various remote measurements. Note that the 3D SDF exponent ($-\beta$, in the figure) is related to the PSD exponent by $\beta=\alpha+2$. Reproduced from \citet{armstrong1995} and \citet{chepurnov2010}.}\label{fig:big_power_law}
\end{wrapfigure}
The interstellar medium has been long recognized to be turbulent on scales up to $\sim 1000 $\,pc \citep{lee1976,armstrong1981}. A signature of turbulence is the impressive power-law spectrum of density fluctuations  (dubbed ``The Big Power Law'') obtained by \citet{armstrong1995} and extended later by \citet{chepurnov2010} (see Fig.~\ref{fig:big_power_law}). The density spectral index is between $\sim -11/3$ and $-4$, which is consistent with the presence of  Kolmogorov-like turbulence on scales from $10^6$\,m to $10^{18}$\,m, but does not exclude the presence of shock-like structures. Such insights on the ISM irregularities have been obtained via remote observations, such as radio scattering and scintillations, dispersion measures, Faraday rotation measures, spectroscopy, etc. In the ISM, energy is injected over a wide range of scales from different processes, and turbulence is certainly not statistically homogeneous, as discussed in the extensively by \citet{ferriere2001}, \citet{elmegreen2004}, \citet{scalo2004}, see also \citet{redfield2004}. Recent findings on inhomogeneities in the  Local Interstellar Cloud (LIC) in which the heliosphere is immersed, are discussed by \citet{Linsky_EA_this_volume} in this volume. 

The LISM properties are affected by the presence of the heliosphere over distances of the order of several hundred \au \citep{holzer1989,zank2015}, or exceeding $1000 $\,\au, from the perspective of  TeV GCRs \citep{zhang2020}. It is reasonable to believe that the properties of interstellar turbulence  may change with the distance from the HP. 
Therefore, is should be clarified that the in situ observations are likely not representative of the entire VLISM,  but of a narrow, very dynamic region near the HP.

The first in situ observations of interstellar turbulence have been made after the crossing of the heliopause (HP) by the \voy1 spacecraft in 2012 August (122 \au)  and by \voy2 in 2018 November (119 \au) \citep[e.g.,][]{stone2013,burlaga2019nature}. A currently accepted hypothesis is that turbulence in this region consists of a superposition of the unperturbed LISM turbulence and fluctuations induced by heliospheric processes on scales ranging from the $\sim11$ year solar cycle to plasma kinetic scales \citep{burlaga2015,burlaga2018,zank2017,fraternale2019a,matsukiyo2019,lee2019,lee2020,fraternale2021a}. 

Major shock/compression waves have been observed at \voy1 
\citep{burlaga2013c,burlaga2016,burlaga2017,burlaga2021b}. 
They are transmitted into the VLISM when SW perturbations such as GMIRs hit the HP on the inner side, as was first suggested by \citet{gurnett1993} and later demonstrated in several simulations  \citep{steinolfson1994b,pogorelov1995,pogorelov2000,zank2003,washimi2011,borovikov2012,pogorelov2013a,fermo2015,kim2017b}. Compressible, finite-amplitude waves may undergo nonlinear steepening and interactions and can occasionally merge to produce larger waves \citep[e.g.,][]{pogorelov2021}. These shock-like perturbations are associated with plasma wave events and 2--3\,kHz radio emissions first detected at 1\,\au by  \cite{kurth1984,gurnett1993} and recently observed in situ by \textit{Voyager}'s Plasma Wave Subsystem (PWS)  \citep{gurnett2015,gurnett2019nature,kurth2020}. It is believed that radio emissions can be excited by interstellar shocks provided that primed nonthermal electrons exist in that region. In fact, electrons can be primed via resonant acceleration by a
lower hybrid wave (LHW) mechanism which, in turn, may be driven by instabilities of ring-beam PUI distributions  \citep{omelchenko1989, cairns2002}.  In addition, shocks and compression waves may induce the observed anisotropy of GCR proton fluxes \citep{gurnett2015,rankin2019}. Details are discussed in  \citet{Mostafavi_EA_this_volume} and \citet{Richardson_EA_this_volume} in this volume. 

An important contribution to building a theory of interstellar (Alfv\'enic) turbulence was given by \citet{sridhar1994,goldreich1995}. A model of VLISM turbulence has not been developed yet. A number of physical processes and factors should be considered to understand and model turbulence and dissipation in the VLISM.  
These include:
\begin{enumerate}
	\item Motion of the HP, i.e. the ``forcing effect'' associated with the solar activity and HP instabilities;
	\item Bow wave/shock effect on the VLISM turbulence;
	\item Shock/turbulence interactions, at MHD and kinetic scales;
	\item Coulomb collisions;
	\item Charge exchange collisions, kinetic instabilities due to the presence of supra-thermal particles;
	\item GCR/turbulence interaction;
	\item Partial ionization of the medium;
\end{enumerate}

\subsection{In situ observations of magnetic field turbulence in the VLISM}
\label{sec:VLISM-evidence}

\citet{burlaga2015} presented the first in situ observations of magnetic turbulence in the VLISM from \voy1 MAG data at the 1\,day resolution.  \citep{burlaga2015} and \citet{burlaga2018} investigated two relatively ``quiet'' intervals between shocks at a distance of $\sim125$\,\au and $\sim135$\,\au, respectively.  It was shown that (i) the intensity of turbulence is low, $\langle |\delta {\vct{B}}| / B_0 \rangle\sim 0.02$, which is close to the estimated level of systematic uncertainties ($\sim 0.03 $\,nT or larger for the $B_R$ component) and to the magnetometer's limit  \citep{burlaga2014c}.
(ii) the trace power spectra exhibit a power law with a Kolmogorov-like spectral index; (iii) on relatively large scales 
($\sim 1$--100\,days), the fluctuations were highly compressive in the 2013--2014 interval, closer to the HP, but (iv) the compressibility decreased in 2015--2016, which suggested an evolution toward an Alfv\'enic state with distance. 

It should be noticed that the terminology ``weak'' is often used improperly to indicate a low intensity of local fluctuations with respect to the average field. This condition is satisfied in the \textit{weak (wave) turbulence} phenomenology \citep[e.g.,][]{nazarenko1992book,galtier2000,schekochihin2012}, but is not sufficient. The strength of nonlinear interactions must also be evaluated \citep[see discussion in, e.g.,][]{oughtonmatthaeus20}, as well as the kind of turbulence forcing. So far, this investigation has not been possible from \textit{Voyager} data in the VLISM. 
To date, detecting weak turbulence regimes in the VLISM turbulence remains an open challenge. 

Another open science question concerns the properties of the unperturbed LISM turbulence. \citet{burlaga2015} considered the possibility that in the 2013 quiet interval turbulence was not much affected by the solar activity, and extrapolated the Kolmogorov spectrum to reach the saturation condition $\delta B/B_\mathrm{LISM}\sim1$, assuming $B_\mathrm{LISM}=0.5 $\,nT. Updated results presented by \citet{burlaga2018}, show  that the outer scale obtained from the extrapolation is
 $\sim0.01 $\,pc. It was suggested that this scale may be interpreted as the outer scale of turbulence in the VLISM. 

\begin{wrapfigure}[18]{l}[0pt]{0.5\textwidth}
	\centering
	\includegraphics[width=1.0\linewidth]{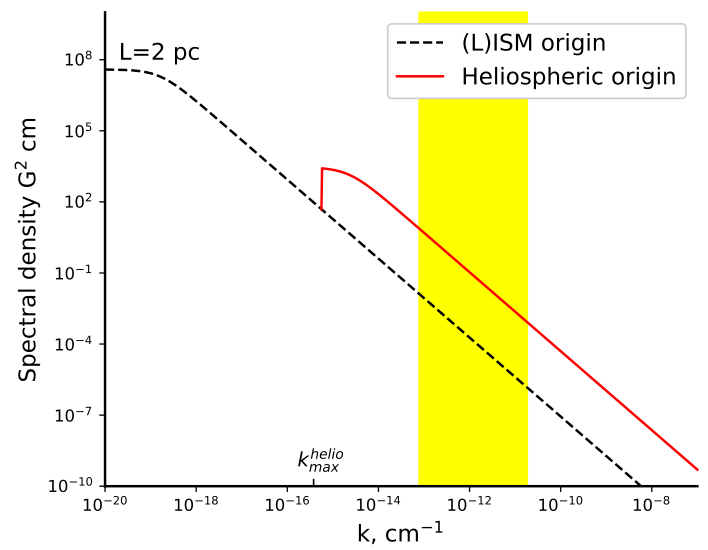}
	\caption{Illustration of the expected form of the incompressible VLISM power spectral density for transverse magnetic field fluctuations. Reproduced from \citet{zank2019}}\label{fig:zank_spectrum}
\end{wrapfigure}
Some recent studies suggested that the observed turbulence is indeed affected by the presence of the heliosphere, as far as 25\,\au from the HP.  In particular, \citet{zank2017} analyzed theoretically the ``radiation''  of small-amplitude waves by the HP into the VLISM. They have shown that IHS fast- and slow-mode waves incident on the HP generate only fast-mode waves that propagate into the VLISM. This result was confirmed by hybrid simulations of \citet{matsukiyo2019}, and provides a plausible explanation for the observed large values of compressibility of VLISM turbulence observed by \citet{burlaga2015}.  Figure \ref{fig:zank_spectrum} from \citet{zank2019} illustrates a possible scenario where the transmitted spectrum is superimposed on a Kolmogorov-type model spectrum \citep{giacalone1999} of LISM turbulence, that would have an outer scale of 2\,pc. Notice that this scale is smaller than the largest ISM scales shown in Fig. \ref{fig:big_power_law}, but may be consistent with the size of the LIC . {Interestingly, the recent investigations of the LIC properties reviewed by \citet{Linsky_EA_this_volume} reveal the presence of temperature inhomogeneities in the LIC on scales smaller than 4,000 \au. According to the scenario  depicted in Fig. \ref{fig:zank_spectrum},} the outer scale of the locally transmitted turbulence may be of the order of $\sim100$\,\au, compatible with frequencies of the order of the solar cycle.

\begin{figure}[t]
	\centering
	\includegraphics[width=0.99\linewidth]{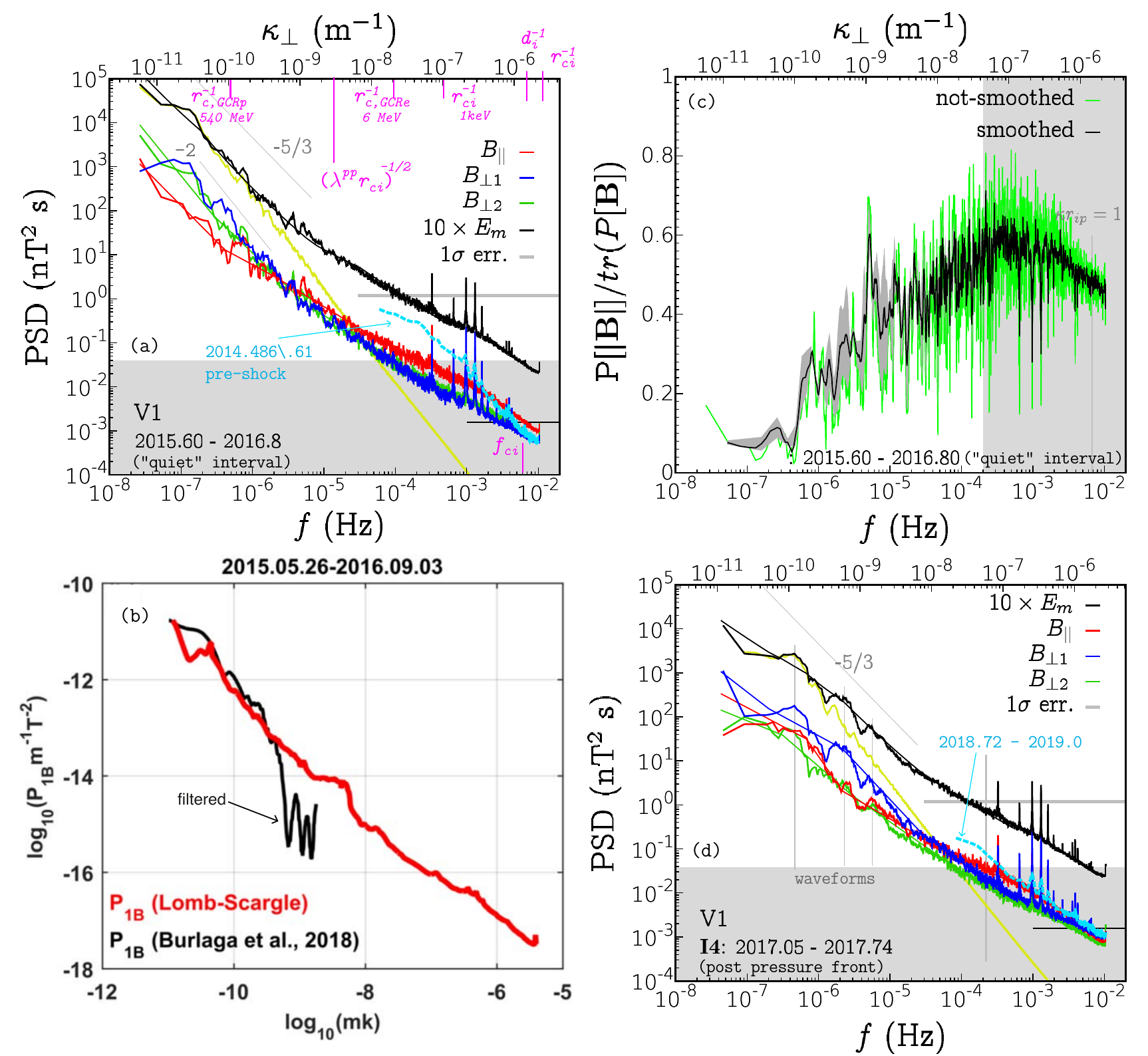}
	\caption{Magnetic field spectra in the VLISM from \voy1 data during a quiet interval (panels a,b, and c) and during a post-shock interval (panel d). Panel (c) shows the scale-dependent magnetic compressibility. Some relevant scales are reported in panel (a). Adapted from \citet{fraternale2021a,lee2020}.}\label{fig:spectraVLSIM}
\end{figure}

\citet{fraternale2019a} analyzed four \voy1 intervals out to 135\,\au, including two new periods in post-shock, ``disturbed'', regions. They used 48\,s resolution data and  were able to detect the presence of intermittency in the frequency range $10^{-6}\lesssim f_\SC \lesssim 10^{-4} $\,Hz, which is mostly observed in the transverse components of $\vct{B}$.  Spectral indices, structure function exponents, and the turbulence intensity for the parallel and perpendicular components are summarized in Table 5 of \citet{fraternale2019a}.  It was noted that a common feature of all periods is the high magnetic compressibility in the high frequency regime. For instance, see panel (d) in Fig.~\ref{fig:spectraVLSIM}. This is relevant because (i) it may indicate that physically relevant information may be obtained using high resolution \textit{Voyager} data, despite the high level of noise, and  (ii)  new questions arise about the turbulent injection and dissipation of fine scale compressible modes. It was also suggested that interstellar shocks might affect the properties of turbulence.

\citet{zank2019} investigated theoretically the nonlinear interaction of linear wave modes in the VLISM, in the NI MHD framework. In particular, two cases were considered. The first case is that a fast-mode wave with a  2D, zero-frequency mode to generate a slow-mode wave. In this scenario, the wavenumber of the latter mode would be larger by a factor $\sim V_\mathrm{f}/V_\A$. The second case is that a fast mode wave plus a zero-frequency mode generates an Alfv\'en mode with similar wavenumber. Here, the predominantly fast-mode spectrum transmitted at the HP would be replaced by an Alfv\'en-mode spectrum within a certain distance from the HP, estimated to be $\sim 10$\,\au by \citet{zank2019}. Both mechanisms may play a role in the VLISM.

\begin{figure}[t]
	\centering
	\includegraphics[width=0.66\linewidth]{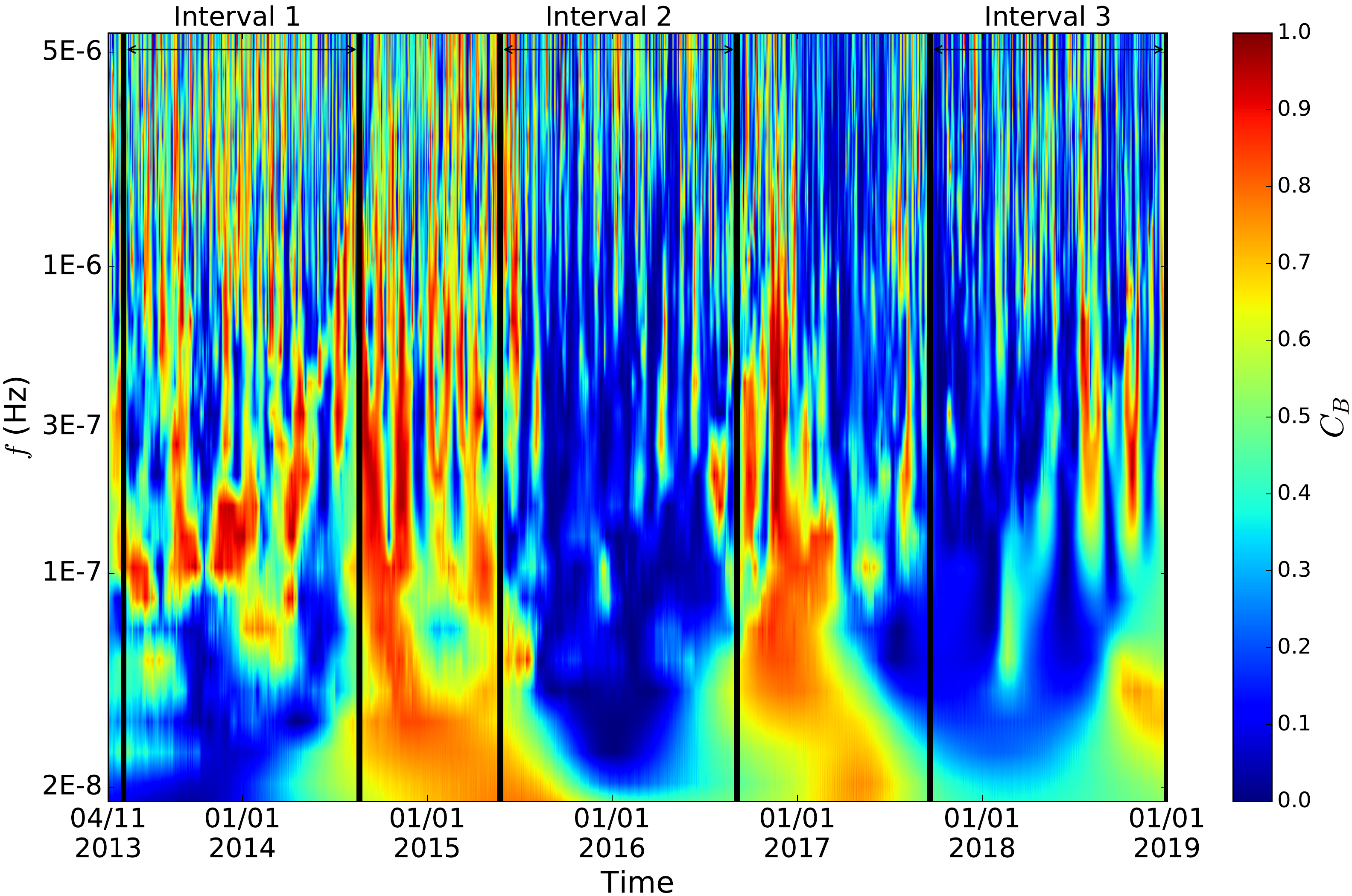}
	\caption{The wavelet spectrogram of the magnetic compressibility for \voy1 data from 2013 to 2019. Adapted from \citet{zhao2020}.}\label{fig:wavelet0}
\end{figure}

\citet{fraternale2020a} provided the first evidence of fine-scale intermittent turbulence with $|\delta \vct{B}|\sim 0.1 B$ in the precursor of the shock wave that crossed \voy1 on day 236 of 2014. In this case, the frequency range of scales of interest is  $f_\SC\gtrsim10^{-4} $\,Hz ($10^{-2} \lesssim k_\perp r_\ci \lesssim 1 $, where $r_\ci \sim 345$\,km was estimated using $T=30,000$\,K). The turbulent cascade exhibits coherent fine-scale structures compatible with a filamentary topology, and an unexpectedly steep spectrum for the transverse fluctuations.  A relevant implication of this study is that the HBL is not featureless on scales smaller the thermal proton-proton Coulomb collisional mean free path ($\lambda^\mathrm{pp}\sim 0.5-4$\,\au for $T\sim10,000$--30,000\,K). Moreover, a question is raised about if the weak subcritical shocks in the VLISM are capable of accelerating ions in their foreshock. A detailed discussion of this specific turbulence event is presented in Sec.~4.4 of \citet{Mostafavi_EA_this_volume} in this volume. Here, the trace PSD is shown in panel (d) of Fig. \ref{fig:spectraVLSIM}.

\begin{figure}[t]
	\centering
	\includegraphics[width=\linewidth]{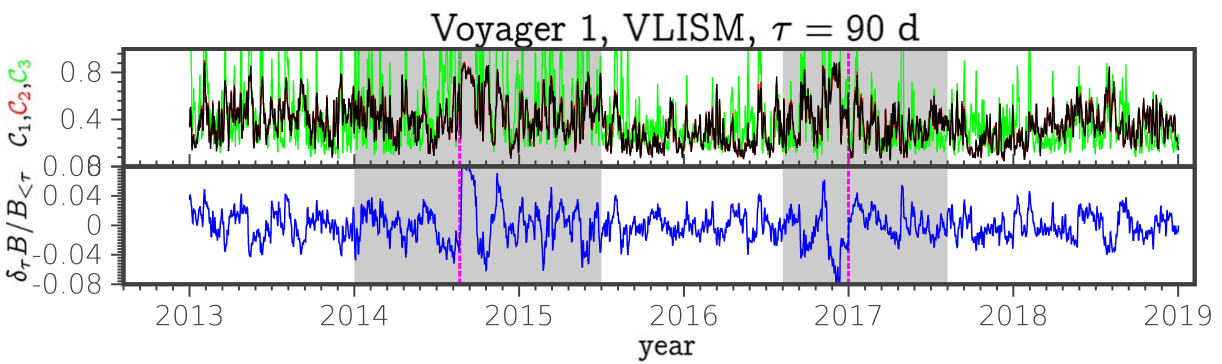}
	\caption{Time-domain magnetic compressibility from high-pass filtered data (the filter window size is 90\,days) along the \voy1 trajectory. ({Top panel}) Magnetic compressibility in the time space obtained from three different proxies (Eqs.~3--5 in \citet{fraternale2021a}). ({Bottom panel}) Normalized intensity of the fluctuations of $|\vct{B}|$. The vertical lines indicate the 2014 shock wave and the 2017 pressure front. Adapted from \citet{fraternale2021a}.}\label{fig:compressFrat}
\end{figure}

\citet{zhao2020} performed a wavelet and Hilbert spectral analysis of three ``quiet'' periods using daily data, focusing on the evolution of the magnetic compressibility. Figure \ref{fig:wavelet0} from that study shows the wavelet spectrogram of the magnetic field strength, normalized by the trace spectrogram. \citet{zhao2020} concluded that a conversion from compressible to incompressible turbulence occurs within the first interval, in particular the compressive fluctuations are confined within a spatial region of $\sim 2$\,\au from the HP. This seems in agreement with \citet{zank2019} for wavenumbers in the range $k \in [10^{-11}, 10^{-10}] \, \mathrm{m}^{-1}$. Such a decrease of compressibility with distance was also addressed by \citet{burlaga2020b}. However, some questions are still open regarding this point.  \citet{fraternale2021a} pointed out that the decrease of compressibility with distance is not smooth, and is largely due to an  increase of $\delta B_\perp$, rather than a decrease of $\delta B_\parallel$. During 2018--2019, the level of compressibility was larger than during 2015--2016 due to the presence of compressible waveforms with similar amplitude and periodicity of those observed in 2013--2014 (see Fig.~\ref{fig:compressFrat}). Compressible wave-like structures have been observed by \voy1 during 2021 out to $\sim$150 \au from the Sun \citep{burlaga2021b}. The data analyses of post-shock intervals by both \citet{fraternale2019b} and \citet{zhao2020} suggest that time dependent effects and shock-wave interactions may both affect \Voyager observations, in addition to the mode conversion processes described by \citet{zank2019}.  Data-driven, global simulations of  \citet{kim2017b} have shown that the largest compressible perturbations can reach far distances, beyond 200\,\au. An interesting aspect not yet fully understood is the scale-dependent evolution of the magnetic compressibility. In fact, $C_\mathrm{m}$ is found to grow with the frequency, reaching a maximum of $\sim0.55$ in the high-frequency regime near MHD/ion-kinetic transitional scales (see Fig. \ref{fig:spectraVLSIM}(c)), which is unlikely to be an artifact of noise.

\citet{lee2020} and \citet{fraternale2021a} performed a detailed spectral analysis of magnetic turbulence observed by \voy1 until 2019.  They described the fluctuations in the frequency range $10^{-8} \, \mathrm{Hz}\lesssim f_\SC < 10^{-2} $\,Hz ($10^{-11}\,\mathrm{m}^{-1}\lesssim k<3 \times 10^{-6}$\,m$^{-1}$) using 48\,s data at \voy1, providing further evidence of the heliospheric influence on the observed VLISM turbulence out to 150\,\au. Examples of magnetic energy spectra are shown in Fig. \ref{fig:spectraVLSIM}. Characteristic scales of the thermal plasma and energetic particles are reported in panel (a), including also the proton and electron GCR gyroscales obtained by \citet{rankin2020}. 
\citet{lee2020} also performed a combined analysis of magnetic field and plasma wave data and analyzed short time intervals in the wavenumber range $10^{-10} \mathrm{m}^{-1}<k<10^{-9}$\,m$^{-1}$. By comparing  observations and the theoretical dispersion relations of MHD modes, they concluded that at these scales no linear modes can explain \Voyager observations, and turbulence might be associated with arc/spherically polarized Alfv\'en wave modes. 

\citet{fraternale2021a}  have shown that the spectrum of fluctuations on frequencies $f_\SC\lesssim10^{-6} $\,Hz is dominated by quasi-periodic waveforms with the period of $\sim$10--100\,days ($\ell_\perp \sim 0.2$--2\,\au). 
Such oscillations determine the breaks in multi-scale statistics. Often, they have a mixed compressive/transverse nature and display ``N-wave'' profiles that may be ascribed to nonlinear steepening \citep{whitham1974,webb1993}. Wave trains are particularly intense in post-shock intervals, as noted by \citet{burlaga2016,burlaga2019}, and also by \citet{burlaga2021b}. From an estimate of the turbulent Alfv\'en Mach number ($M_\mathrm{A, {turb}} \sim 0.025$--0.1), \citet{fraternale2021a}  argued that such fluctuations may have been amplified locally due to the encounter with one or more shocks but their nature is still unclear. They estimated that the magnetic pressure fluctuations in quiet intervals, $\delta P_\mathrm{m}\sim1.5\times10^{-14} $\,Pa, is a significant fraction (up to 50\%) of the pressure jump associated with the shocks. In analogy with well known results of shock/turbulence interaction in fluids \citep[e.g.,][]{lele2009} or collisionless plasmas \citep[e.g.,][]{trotta2021}, this suggests that the structure and propagation of interstellar shocks/compression waves may be affected by turbulence (and viceversa).

Panels (a), (b), and (c) of Fig.~\ref{fig:spectraVLSIM} show results for the 2015.5--2016 quiet interval at \voy1. Panel (b) is taken from \citet{lee2020}, and also shows the PSD obtained by \citet{burlaga2018}.  Panel (d) shows spectra during 2017--2018. This interval follows a large pressure pulse that crossed \voy1 in 2017.0. At frequencies $f\SC\gtrsim5\times10^{-6} $\,Hz ($k_\perp\sim10^{-9} \, \mathrm{m}^{-1}$), a Kolmogorov-like spectral form is generally observed. However, \citet{fraternale2021a} pointed out that interpreting fluctuations as a classic Kolmogorov turbulence may be too simplistic a view, at least in the HBL. Both spectral index anisotropy and variance anisotropy are observed.  Occasionally, the spectral index is close to -2, which may be partially ascribed to a coherent-cascade scenario, an example of which is the Burgers' turbulence phenomenology.   A common feature in all \voy1 intervals is the flattening at $f\SC\gtrsim5\times10^{-6} $\,Hz. Certainly, the presence of noise in MAG data and the may affect this range of frequencies \citep{burlaga2014d}. \citet{burlaga2018} applied a low-pass filter to exclude this range.  Being aware of that, one can notice that significant anisotropy and variations of the power laws are observed in the high frequency MHD regime. Since these features are not typical of random noise, recent studies suggest that physically relevant fluctuations in the microscale regime 
($\ell \sim 1$--1000\,$d_\mathrm{i}$) may be present in \textit{Voyager} data.  A striking example is the intense foreshock turbulence event (see the light-blue curve in Fig.~\ref{fig:spectraVLSIM}(a)). A similar intensification of fine-scale  turbulence was also observed in late 2018 (panel d). Therefore, a question arises regarding whether local mechanisms may be responsible for the injection of wave power in this regime.

A relevant physical aspect that may affect turbulence in the VLISM is the plasma collisionality with respect to Coulomb collisions. Recent estimates for the transport coefficients parameters can be found in \citet{baranov2013,mostafavi2018,fraternale2021a}. The collisional transport theory holds if $\ell_\perp \gtrsim \sqrt{\lambda^{pp}\, r_\mathrm{ci}}$, and $\ell_\parallel\gtrsim \lambda^{pp}$, where we indicate with $\ell_\perp$ ($\ell_\parallel$)  the fluctuation scale in the direction perpendicular (parallel) to $\vct{B}$, {and with $\lambda^{pp}$ the proton--proton Coulomb collisional mean free path.}
Since \voy1 is likely sampling mostly $\ell_\perp$, \citet{fraternale2021a} argued that part of observed fluctuations in the high frequency range of the spectrum do not satisfy the first condition. The cutoff perpendicular scale would be in the range of 
$\sim 1$--$3\times10^{-3}$\,\au for 
$ T \approx 7,500$--30,000\,K ($\lambda^\mathrm{pp}\approx 0.5$--4\,\au). This is reported in Fig. \ref{fig:spectraVLSIM}(a). According to \citet{baranov2013}, the Reynolds number based on the bulk VLISM flow and a reference length of 100\,\au is relatively low, $\mathrm{Re} \sim 100-1000$, while the magnetic Reynolds number is $\mathrm{Re}_\mathrm{m}\sim10^{14}-10^{15}$. The later estimates of \citet{fraternale2021a} confirm the huge separation of viscous and resistive scales (large magnetic Prandtl number, $\mathrm{Pr}_\mathrm{m}\gg 1$). These peculiar conditions have been investigated by means of numerical simulations of compressible turbulence by \citet{cho2002}, and a model that was proposed by \citet{cho2003b} \citep[see also ][and references therein]{biskamp2003book}. They found the existence of a viscosity-damped regime where magnetic field fluctuations follow a shallower, $\sim k^{-1}$, spectral {form at scales smaller than viscous cutoff,  where the velocity field is smooth. Such spectral behavior resembles Batchelor's (\citeyear{batchelor1959}) viscous–convective subrange for a \emph{passive} scalar in hydrodynamics. 
However, the magnetic field is dynamically important and the analogy with the passive scalar cascade may not always be appropriate}.  The relevance of these finding to the observed VLISM turbulence is the subject of current investigation.   An estimate of the effective Reynolds number \citep{matthaeus2005}  based on the observed correlation scale and the ion inertial length gives $\mathrm{Re}_\mathrm{eff}\equiv(\ell_\mathrm{c}/d_\mathrm{i})^{4/3}\sim10^7$ \citep{fraternale2021a}.

Intermittency in the VLISM has been observed by \citet{fraternale2019a}, and a detailed investigation of quiet intervals was carried out by \citet{burlaga2020} using 1\,h increments.  They have shown that the intermittency in the transverse component is dominant in the VLISM beyond several \au from the HP. The kurtosis reaches values up to 7.  \citet{burlaga2020,burlaga2020b} also provide the first observation of VLISM turbulence at \voy2. Here, intermittency is significantly larger than at \voy1, the compressible fluctuations are dominant at $f_\SC\lesssim10^{-5} $\,Hz and the intermittency is greater for the compressible component than for the transverse one, with kurtosis values exceeding 12. In the high-frequency regime, it is typically very difficult to observe intermittency due to noise in the measurements. We note that so far such observations have only been made in the regions of enhanced fluctuations described previously, and that the actual levels of intermittency may be higher than the observed ones.  
\begin{figure}[t!]
	\centering
	\includegraphics[width=0.66\linewidth]{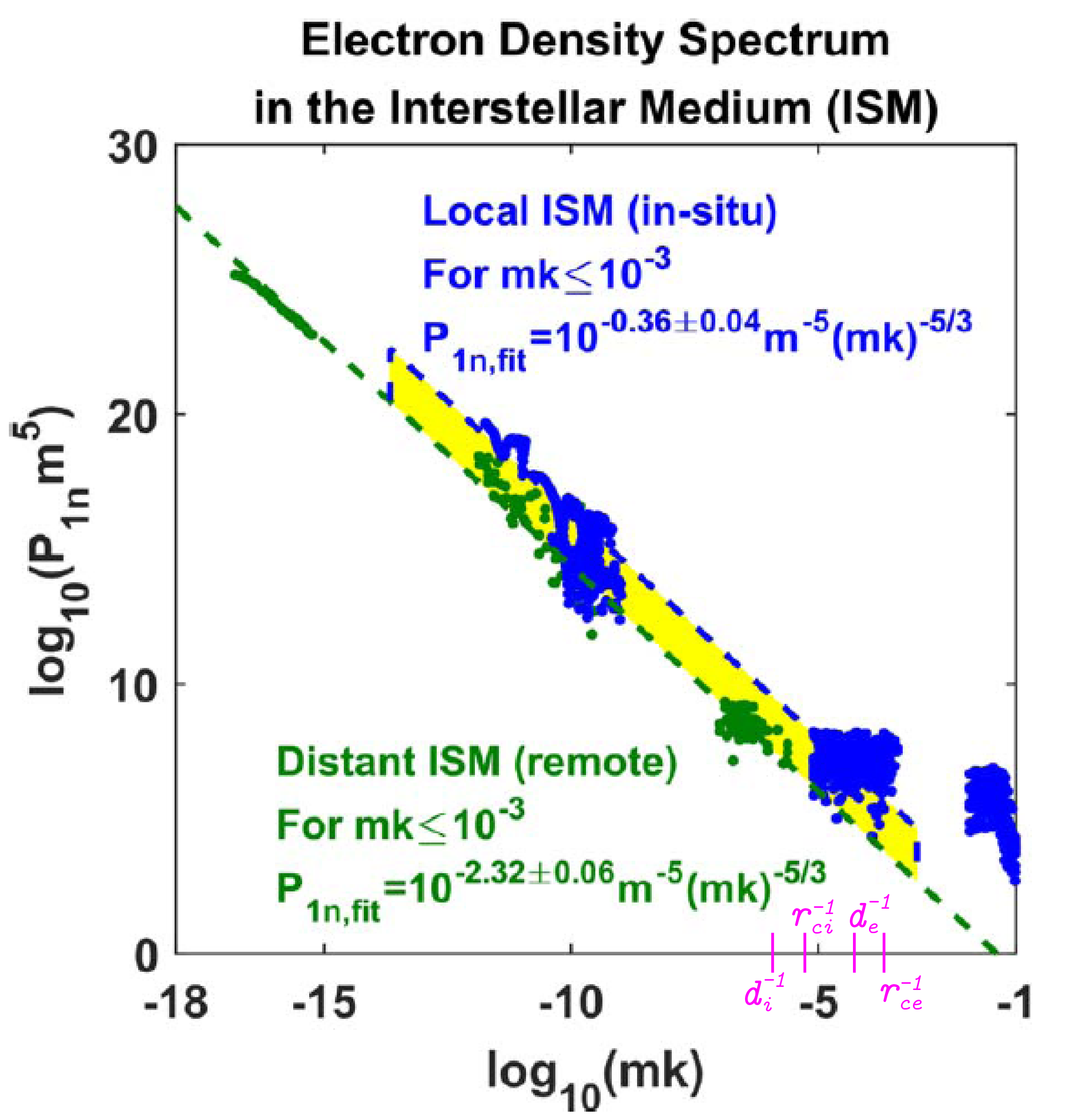}
	\caption{Combined 1D spectral density of electron density fluctuations obtained from in situ  measurements by \voy1 PWS (blue) and remote observations (green). Adapted from \citet{lee2020}.}\label{fig:elecPSD_Lee}
\end{figure}
\newpage
\subsection{In situ observations of electron density fluctuations in the VLISM}
\label{sec:VLISM-elec}

\begin{wrapfigure}[29]{r}[0pt]{0.5\linewidth}
	\centering
	\includegraphics[width=1.0\linewidth]{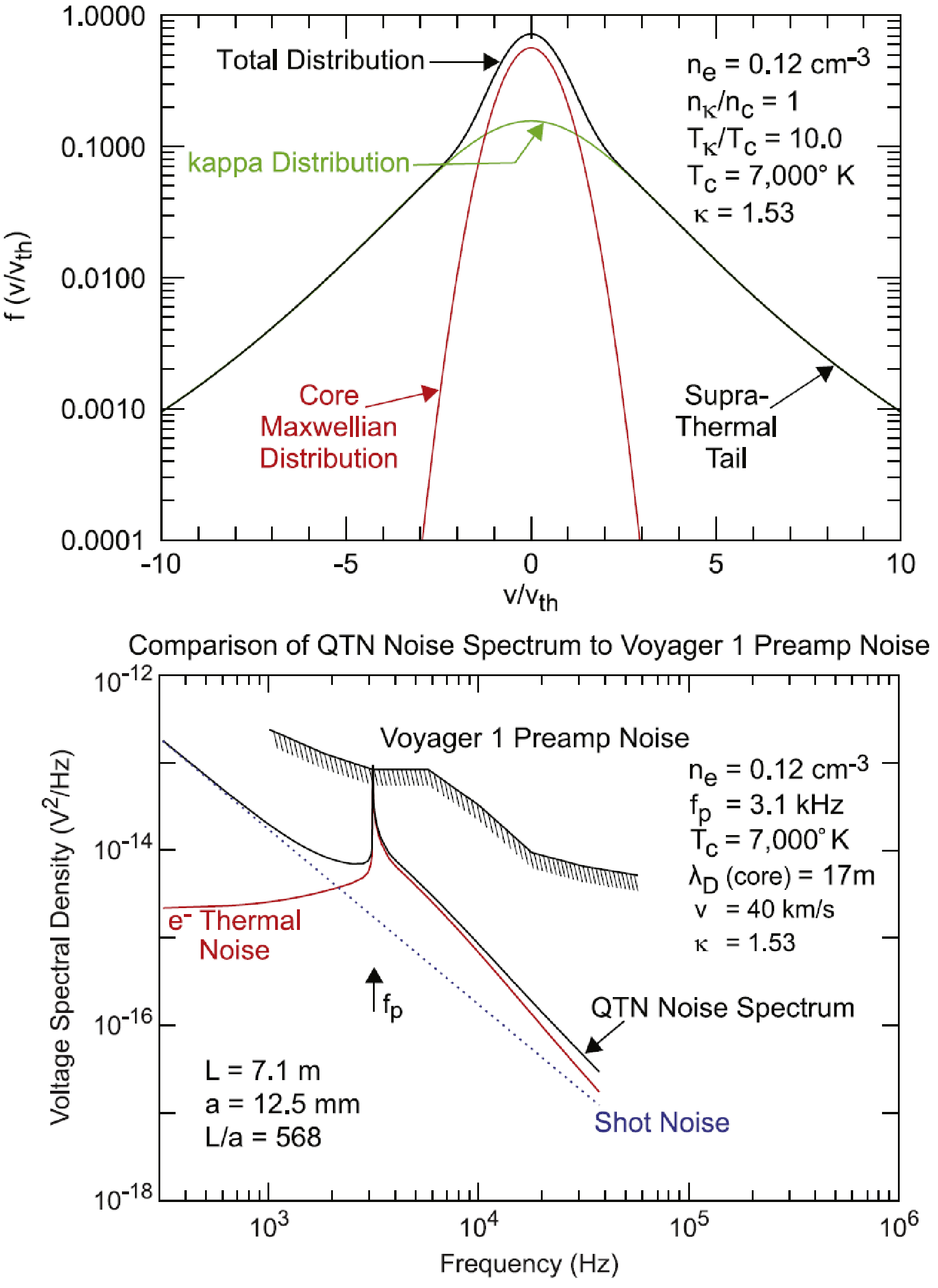}
	\caption{(Top panel) The model of electron velocity distribution used to reproduce the weak plasma wave emission line in \voy1 PWS measurements. (Bottom panel) Simulated QTN and comparison with the \voy1 PWS preamp noise. Adapted from \citet{gurnett2021}.}\label{fig:gurnettQTN}
\end{wrapfigure}

Recently, \citet{lee2019}, \citet{lee2020} and \citet{ocker2021b}  computed spectral density functions (SDFs) of electron density fluctuations using \voy1 PWS data, and compared them with the far ISM spectrum. Both enhanced plasma wave events and the recently discovered continuous emission line \citep{ocker2021a} have been used for the analysis. To date, these are the only available observations in the kinetic regime, at scales as small as $\sim 1000$\,m. The physical mechanisms responsible for the persistent plasma waves is not entirely clear, but \citet{gurnett2021} suggested that the emission is driven by suprathermal electrons that excite Langmuir waves, with minimum wavelength $\lambda \sim 700$--2300\,m and phase speed $\sim 3000$\,km\,s$^{-1}$. These oscillations are comparable to the quasi-thermal noise (QTN) but, according to \citet{gurnett2021}, QTN from a Maxwellian electron velocity distribution (with either $T=7000$\,K or 30,000\,K) cannot be detected by the effective 7.1\,m dipole antenna of \Voyager, since the Debye length is estimated to be 35\,m. Figure \ref{fig:gurnettQTN}(top) from their study shows that a core-Maxwellian and kappa distribution may explain \voy1 PWS observations (bottom panel). In this case, suprathermal electrons may contribute significantly to the pressure in the VLISM. {\citet{meyer2022} instead predict that a minute concentration of suprathermal electrons, may be sufficient to explain the observed \voy1 emission line}. What is the origin of the suprathermal tail? Is the narrowband emission caused by Brownian-motion-like fluctuations in the phase-space density of the high-energy tail, or it is due to fluctuations caused by the a turbulent cascade? These questions are subject to current scientific debate.

The SDF obtained by \citet{lee2020} is shown in Fig.~\ref{fig:elecPSD_Lee}.  At kinetic scales,  a bulge of enhanced power is observed at  $k\gtrsim 10^{-5}$\,m$^{-1}$ ($k\, d_\mathrm{i}\gtrsim 1$), which can indicate the presence of high frequency  wave activity that may include Langmuir waves, ion and electron cyclotron waves, mirror waves, or compressive kinetic Alfv\'en waves. The shallower spectral slope at such scales was also found by \citet{ocker2021b}, who also warn about the possibility that quantization of the PWS data in this regime might bias the density fluctuations. At MHD scales ($k\lesssim 10^{-10}$\,m$^{-1}$), the in situ spectrum obtained in the VLISM by \citet{lee2020} follows a Kolmogorov-like power law ($P_\mathrm{1n,in situ}=10^{-0.36\pm0.04} \mathrm{m}^{-20/3}k^{-5/3}$), and its intensity  is higher than the remote spectrum of \citet{armstrong1995}  by a factor of $\sim 90$ ($P_\mathrm{1n,remote}=10^{-2.32\pm0.06} \mathrm{m}^{-20/3}k^{-5/3}$). The power excess of in situ data was later confirmed by \citet{ocker2021b} and is consistent with the presence of compression waves. \citet{ocker2021b} have also investigated the electron density turbulence near the Guitar Nebula and other stellar bow shocks, providing evidence that density fluctuations near such bow shocks may be significantly amplified with respect to the diffuse, warm ionized medium.

\subsection{Microinstability of PUI distributions and the \IBEX ribbon}
\label{sec:VLISM-pui}

A kinetic process expected to take place in the VLISM is the instability of the ring-beam distributions of PUIs born in the VLISM by charge exchange between SW neutrals and VLISM ions. The issue of PUI isotropization in the VLISM is still highly debated.  Models predict that the ring-beam distribution is subject to a number of instabilities such as beam-driven Alfv\'en ion cyclotron (AIC), mirror and  ion Bernstein (IB), magnetosonic, ion-ion beam-core modes, reviewed by \citet{gary1991,gary1993book,gary1984}.  Hybrid simulations and instability analysis have been presented, e.g., by \citet{florinski2010,summerlin2014,florinski2016,min2018,roytershteyn2019,mousavi2020}. \citet{florinski2016} discovered a stability gap for the AIC instability. However, here mirror and IB modes may be unstable, as shown by \citet{min2018}. \citet{roytershteyn2019} simulated realistic pickup ion distributions inferred from the MS-FLUKSS global model. They found that the instability reaches the saturation condition on short time scales of hours or days, with low magnetic fluctuation amplitudes, but no isotropization was achieved on the time scales accessible to the simulations. Recently, \citet{mousavi2020} explored the excitation of possible oblique instabilities using the same  distributions of \citet{roytershteyn2019} and a larger 2D simulation domain. They have shown that mirror mode waves are also unstable, initially with a larger growth rate than AIC waves. The mirror modes were reported to produce additional scattering of PUIs and to eventually result in additional growth of AIC waves that dominate late in the simulations. It is important to emphasize that majority of existing simulations report behavior on relatively short time scales (compared to the charge exchange time in the VLISM, see below). A longer time-evolution is likely accessible only to quasi-linear analysis of the type undertaken by \citet{min2018}, rather than  to direct kinetic simulations.
\begin{figure}[t]
	\centering
	\includegraphics[width=\linewidth]{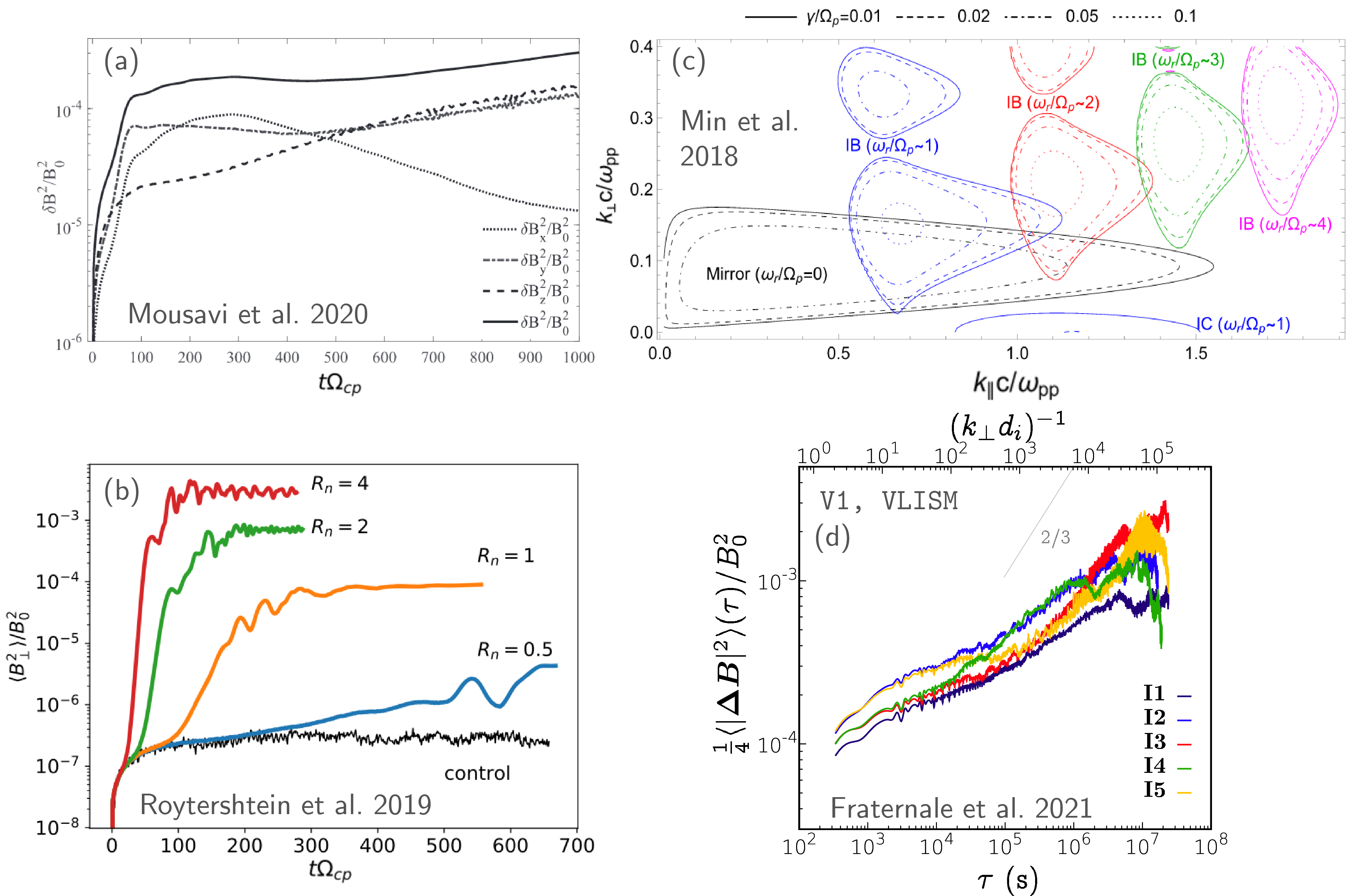}
	\caption{Amplitudes of magnetic field self-generated turbulence due to PUI instability from hybrid simulations and comparison with \voy1 observations in the VLISM. (Panels a,b) Results from 1D hybrid simulations of \citet{roytershteyn2019} and 2D simulations of \citet{mousavi2020}. The realistic PUI density in panel (b) corresponds to the density amplification factor $R_n=1$. (Panel c) growth rate contours of three instabilities in the $k_\parallel-k_\perp$ space from the linear stability analysis of \citet{min2018} (case $(v^\mathrm{ring}_{\mathrm{th},\parallel}/V_\mathrm{A})^2=0.01$). (Panel d) Normalized 2\textsuperscript{nd}--order structure functions of magnetic field fluctuations at \voy1 in the VLISM, in different periods between 2013 and 2019 (labeled as I1--I5), adapted from  \citet{fraternale2021a}.}\label{fig:PUIinstab}
\end{figure}
Can \Voyager observe signatures of PUI self-generated turbulence? The above studies agree in that the range of parallel wavenumbers of interest for the instability is $0.1\lesssim k_\parallel\, d_\mathrm{i}\lesssim 1$. 

As shown in Fig. \ref{fig:spectraVLSIM}(a) and Fig. \ref{fig:PUIinstab}(d), such a wavenumber range  may have been partially observed by \voy1. The above studies agree that the instability amplitudes at saturation conditions are in the range of $\delta B/B_0\sim 5\times10^{-5}-5\times10^{-4}$, for realistic PUI densities.   \citet{burlaga2014} and \citet{florinski2016} compared the instability amplitudes with \voy1 observations and estimated that the instability amplitude should be much larger than the level of turbulence observed by \Voyager. However, this conclusion had to be revised because of a normalization error in the PSDs of \citet{burlaga2015}. A new comparison has been made recently by \citet{fraternale2021a}. Interestingly, the normalized intensity of magnetic field fluctuations at \voy1 in the VLISM seems to be compatible with the expected amplitudes of PUI waves. Panels (a) and (b) of Fig.~\ref{fig:PUIinstab} report the normalized amplitude of self-generated turbulence in the hybrid simulations of  \citet{roytershteyn2019} and \citet{mousavi2020}. Panel (d) shows the normalized 2\textsuperscript{nd}--order structure functions of magnetic field fluctuations in different intervals at \voy1 from 288\,s averaged data. 
Since \Voyager MAG measurements in the microscale regime lie in the noisy band and are very close to the sensitivity of the magnetometers,  further investigations are needed to draw conclusions on this topic.

The VLISM turbulence and the instability process of PUI distributions are both key ingredients of the existing theories for the formation of the \IBEX ``ribbon'' of enhanced ENAs, as discussed in Sec.~3.2.3 of \citet{Galli_EA_this_volume} in this volume. 

Essentially, the fundamental and still open question is whether a quasi-anisotropic PUI pitch angle distribution in space can be reached in the VLISM, what is the role of preexisting turbulence and kinetic instabilities in the isotropization of the PUI distribution, and what is the typical time scale of the isotropization, as compared to the charge exchange collision rate.  

Both the weak scattering \citep[e.g.,][and references therein]{zirnstein2018} and the strong scattering limits have been considered. 
In the latter case, one theory predicts that the ribbon is formed within a ``retention region'', in directions  perpendicular to the ISMF, where newly ionized atoms are temporarily contained due to the enhanced scattering by the local turbulence \citep{schwadron2013b,isenberg2014,schwadron2018}. \citet{giacalone2015} suggested that magnetic mirroring may play an important role in trapping particles in pre-existing, small-amplitude compressible turbulence. \citet{zirnstein2020} extended the work of  \citet{giacalone2015} by simulating the 3D transport of PUI in the VLISM in the presence of Kolmogorov-like, homogeneous, and isotropic turbulence with a power spectrum that fits \voy1 observations by \citet{burlaga2018}. Interestingly, it was found that the structure of the ribbon can be well reproduced only if the magnetic turbulence correlation scale is smaller than 50\,\au.  This result is consistent with the idea that the ribbon source lies within a few tens of \au from the HP \citep{swaczyna2016}, and fits well the turbulence scenario the that is emerging from the recent studies.

\FloatBarrier


\section{Magnetic reconnection}\label{sec:reconnection}
\begin{figure}[t]
	\centering
	\includegraphics[width=\textwidth]{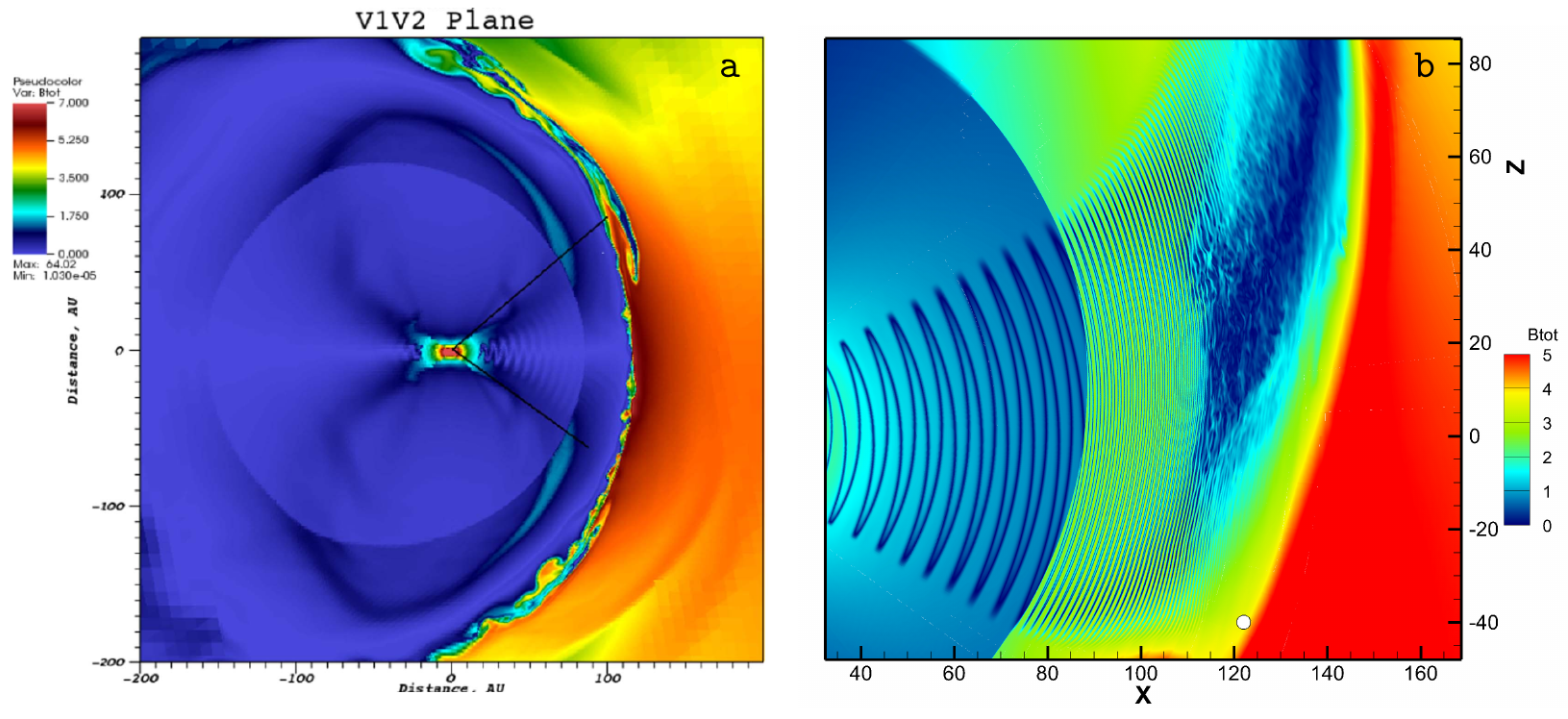}
	\caption{(Left panel) Instability and reconnection at the HP, visualization of B in the V1--V2 plane \citep[from][]{pogorelov2017b}. (Right panel) Transition to chaotic behavior in the IHS \citep[from][]{pogorelov2013a}. Magnetic field distribution (in $\mu$G) is shown in the meridional plane defined by the Sun's rotation axis and the LISM velocity vector $\mathbf{V}_\infty$. The angle between the Sun's rotation and magnetic axes is $30^\circ$.\label{fig:HCS}}
\end{figure}

It has long been suggested that magnetic reconnection at the heliopause may play a significant role in determining the overall morphology of the magnetic field, as well as the plasma transport across the heliopause~\citep[e.g.][]{fahr1986}. Indeed, simulations of \citet{pogorelov2017b} confirmed that the HP is subject to MHD instabilities and magnetic reconnection (see Fig.~\ref{fig:HCS}, left panel). Further, the sector structure of the HMF in the IHS is inherently favorable to magnetic reconnection due to both the compression of sectors and the increase of $d_i$ with distance (right panel in Fig.~\ref{fig:HCS}), which  may play a role in accelerating ACRs~\citep[e.g. ][]{lazarian2009,drake2010,opher2011}.

{As discussed by \citet{burlaga2020}, there is no compelling direct evidence in \Voyager data for magnetic reconnection, even in the distant IHS. On the other hand, observations alone are not sufficient to conclude that magnetic reconnection does not occur and, arguably, reconnection cannot really be separated from turbulence \citep{lazarian1999,eyink2011,eyink2015}. 
Some of the principal points of discussion in the literature on magnetic reconnection in the IHS include: 
\begin{enumerate}
    \item The magnetic flux $R B V_R$ apparent decrease at \voy1, not observed at \voy2 \citep{burlaga2019}. We emphasize that this flux should be conserved only for a radial, steady flow, and azimuthal magnetic field, in absence of dissipative processes. None of these conditions are actually satisfied in the IHS. On the other hand, \citet{drake2017} have shown that the magnetic flux is not necessarily reduced in a reconnecting current layer.
    \item A few ``D-sheets'', supposed to be a manifestation of reconnection in the supersonic SW, have been observed in the IHS \citep{burlaga2006,burlaga2009,burlaga2017}. Such events are rare in the data, but it may not be surprising because the spacecraft sample a tiny portion of space and time.
    \item Less HCS crossings than expected have been detected at \voy1 \citep{richardson2016}
    \item Anticorrelation of magnetic field and density fluctuations was occasionally observed at \voy2 \citep{drake2010}, consistently with reconnection. However, as discussed in Sec. \ref{sec:IHS-fine}, other physical mechanisms have been proposed.
    \item The HMF magnitude near the HP obtained from global simulations is in agreement with observations only in time dependent simulations where the HCS is allowed to dissipate (by numerical viscosity). In the assumption of unipolar HMF, its strength in the IHS is 5-7 times stronger than observed \citep[see][]{pogorelov2021,Kleimann_EA_this_volume}. This suggests that HMF dissipation occurs in the real system but quantifying the dissipation rate and pathways remains a challenge.
\end{enumerate}
}

Description of magnetic reconnection in the IHS and at the HP is an extremely difficult problem in both  global and kinetic models. Indeed, significant progress has been achieved in recent decades in understanding of magnetic reconnection in weakly collisional plasma, such as the solar wind~\citep[e.g.][]{yamada2010,ji2011,burch2016}. At the most fundamental level, the reconnection in such regimes proceeds by compressing current sheets---the regions of very large magnetic shear---all the way down to electron scales. However, in high Reynolds number systems turbulence, either self-generated or pre-existing, plays an essential role~\citep[e.g.][and references therein]{lazarian1999, eyink2011, eyink2015, daughton2012,lazarian2020} to enable magnetic reconnection on global scales. Such a coupling between microscopic and macroscopic dynamics is impossible to describe in global models. 

Even the influence of turbulence may be difficult to capture properly due to the limited resolution of such models. The only feasible approach consists of identifying potential reconnection sites in global models based on simple analytical criteria (such as large shear of magnetic field) and then studying these regions using local models with boundary and driving conditions corresponding to the regions of interest. It is beneficial to employ kinetic formalism in the local simulations and utilize fully kinetic and hybrid kinetic (kinetic ions + fluid electrons) particle-in-cell codes to study magnetic reconnection on microscopic (below 100 $d_i$) and mesoscopic scales (up to $\sim1000~d_i$).  

Even though the PUIs have relatively low density in the regions of interest, they are energetically dominant beyond the HTS. Meanwhile, previous investigations of magnetic reconnection in the IHS and at the HP either ignored the presence of PUIs, or modeled them in highly simplified manner (e.g. as an isotropic Maxwellian population with higher temperature than the background~\citep[e.g. as an isotropic Maxwellian population with higher temperature than the background,][]{drake2010}. Some relevant clues to the possible influence of PUIs are offered by the previous investigations of the role played by heavy ions, in particular O+, in the reconnection process in the Earth's magnetosphere. Such investigations revealed existence of a separate, larger, O+ diffusion region in addition to proton and electron diffusion regions, as well as offered evidence that the presence of O+ may significantly affect reconnection dynamics~\citep{liu2015}. As such, one might expect that PUIs also have a substantial effect on the structure of reconnection regions, onset of reconnection, and other important properties of the reconnection regions.

\FloatBarrier


\section{Concluding Remarks}\label{sec:conclusions}
Turbulence is critical to various heliospheric processes accompanying the SW--LISM interaction on scales ranging from $\sim 100 $\au to the electron kinetic scales. 

In this paper, we attempted to give an overview of the evidence of turbulence in the distant, supersonic SW,  heliosheath, and VLISM regions.  A long-standing problem of identifying the channels of energy transfer and turbulent heating rates of ions and electrons is perhaps the most fundamental open challenge in all regions.  Understanding the energy reservoirs in the outer heliosphere is also critical. Energetically dominant PUIs inject turbulent energy at small scales through several instability processes, but large-scale structures, instabilities, and shock waves are also present. 
At inertial range scales,
the study of variance anisotropy measurements 
at large heliocentric distances
have not provided sufficient insight into the geometry of turbulence, which raises a question about the applicability of the ``standard'' paradigms based on near-Earth observations. 
 
In the IHS and VLISM, the best evidence of turbulence so far has been the observation of intermittency of magnetic field fluctuations, power-law regimes for spectra and higher order structure functions, and multifractal statistics. The prominent feature of turbulence in these regions is its compressible nature. This also constitutes the major challenge from the modeling perspective.  Recent theoretical and numerical investigations have provided us with convincing explanations on  how the termination shock and the heliopause can generate compressible turbulence in the IHS and VLISM. 

However, the mechanisms responsible for the {in situ} origin of microscale compressible fluctuations  are less clear. In the IHS, they have been associated with local temperature-anisotropy-driven instabilities. What is the contribution of these structures to the overall cascade rate? How are large-scale traveling wave modes coupled to turbulence? How do turbulence and reconnection operate to dissipate the HCS, especially near the HP? To date, these questions remain largely unsolved.

Recent studies suggest that the turbulence observed in the VLISM is a superposition of the fluctuations emanated by the HP and the background LISM turbulence.  However, discriminating between these two components has not been possible so far, which leaves the question about the properties of the unperturbed LISM turbulence open. Some features of the VLISM turbulence remain obscure. For instance, what is the dissipation rate of compressible turbulence? What is the origin of quasiperiodic oscillations and N-wave profiles observed at MHD scales? What is the role of turbulence in the VLISM shock formation, structure and propagation? What processes are dominant at ion scales and what are the isotropization time scales of suprathermal particle distributions? What is the origin of suprathermal electrons and Langmuir waves? What specific turbulence cascade phenomenology can describe the VLISM turbulence?

We only have partial answers to these questions. 
Pursuing the investigation of turbulence in the outer heliosphere will allow the space physics community to make a step forward in the understanding of turbulence as a basic physical process.  Such progress should shed light on the mechanisms of magnetic field dissipation in the IHS and on the dynamical effects of turbulence on the SW flow.


\begin{acknowledgements}
This work was made possible by the International Space Science Institute
and its interdisciplinary workshop ``The Heliosphere in the Local Interstellar Medium''\footnote{\url{www.issibern.ch/workshops/heliosphere}}.  FF was supported by NASA grants 80NSSC19K0260, 80NSSC18K1649, 80NSSC18K1212.
C.W.S.\ is supported by Advanced Composition Explorer Mission as well as NASA grants NNX17AB86G, 80NSSC17K0009, and 80NSSC18K1215.
H.F., J.K., and S.O. acknowledge support for their work within the framework of the DFG (Deutsche Forschungsgemeinschaft) grant FI~706/23-1. 
NP was supported, in part, by NASA grants 80NSSC19K0260, 80NSSC18K1649, 80NSSC18K1212, NSF-BSF grant PHY-2010450, and by the IBEX mission as a part of NASA’s Explorer program.
GPZ, LA, and LZ acknowledge the partial support of the NSF EPSCoR RII-Track-1 Cooperative Agreement OIA-1655280.

\end{acknowledgements}

%
%

\bibliographystyle{spbasic}      


\end{document}